\documentclass[pra,twocolumn,10pt,floatfix,longbibliography]{revtex4-1}
\usepackage[english]{babel}
\usepackage{amsmath,amssymb,bm,enumerate,textcomp}
\usepackage{graphicx,xcolor,xspace,hyperref}
\xspaceaddexceptions{]\}+}

\newcommand{\R}{\mathbb{R}}
\newcommand{\Z}{\mathbb{Z}}
\renewcommand{\H}{\mathcal{H}}

\newcommand{\M}{\mathcal{M}}

\newcommand{\Var}{\mathrm{Var}}

\renewcommand{\d}{\mathrm{d}}
\newcommand{\od}{\mathrm{od}}
\newcommand{\xy}{\mathrm{xy}}
\newcommand{\I}{\ensuremath{\mathrm{I}}\xspace}
\newcommand{\II}{\ensuremath{\mathrm{II}}\xspace}
\newcommand{\X}{\ensuremath{\mathcal{X}}\xspace}
\newcommand{\Y}{\ensuremath{\mathcal{Y}}\xspace}
\newcommand{\m}{\ensuremath{\mathfrak{m}}\xspace}
\newcommand{\eP}{\ensuremath{\hat{e}_+}\xspace}
\newcommand{\eM}{\ensuremath{\hat{e}_-}\xspace}
\newcommand{\p}{\mathrm{Prob}}
\newcommand{\e}{\mathrm{e}}
\renewcommand{\b}{\mathrm{b}}
\newcommand{\w}{\mathrm{w}}
\newcommand{\sw}{\mathrm{sw}}
\newcommand{\td}{{2\mathrm{d}}}

\newcommand{\U}[1]{\ensuremath{\operatorname{U(}\!#1\!\operatorname{)}}\xspace}
\newcommand{\SU}[1]{\ensuremath{\operatorname{SU(}\!#1\!\operatorname{)}}\xspace}

\newcommand{\ev}[1]{\langle#1\rangle}

\begin{document}

\title{Infinite randomness with continuously varying critical exponents in the random XYZ spin chain}

\date{\today}

\author{Brenden Roberts}
\author{Olexei I.~Motrunich}
\affiliation{Institute for Quantum Information and Matter,\\California Institute of Technology, Pasadena, CA 91125}

\begin{abstract}
We study the antiferromagnetic XYZ spin chain with quenched bond randomness, focusing on a critical line between localized Ising magnetic phases.
A previous calculation using the spectrum-bifurcation renormalization group, and assuming marginal many-body localization, proposed that critical indices vary continuously.
In this work we solve the low-energy physics using an unbiased numerically exact tensor network method named the ``rigorous renormalization group.''
We find a line of fixed points consistent with infinite-randomness phenomenology, with indeed continuously varying critical exponents for average spin correlations.
A self-consistent Hartree--Fock-type treatment of the $z$ couplings as interactions added to the free-fermion random XY model captures much of the important physics including the varying exponents; we provide an understanding of this as a result of local correlation induced between the mean-field couplings.
We solve the problem of the locally-correlated XY spin chain with arbitrary degree of correlation and provide analytical strong-disorder renormalization group proofs of continuously varying exponents based on an associated classical random walk problem.
This is also an example of a line of fixed points with continuously varying exponents in the equivalent disordered free-fermion chain.
We argue that this line of fixed points also controls an extended region of the critical
interacting XYZ spin chain.
\end{abstract}

\maketitle

\section{Introduction}
\label{sec:introduction} 

In many situations, phases of many-body quantum systems are stable under weak static, or ``quenched,'' disorder in the presence of a gap,
and the disorder average of certain quantities can be calculated in a related clean system via either the replica trick or supersymmetry arguments for non-interacting models \cite{mezard1987spin,efetov2010supersymmetry}.
However, these methods are not suitable for relevant disorder, or disorder along with interactions, which together produce a rich variety of behaviors.
In contrast, real-space thinking should be suitable for directly accounting for spatial inhomogeneity.
Interestingly, strong disorder causes certain classes of disordered systems to become tractable on long scales, making real-space renormalization group (RG) approaches amenable to analytical treatments controlled by the flow to infinite randomness.
In this work we investigate a modern application of real-space RG to a random XYZ spin chain \cite{fisher1994random, slagle2016disordered}, where we use exact numerics to perform unbiased exploration and validation, and also use the strong-disorder renormalization group (SDRG) to demonstrate and characterize such fixed points using the language of random walks.

The original development of a real-space RG appropriate for strong-disorder physics in one dimension (1d) is due to Ma, Dasgupta, and Hu \cite{ma1979random,*dasgupta1980low}.
The feature distinguishing SDRG from, e.g., spin blocking, is that effective degrees of freedom are explicitly associated with an energy scale rather than with a spatial grouping.
In this way the disorder realization determines the pattern of integrating out fluctuations.

Such an approach is now understood to be well-motivated by the idea of an \emph{infinite-randomness fixed point} (IRFP), a stable solution of the SDRG equations discovered by Fisher in Refs.~\cite{fisher1992random,fisher1994random,fisher1995critical} at which effective disorder strength grows with the scale without bound, and SDRG predictions become asymptotically exact.
In an IRFP, disorder dominates the low-energy physics and physical observables are not self-averaging; average behaviors are instead often determined by rare regions within a disorder realization.
Interestingly, although such fixed points lack conformal symmetry, the phenomenology can resemble that of CFT fixed points: for instance, the scaling of average entanglement follows the conformal form with an effective central charge which in some cases is related to the central charge of the clean theory (but does not obey the same rules under RG) \cite{refael2004entanglement,bonesteel2007infinite,fidkowski2008c}.

Since its introduction, the SDRG has been specialized to a variety of classical and quantum systems, and the original scheme has seen many generalizations; see recent reviews \cite{igloi2005strong,*igloi2018strong}.
For example, applications in two-dimensional (2d) random models also yield IRFPs in these settings \cite{senthil1996higher,pich1998critical,fisher1999phase,motrunich2000infinite,motrunich2002particlehole,sanyal2016vacancy,bhola2020dulmage}.
In another direction, SDRG methods were extended to treat all eigenstates of a quantum Hamiltonian \cite{pekker2014hilbert,vasseur2015quantum,you2016entanglement,monthus2018strong}, in order to assess the possibility of many-body localization (MBL) of excited states.
(There are by now multiple reviews of MBL, for instance see Refs.~\cite{nandkishore2015many,abanin2019colloquium}.)
The many-body extended SDRG procedures do not perform an iterative targeting of the low-energy space, but instead tabulate emergent conservation laws corresponding to the local integrals of motion of an MBL phase; nevertheless, the equations are formally quite similar to the original picture implementing a more traditional RG.

One of the extended many-body SDRG procedures, the ``spectrum bifurcation renormalization group'' (SBRG) developed in Ref.~\cite{you2016entanglement} for Hamiltonians comprising Pauli strings, was applied to the random XYZ spin chain by \citet{slagle2016disordered}.
There, along a phase boundary between localized Ising antiferromagnets (proposed to be MBL), disorder- and energy-averaged Edwards--Anderson spin correlations were found to decay as power laws with continuously varying critical exponents.
Average entanglement entropy scaling also exhibited a stable effective central charge.
The phase transition was conjectured to be ``marginal MBL,'' meaning that eigenstates do not thermalize but exhibit a logarithmic violation of the area law.
However, it has recently been argued that such marginal MBL Hamiltonians are perturbatively unstable to ergodicity at finite energy density due to resonances \cite{moudgalya2020perturbative,ware2021perturbative}.
As is true of all excited-state SDRG schemes, Refs.~\cite{you2016entanglement,slagle2016disordered} rely on MBL for validity, and these recent arguments call this assumption into question.

In the present work we investigate the SBRG findings using unbiased numerics for the ground state and low-energy excited states.
We emphasize that our focus is entirely on low-energy properties, and we will not have anything to say about MBL physics at arbitrary energy density.
However, we find the possibility of continuously varying power laws in IRFPs already very interesting and worth further study.
The random XYZ chain---while suspected to support infinite-randomness phenomenology in Fisher's original work,
Ref.~\cite{fisher1994random}---has eluded understanding due to the lack of a closed-form SDRG solution, and developing a stronger grasp of such instances would constitute an important advance.

Strongly disordered models pose an especially difficult challenge for unbiased numerics, and have long been recognized as among the only 1d models to be resistant to standard methods, chiefly the density matrix renormalization group (DMRG).
We apply a relatively new tensor network numerical method named the \emph{rigorous renormalization group} (RRG) to this problem, as it has already been shown to be effective in the related random XY model \cite{roberts2017implementation}.
Our goal for the unbiased tensor network computations is to test the findings of Ref.~\cite{slagle2016disordered}, and better understand the disordered fixed points associated with the critical line.

As a brief overview of our results, the data found by RRG are in support of both infinite-randomness physics as well as continuously varying critical indices for disorder-averaged correlations.
These conclusions are based on direct measurements in MPS, along with scaling of low-energy spectral gaps, which we solve for in the various symmetry sectors of the model up to systems of length $N=80$ spins.
Our findings are in general agreement with the SBRG results, namely, that critical indices controlling decay of correlations, as well as long-range mutual information, vary along the critical line, while the ``central charge'' is fixed.
We additionally study the critical exponent $\psi$, which characterizes IRFP dynamics through the relationship $\log(1/E) \sim L^\psi$
between energy scale and length,
and find that its value is close to, but may be varying away from, the free-fermion fixed point with $\psi = \frac12$.

\begin{figure}[ht]
\includegraphics[width=0.9\columnwidth]{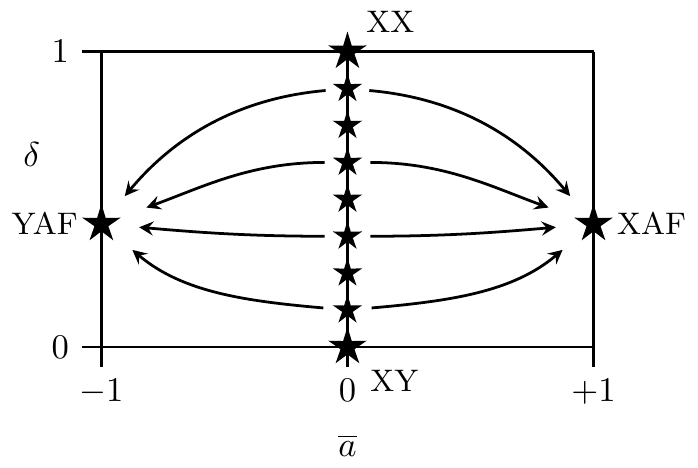}
\caption{\label{fig:rg_cartoon1}
Shown is an updated version of the schematic RG flow of XY antiferromagnets in Fig.~4 of Ref.~\cite{fisher1994random}.
In this work we prove the line of fixed points along the exactly marginal direction $\delta$, which describes the degree of correlation between bond terms $J^x$ and $J^y$, in the notation of Eq.~\eqref{eq:Hxyz}. 
(Note that $\delta = 1$ corresponds to $\sigma_a^2 = 0$ in Fisher's notation.) 
The average anisotropy $\overline a$ is as defined in Ref.~\cite{fisher1994random}; in the present work we consider only the line $\overline a = 0$.
}
\end{figure}

These numerical results for the critical line are captured reasonably well by a self-consistent Hartree--Fock mean-field that treats $J^z$ couplings as interactions added to the free-fermion XY chain 
[throughout, $J_j^{x,y,z}$ refer to terms in the XYZ chain as in Eq.~(\ref{eq:Hxyz})]; the Hartree-Fock also apparently produces continuously varying exponents.
This finding motivates study of a ``locally-correlated'' XY chain with correlations only between terms on the same link of the lattice.
The locally correlated model again exhibits similar behavior, and its SDRG structure has an advantageous mathematical connection to the theory of random walks.
Within this setting we write rigorous bounds fully determining the critical exponent for power-law decay of a certain average spin correlation function.
This exponent indeed varies continuously, proving that the free-fermion critical line of the locally-correlated model is marginal, and is described by a line of IRFPs.
This result resolves a question posed by \citet{fisher1994random}, as illustrated in Fig.~\ref{fig:rg_cartoon1}.
In this figure, we parameterize correlations between $J_j^x$ and $J_j^y$ by a generic parameter $\delta$ varying between $\delta = 0$ (completely uncorrelated or XY model) and $\delta = 1$ (completely correlated or XX model) [for a specific example, see Eq.~(\ref{eq:sdrg_xy})]; deviation of $\delta$ from $1$ can also be viewed as introducing random anisotropy to the XX model.

\begin{figure}[ht]
\includegraphics[width=0.9\columnwidth]{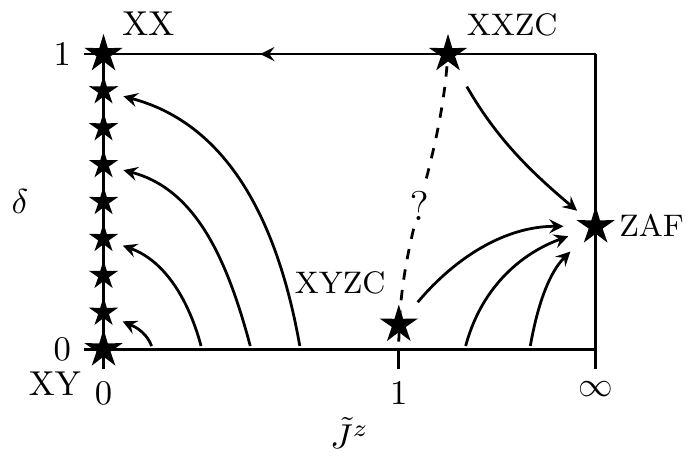}
\caption{\label{fig:rg_cartoon2}
We propose the following schematic flows for the XYZ antiferromagnet, where $\delta$ is the degree of correlation between local $J^x$ and $J^y$ couplings [as defined in Eq.~\eqref{eq:Hxyz}] and $\tilde J^z$ is the bandwidth of the $J^z$ distribution, with statistical isotropy corresponding to $\tilde J^z = 1$.
The line of fixed points at $\tilde J^z = 0$ is the same as in Fig.~\ref{fig:rg_cartoon1}, and $\tilde J^z$ is argued to be perturbatively irrelevant.
We conjecture that any $\tilde J^z < 1$ is irrelevant at $\delta=0$, but through generation of finite $\delta$ flows to the line of non-interacting IRFPs.
The methods we employ cannot access the statistically isotropic XYZC or \U1-symmetric XXZC fixed points, but XXZC was previously described by \citet{fisher1994random}.
The flows on the dashed line between XYZC and XXZC lie on a manifold separating the basins for XY and ZAF, which is not well described by this slice through parameter space.
We avoid any specific conjecture on this matter but remark that it is an interesting topic for further study.
}
\end{figure}

Returning to the interacting model, based on the above understanding of the noninteracting case and the RRG numerical data, we conjecture that at least in the neighborhood of the free-fermion model, interactions are irrelevant and the local correlations generated in the SDRG drive the interacting theory to the line of noninteracting IRFPs at long distances.
This scenario is presented in Fig.~\ref{fig:rg_cartoon2} and represents our conjectured explanation for the continuously varying critical exponents in the XYZ chain.

The outline of this paper is as follows.
In Sec.~\ref{sec:random_xyz} we present the XYZ spin model and summarize the history of its SDRG, along with explicitly developing the RG rules in the many-body language.
In Sec.~\ref{sec:unbiased_study} we perform an unbiased study of the ground state using RRG.
In Sec.~\ref{sec:mf}, based on our numerical results, we develop both a Hartree--Fock mean-field theory and the free-fermion locally correlated effective model.
In Sec.~\ref{sec:sdrg_correlated}, we use a picture of the SDRG procedure in terms of random walks to prove continuously varying critical exponents in the locally correlated effective model.
In Sec.~\ref{sec:fp_interacting} we conjecture a possible long-distance fate of the RG flow for the critical XYZ spin chain, and finally in Sec.~\ref{sec:discussion} we discuss the implications of all of these results taken together.

\section{Random XYZ model and review of previous SDRG results}
\label{sec:random_xyz}

\subsection{Spin chain Hamiltonian}
\label{sec:spin_chain}

As our most general model we consider the antiferromagnetic XYZ spin chain with quenched randomness in all couplings; that is,
\begin{equation}
H = \sum_{j=1}^{N-1} \left( J^x_j \sigma^x_j \sigma^x_{j+1} + J^y_j \sigma^y_j \sigma^y_{j+1} + J^z_j \sigma^z_j \sigma^z_{j+1} \right) ~.
\label{eq:Hxyz}
\end{equation}
The couplings $J^\alpha_j > 0$, $\alpha = x,y,z$, are independent.
This model generically has a $\Z_2\times\Z_2$ global symmetry, with generators given by the Ising-type operators $g_x = \prod_{j=1}^N \sigma^x_j$ and $g_y = \prod_{j=1}^N \sigma^y_j$.
In particular, local field terms are excluded by this symmetry.
This model also respects time reversal on the spins, which we implement as $g_y \mathcal K$, where $\mathcal K$ is complex conjugation in the $z$ basis.

We impose the same functional form on the disorder distributions of $J_j^x$, $J_j^y$, and $J_j^z$ (though delay specification until Sec.~\ref{sec:unbiased_study}), with bandwidths specified by a set of parameters $\tilde J^x, \tilde J^y, \tilde J^z > 0$.
If the value of any one of these is larger than the other two, the ground state of the model displays Ising antiferromagnetic (AFM) order.
As we are considering strong disorder, we anticipate that these phases are localized.
If two bandwidths are equal and of the largest magnitude, the model lies on a boundary between localized phases with distinct types of magnetic order; we will primarily consider this case.
If all three disorder bandwidths are equal, the model has a statistical $S_3$ permutation symmetry and sits at a tricritical point in the phase diagram \cite{fisher1994random,slagle2016disordered}.

Many exact results are known for phases of the Hamiltonian Eq.~\eqref{eq:Hxyz} in certain limits, and we provide a brief recap here.
The SDRG was in fact originally introduced by Ma, Dasgupta, and Hu in order to study the random Heisenberg antiferromagnet with \SU2 symmetry \cite{ma1979random,*dasgupta1980low}, achieved in the present notation by fixing $J^x_j=J^y_j=J^z_j$ for all bonds $j$.
These works argued for the asymptotic development of a power-law singularity in the distribution of couplings and computed leading contributions to critical indices, which vary slowly along the flow.

\citet{fisher1994random} generalized this analysis to account for anisotropy and performed a thorough study of the resulting phase diagram.
The SDRG rules for the random XX model ($J^x_j = J^y_j$ and $J^z_j = 0$ for all $j$), which breaks the \SU2 spin rotation symmetry to a \U1 subgroup, are very similar to those of the isotropic model, and in particular both realize \emph{random-singlet} (RS) phases \cite{bhatt1982scaling}.
In the ground state the microscopic spins are paired up into singlet states at arbitrarily long scales.
Correlations between the spins in a singlet are of order unity, and are strongly suppressed with the rest of the system.
Thus typical spin correlations are short-ranged, whereas the average correlations are dominated by rare paired spins.
This is one hallmark of an IRFP: that a distribution which is broad on a logarithmic scale leads to exponential separation between typical and averaged properties of the state.
From the density of paired spins one finds that average spin correlations exhibit power-law decay, scaling as $r^{-2}$ for separation $r$.
This defines the XX fixed point exponents $\eta_x = \eta_y = \eta_z = 2$.
The characteristic energy scale of the singlets in the RS phase follows
\begin{equation}
\log (1/E) \sim L^\psi~,
\label{eq:psi_def}
\end{equation}
where $\psi = \frac 12$.
As a consequence for the density of states, the dynamical exponent is formally infinite.

The random XY chain (i.e., independent $J_j^x$ and $J_j^y$ but with $\tilde J^x = \tilde J^y$, $\tilde J^z = 0$), in contrast, does not realize the RS phase.
With the mean in-plane anisotropy $\tilde J^x - \tilde J^y$ serving as the quantum control parameter, \citet{fisher1994random} computed the critical exponents $\nu = 2$ and $\beta = 3 - \sqrt 5$ for the transition separating Ising $x$- and $y$-AFM phases.
This was accomplished through a lattice duality mapping to two decoupled copies of the random transverse-field Ising model (RTFIM), whose SDRG equations are also well-studied \cite{fisher1992random,fisher1995critical,fisher1998distributions}.
Translating the RTFIM results to the present XY chain, at the phase transition the critical exponent for the decay of $x$ and $y$ components of spin correlations is $\eta_x = \eta_y = 4 - 2 \phi$, where $\phi = \frac{1+\sqrt 5}{2}$ is the golden ratio.

Starting from the opposite limit of the XX model, with $J_j^x = J_j^y$ for all $j$, it was also found by \citet{fisher1994random} that weak random in-plane anisotropy, which moves along the phase transition toward the XY point, is a marginal perturbation.
It was not clear whether this is the case along the entire phase boundary, and we will in fact be led to take up this question in some detail in Sec.~\ref{sec:sdrg_correlated}.

The set of exponents for disorder-averaged spin correlations can be completed using the mapping of the XX and XY models to free fermions \cite{igloi2000random}.
For the anisotropic model with $S_2$ permutation symmetry, $\eta_z = 4$.
In a chain with open boundaries, consideration of the form of the surface magnetization leads to the scaling of the end-to-end spin correlations $\eta^\e_x = \eta^\e_z = 1$ for the XX model and $\eta^\e_x = 1$, $\eta^\e_z = 2$ for the XY model.

Focusing on a different type of spin chain, \citet{damle2002permutation} studied permutation-symmetric multicritical points arising from effective low-energy theories of partially dimerized spin-$S$ models with \SU2 symmetry.
They performed a fixed-point analysis of the SDRG equations for degrees of freedom localized at the boundaries between distinct domains of $n=2S+1$ different types of local order (i.e., topological phases distinguished by the properties of edge modes localized near the ends of open chains).
Their primary result is a generalization of the $n=2$ random-singlet criticality to a countably infinite set of IRFPs with critical exponents $\psi = \frac 1n$ and $\nu = \frac{2n}{\sqrt{4n+1}-1}$.
The permutation symmetry refers to the interchange of distributions for the different types of order, which mediate effective couplings between the domain walls.
While the permutation-symmetric tricritical point at $\tilde J^x = \tilde J^y = \tilde J^z$ in our model shares the statistical symmetry of these theories for $n=3$, its microscopic details are dissimilar and it is not clear {\itshape a priori} whether this category of universality applies.
Indeed, our estimates of the exponent $\psi$ at the XYZ tricritical point in Sec.~\ref{subsubsec:gap} appear to rule out the applicability of the Damle--Huse universality in this case.

\subsection{Majorana representation}
\label{sec:majoranas}

Aspects of this problem become more evident in the language of fermions, for which we use the Jordan--Wigner transformation.
Equation \eqref{eq:Hxyz} maps to a spinless $p$-wave superconductor with density-density interactions:
\begin{align}
H &= \sum_{j=1}^{N-1} (t_j c^\dag_j c_{j+1} + \Delta_j c^\dag_j c^\dag_{j+1} + \mathrm{H.c.}) \nonumber \\
&\qquad\qquad\qquad+ J^z_j (2n_j - 1) (2n_{j+1} - 1)~,
\end{align}
which has position-dependent hopping $t_j = J^x_j + J^y_j$ and pairing potential $\Delta_j = J^x_j - J^y_j$.
Following the idea of \citet{kitaev2001unpaired,motrunich2001griffiths}, it is enlightening to introduce two species of Majorana fermion,
\begin{equation}
\eta_j = c^\dag_j + c_j~~\text{and}~~\zeta_j = \frac{1}{i}(c^\dag_j-c_j)~.
\label{eq:majoranas}
\end{equation}
The $\eta_j$ and $\zeta_j$ are Hermitian, and normalized so that $(\eta_j)^2 = (\zeta_j)^2 = 1$.
In terms of these operators the Hamiltonian is written
\begin{equation}
H = \sum_{j=1}^{N-1} i J^x_j \zeta_j \eta_{j+1} - i J^y_j \eta_j \zeta_{j+1} - J^z_j \eta_j \zeta_j \eta_{j+1} \zeta_{j+1}~.
\label{eq:Hmaj}
\end{equation}

The symmetry group of the problem is somewhat more expressive in the Majorana language.
In the following we specialize to even system sizes $N \in 2\Z$.
The generators of the global symmetry translate to
\begin{align}
g_x &= i^{N/2} \zeta_1 \eta_2 \zeta_3 \cdots \eta_N~, \\
g_y &= (-i)^{N/2} \eta_1 \zeta_2 \eta_3 \cdots \zeta_N~.
\end{align}
The symmetries measure fermion parity on two disjoint sets partitioning the Majorana orbitals.
The Hamiltonian Eq.~\eqref{eq:Hmaj} takes the form of separate ``imaginary random hopping'' problems (see Ref.~\cite{motrunich2001griffiths}) on these two chains of Majoranas of length $N$, which we denote $\X = \{\zeta_1, \eta_2, \zeta_3, \ldots, \eta_N \}$ and $\Y = \{\eta_1, \zeta_2, \eta_3, \ldots, \zeta_N \}$.
On each chain the coefficients of the Majorana hopping terms---which are fermion parity measurements on adjacent orbitals within a chain---alternate between $iJ^x_j$ and $-iJ^y_j$.
There are also inter-chain coupling terms with coefficients $-J^z_j$.
A single ``rung'' term $i \eta_j \zeta_j$ is odd under the parity symmetries, and $H$ instead includes the double-rung interactions $-\eta_j \zeta_j \eta_{j+1} \zeta_{j+1}$.

The anti-unitary symmetry $\mathcal K$ (i.e., complex conjugation in the $\sigma^z$ basis) acts on the Majoranas as $\{i,\eta_j,\zeta_j\} \mapsto \{ -i , \eta_j , -\zeta_j \}$.
This symmetry prohibits nonzero expectation values of the form $\ev{i\eta_j \eta_k}$ or $\ev{i\zeta_j \zeta_k}$, even when these orbitals belong to the same Majorana chain.

Constraining $J^z_j = 0$ for all $j$, the resulting Hamiltonian $H_\xy \equiv H[\tilde J^x,\tilde J^y,\tilde J^z=0]$ is quadratic and can be solved for any particular disorder realization by diagonalization of the auxiliary Bogoliubov--de Gennes (BdG) matrix in the particle-hole basis.
The mapping to the Majoranas in Eq.~\eqref{eq:majoranas} transforms the BdG matrix into a particular form decoupling the two Majorana chains \X and \Y.
This further simplifies the solution for the single-particle eigenstates to diagonalization of a pair of $N \times N$ tridiagonal matrices.

As we are considering boundaries between Ising ordered phases, the natural observables are the corresponding magnetic order parameters $\sigma^\alpha$, $\alpha=x,y,z$.
Written in terms of fermion operators, the spin correlation functions $C^\alpha(j,k) = \ev{\sigma^\alpha_j \sigma^\alpha_k}$ are
\begin{align}
C^x(j,k) &= \ev{i \zeta_j (i \eta_{j+1} \zeta_{j+1}) \cdots (i \eta_{k-1} \zeta_{k-1}) \eta_{k}}~, \label{eq:Cx}\\
C^y(j,k) &= \ev{-i\eta_j (i \eta_{j+1} \zeta_{j+1}) \cdots (i \eta_{k-1} \zeta_{k-1}) \zeta_k}~, \label{eq:Cy}\\
C^z(j,k) &= \ev{-\eta_j \zeta_j \eta_k \zeta_k}~. \label{eq:Cz}
\end{align}
From Wick's theorem, in the ground state of any specific disorder realization $C^x(j,j+r)$ and $C^y(j,j+r)$ can be computed as Pfaffians of antisymmetric $2r \times 2r$ matrices, and the calculation further simplifies due to the separation into two Majorana chains.
We focus on this case and consider the angle brackets $\ev\cdot$ as denoting expectation values measured in the ground state, although the expressions Eqs.~(\ref{eq:Cx})--(\ref{eq:Cz}) apply more generally.
We will be discussing disorder-averaged correlations $\overline{C^\alpha(j,j+r)}$ and when this is clear we will drop the overline.
In the following we work exclusively along the line with statistical symmetry between $J_j^x$ and $J_j^y$ and will often collectively refer to $C^{x,y}(j,j+r)$, as $C^\perp(j,j+r)$.

\subsection{Strong-disorder renormalization group}
\label{sec:sdrg}

\subsubsection{Decoupled Majorana chains}
\label{sec:sdrg_majorana}

Examining the Hamiltonian on Majorana chains \X and \Y also clarifies the form of the analytic SDRG.
In the decoupled model $H_\xy$, the RG proceeds independently on each of the chains, which are endowed with parity conservation.
The SDRG for a single such chain was developed explicitly in the single-particle spectrum language by \citet{motrunich2001griffiths} and in the many-body Hamiltonian language by \citet{monthus2018strong}.
We review the result here, specialized to our case, in the many-body language, which naturally extends to the interacting problem \cite{monthus2018strong}.
For now we consider only a single Majorana chain, and relabel the orbitals as $\gamma_n$, $n = 1,\ldots,N$.
The Hamiltonian acting on this chain is $H_\M = \sum_{n=1}^{N-1} i h_n \gamma_n \gamma_{n+1}$.
Suppose that the largest energy scale is set by $H_0 = i h_k \gamma_k \gamma_{k+1}$ for some $k \in [1,N-1]$.
$H_0$ measures fermion parity on the two orbitals, with eigenvalues $\pm h_k$ associated with the two parity states; denote the splitting by $\Omega = 2 h_k$.
Accordingly, this term is diagonalized by the complex fermion mode \mbox{$f^\dag_0 = \frac 12(\gamma_k + i \gamma_{k+1})$}, which has projectors $\pi^+ = f_0 f^\dag_0 $ and $\pi^- = 1-\pi^+ = f^\dag_0 f_0$ into the even and odd parity sectors, respectively.
In terms of the projectors we have $H_0 = (\Omega/2) (\pi^+ - \pi^-)$.

The rest of the terms in $H_\M \equiv H_0 + V$ can be treated as a perturbation if the nearby couplings are much smaller than the local gap $|\Omega|$.
Although this condition may not be satisfied initially, the validity of the assumption improves during the RG flow because the SDRG generates an effective disorder distribution with increasingly broad logarithm.
The rest of the Hamiltonian can be divided into diagonal and off-diagonal components with respect to $H_0$; specifically, $V = V_\d + V_\od$, where
\begin{align}
V_\d &= \pi^+ V \pi^+ + \pi^- V \pi^-~, \\
V_\od &= \pi^- V \pi^+ + \pi^+ V \pi^- = \pi^- H_\M \pi^+ + \pi^+ H_\M \pi^-~.
\end{align}
Note that $V_\od$ contains only a constant number of local terms.
We denote the small scale of these terms relative to $H_0$ by the parameter $\epsilon$.
The effective Hamiltonian with emergent good quantum number $\ev{f_0^\dag f_0}$ is found by a Schrieffer--Wolff transformation eliminating $V_\od$ up to $O(\epsilon^2)$ \cite{schrieffer1966relation,macdonald1988t,bravyi2011schrieffer,lin2017quasiparticle}.
That is, $H'_\M = e^{iS} H_\M e^{-iS}$, where the Hermitian generator of the rotation can be expanded in powers of $\epsilon$ as $S = S^{[1]} + S^{[2]} + \cdots$.
The conditions on the rotation are that $S^{[1]}$ is off-diagonal and satisfies $V_\od = [H_0,iS^{[1]}]$, and $S^{[2]}$ eliminates off-diagonal terms at $O(\epsilon^2)$ (but we will not need to write it explicitly).
A suitable generator is $iS^{[1]} = \frac 1\Omega (\pi^+ H_\M \pi^- - \pi^- H_\M \pi^+)$, 
\begin{align}
H'_\M &= e^{iS} H_\M e^{-iS} \\
&= H_\M + [i S,H_\M] +\frac 12 [iS,[iS,H_\M]] + \cdots \\
&= H_0 + V_\d + \frac 12 \sum_{\iota = \pm} \pi^\iota [iS^{[1]},V_\od] \pi^\iota + O(\epsilon^3) \\
&\approx H_0 + V_\d + \frac 1\Omega [\pi^+ H_\M \pi^-,\pi^- H_\M \pi^+ ]~, \label{eq:sdrg_general}
\end{align}
the final line being Eq.~(17) of Ref.~\cite{monthus2018strong}.

The off-diagonal terms are those which share an odd number of Majoranas with $H_0$ and thus anticommute.
Consequently $V_\od = i h_{k-1} \gamma_{k-1}\gamma_k + i h_{k+1} \gamma_{k+1}\gamma_{k+2}$ and
\begin{align}
\pi^+ H_\M \pi^- &= ( i h_{k-1} \gamma_{k-1} + h_{k+1} \gamma_{k+2} ) f_0~, \\
\pi^- H_\M \pi^+ &= ( i h_{k-1} \gamma_{k-1} - h_{k+1} \gamma_{k+2} ) f^\dag_0~.
\end{align}
Finally the rotated Hamiltonian is
\begin{align}
H'_\M &= H_0 + V_\d + \frac{h_{k-1}^2 + h_{k+1}^2}{2 h_k} (i \gamma_k \gamma_{k+1}) \nonumber \\
&\quad + i\frac{h_{k-1} h_{k+1}}{h_k} \gamma_{k-1} \gamma_{k+2} + O(\epsilon^3)~.
\label{eq:sdrg_xy}
\end{align}
This result includes a renormalization of the strength of the $H_0$ term which increases the magnitude of the splitting, in addition to a new term $i h'_{k-1} \gamma_{k-1} \gamma_{k+2}$.
By projecting into the low-energy sector of $H_0$ (which depends on the sign of $h_k$), the Majoranas $\gamma_k$ and $\gamma_{k+1}$ are frozen into one of the definite parity states of the complex fermion mode, and thereby decoupled, or ``decimated,'' from the effective Hamiltonian.
The single effective coupling $h_{k-1}'$ replaces three hopping terms in $H_\M$.
Because the new term maintains the imaginary random-hopping form, the SDRG is closed in this model space and can be iterated, with the flow acting on the disorder distribution of the couplings $\{h_n\}$.
During the RG flow, some of the terms involved in decimations will be themselves renormalized couplings from prior steps; they can be made to fit the present format by re-indexing the chain after every step to remove the decimated Majorana orbitals.
In addition, the specific form of the renormalized coupling $h'_{k-1}$ permits a framing of the SDRG in terms of a classical random walk; this approach will be developed in detail in Sec.~\ref{sec:sdrg_correlated}.

The many-body Hilbert space is therefore decomposed into a tensor product of non-interacting complex fermions in definite parity states.
Returning to the XY model viewed as two decoupled Majorana chains and running the above procedure independently on each of the chains, one can deduce from the signs of the couplings in Eq.~\eqref{eq:Hmaj} that the ground state is even under $g_x$ and $g_y$ if $N \mod 4 = 0$ and odd under $g_x$ and $g_y$ if $N \mod 4 = 2$.
The ground state spin correlations in an eigenstate of the Hamiltonian can also be understood from this picture; see Sec.~\ref{sec:sdrg_correlations}.

As a technical remark, one way to deal with the signs of the couplings in Eq.~\eqref{eq:Hmaj}---needed to deduce $g_x$ and $g_y$ quantum numbers as well as the signs of the correlation functions---is to perform a gauge transformation of the Majorana fermions as $\eta_j = s_j \eta_j^\prime$, where $s_j = 1$ if $j=4n+1$ or $4n+2$ and $s_j = -1$ if $j=4n+3$ or $4n+4$, while $\zeta_j = s_j (-1)^{j+1} \zeta_j^\prime$.
The Hamiltonian written in terms of the primed Majoranas takes the form $\sum_j i J_j^x \zeta_j^\prime \eta_{j+1}^\prime + i J_j^y \eta_j^\prime \zeta_{j+1}^\prime$, i.e., all Majorana hopping amplitudes are positive in the convention where the Majoranas are written in the same order as they appear on the chain: $i h_{nm} \gamma_n^\prime \gamma_m^\prime$ with $n < m$ has $h_{nm} > 0$.
This property is preserved under the SDRG, which simplifies analysis of the signs.
For example, for Majoranas $\gamma_n^\prime, \gamma_m^\prime$ with $n < m$ decimated as a pair we then have $\langle i \gamma_n^\prime \gamma_m^\prime \rangle = -1$ at the zeroth order in the SDRG, and using the non-crossing property of the pairs in each Majorana chain fixes the signs of correlations in Eqs.~\eqref{eq:Cx}--\eqref{eq:Cz} to be $(-1)^{j-k}$.
To avoid confusion, in formulas we keep using the original Majoranas as in Eq.~\eqref{eq:Hmaj}.

\subsubsection{Majorana problem with inter-chain interaction terms}
\label{sec:sdrg_xyz}

In the presence of interactions coupling the two Majorana chains, it is necessary to consider the full Hamiltonian Eq.~\eqref{eq:Hmaj}.
In the notation of the present section we have $H = H_\X + H_\Y + H_\text{int}$, where
\begin{align}
H_\X &= \sum_{n=1}^{N-1} i h^\X_n \gamma^\X_n \gamma^\X_{n+1}~,\\
H_\Y &= \sum_{n=1}^{N-1} i h^\Y_n \gamma^\Y_n \gamma^\Y_{n+1}~,\\
H_\text{int} &= \sum_{n=1}^{N-1} K_n (i\gamma^\X_n \gamma^\X_{n+1}) (i\gamma^\Y_n \gamma^\Y_{n+1})~. \label{eq:sdrg_h_int}
\end{align}

Because all of the terms in $H$ are measurements of fermion parity, the general framework from the previous section---in particular Eq.~\eqref{eq:sdrg_general}---still applies.
Now there are two cases: the largest energy scale can be set by one of either the hopping terms $\{h^\M_n\}$ or the interactions $\{K_n\}$.
While one can in principle consider both cases following Ref.~\cite{monthus2018strong}, for our purposes we will study only the hopping-dominated case.
Suppose that 
$H_0 = ih^\X_k \gamma^\X_k \gamma^\X_{k+1}$.
Now
\begin{align}
V_\od &= i h^\X_{k-1} \gamma^\X_{k-1}\gamma^\X_k + i h^\X_{k+1} \gamma^\X_{k+1} \gamma^\X_{k+2} \nonumber \\
&\quad+ K_{k-1} (i \gamma^\X_{k-1} \gamma^\X_k)(i \gamma^\Y_{k-1} \gamma^\Y_k) \nonumber \\
&\quad+ K_{k+1} (i \gamma^\X_{k+1} \gamma^\X_{k+2})(i \gamma^\Y_{k+1} \gamma^\Y_{k+2})~.
\end{align}
The components appearing in each off-diagonal block of the Hamiltonian are
\begin{align}
\pi^+ H \pi^- &= \Big( \left(h^\X_{k-1} + K_{k-1} (i\gamma^\Y_{k-1}\gamma^\Y_k)\right) i\gamma^\X_{k-1} \nonumber \\
&\quad+ \left(h^\X_{k+1} + K_{k+1}(i \gamma^\Y_{k+1}\gamma^\Y_{k+2})\right) \gamma^\X_{k+2} \Big) f_0 \\
&\equiv ( i h^{\X,\text{int}}_{k-1} \gamma^\X_{k-1} + h^{\X,\text{int}}_{k+1} \gamma^\X_{k+2} ) f_0~, \\
\pi^- H \pi^+ &= \Big( \left(h^\X_{k-1} + K_{k-1} (i\gamma^\Y_{k-1}\gamma^\Y_k)\right) i\gamma^\X_{k-1} \nonumber \\
&\quad- \left(h^\X_{k+1} + K_{k+1}(i \gamma^\Y_{k+1}\gamma^\Y_{k+2})\right) \gamma^\X_{k+2} \Big) f^\dag_0 \\
&\equiv ( i h^{\X,\text{int}}_{k-1} \gamma^\X_{k-1} - h^{\X,\text{int}}_{k+1} \gamma^\X_{k+2} ) f^\dag_0~.
\end{align}
The effect of the interactions in perturbation theory is simply to modify the couplings into operators which we refer to as ``interacting couplings:'' $h^\X_{k\pm 1} \to h^{\X,\text{int}}_{k\pm 1}$.
This is a reasonable shorthand because the interacting couplings commute with each other and all fermion operators appearing in the formula.
Then from the result Eq.~\eqref{eq:sdrg_xy},
\begin{widetext}
\begin{align}
H' &= H_0 + V_\d + \frac{(h^{\X,\text{int}}_{k-1})^2 + (h^{\X,\text{int}}_{k+1})^2}{2 h^\X_k} (i \gamma^\X_k \gamma^\X_{k+1}) + i\frac{h^{\X,\text{int}}_{k-1} h^{\X,\text{int}}_{k+1}}{h^\X_k} \gamma^\X_{k-1} \gamma^\X_{k+2} + O(\epsilon^3) \\
&= H_0 + V_\d + (i \gamma^\X_k \gamma^\X_{k+1})\left(\frac{(h^\X_{k-1})^2 + (h^\X_{k+1})^2+ K_{k-1}^2 + K_{k+1}^2}{2 h^\X_k}  + i \frac{h^\X_{k-1} K_{k-1}}{h^\X_k} \gamma^\Y_{k-1} \gamma^\Y_k + i \frac{h^\X_{k+1} K_{k+1}}{h^\X_k} \gamma^\Y_{k+1} \gamma^\Y_{k+2}\right) \nonumber \\
&\quad+ i \frac{h^\X_{k-1} h^\X_{k+1}}{h^\X_k} \gamma^\X_{k-1} \gamma^\X_{k+2} + \frac{K_{k-1}h^\X_{k+1}}{h^\X_k} (i \gamma^\X_{k-1} \gamma^\X_{k+2}) (i \gamma^\Y_{k-1} \gamma^\Y_k) + \frac{h^\X_{k-1} K_{k+1}}{h^\X_k} (i \gamma^\X_{k-1} \gamma^\X_{k+2}) (i \gamma^\Y_{k+1} \gamma^\Y_{k+2}) \nonumber \\
&\quad+ \frac{K_{k-1} K_{k+1}}{h^\X_k} (i\gamma^\Y_{k-1} \gamma^\Y_k) (i\gamma^\X_{k-1} \gamma^\X_{k+2}) (i\gamma^\Y_{k+1} \gamma^\Y_{k+2}) + O(\epsilon^3)~.
\label{eq:sdrg_xyz_h}
\end{align}
\end{widetext}
Projecting into the low-energy sector sets $i \gamma^\X_k \gamma^\X_{k+1} \to -\text{sgn}(h^\X_k)$ and again decouples the Majorana operators $\gamma^\X_k$ and $\gamma^\X_{k+1}$ from the rest of the system, decimating them by creating a complex fermion mode with definite parity.
As in the non-interacting case, the magnitude of the splitting is increased by renormalization of $H_0$, and a new hopping term $h^{\X\prime}_{k-1}$ is added to the \X chain.
However, the leading-order effect of the interactions, at $O(\epsilon)$, arises from $V_\d$, where the ``degradation'' of the term $K_k (i \gamma^\X_k \gamma^\X_{k+1}) (i \gamma^\Y_k \gamma^\Y_{k+1})$ renormalizes $h_k^{\Y\prime} = h_k^\Y - \mathrm{sgn}(h_k^\X)\,K_k$.
As a result, correlations develop between the hopping terms on the same bond.
This aspect of the perturbation will constitute the basis of a mean-field study of the interacting system, presented in Sec.~\ref{sec:mf}.

The effective Hamiltonian also includes renormalized couplings $h^{\Y\prime}_{k-1}$ and $h^{\Y\prime}_{k+1}$, as well as new four-fermion terms which change the structure of the lattice graph, and a six-fermion term.
The appearance of these terms breaking the form of $H$, as well as the generation of correlations between terms, are an indication that the RG flow cannot be tracked exactly in the interacting model.
However, if the interaction terms already tend to be weak compared to the hopping, the higher-order terms generated by this process will accordingly be weaker still.
This is the situation, at least initially, in the random XYZ model with small $\tilde J^z$; however there is no guarantee at this point that the relative strengths of the different types of couplings are maintained asymptotically. 
We will return to this question more systematically in Sec.~\ref{sec:fp_interacting}, after we understand the non-interacting problem with correlated Majorana hopping amplitudes in the two chains in Sec.~\ref{sec:sdrg_correlated}.

\subsection{XY model spin correlations in SDRG}
\label{sec:sdrg_correlations}

From the controlled SDRG for the random XY model one can deduce that average correlations in the ground state follow power laws---although typical correlations are short-ranged---and even calculate the exponents.
One also obtains a more qualitative picture of the behavior of the spin correlation functions.

Expanding Eq.~\eqref{eq:Cz} in the ground state at distance $r$,
\begin{equation}
C^z(j,j+r) = \ev{i \eta_j \zeta_{j+r}} \ev{i \zeta_j \eta_{j+r}}~.
\end{equation}
Other terms vanish due to symmetry.
One sees immediately that \mbox{$C^z(j,j+r) = 0$} if $r$ is even.
For odd $r$, $C^z(j,j+r)$ assumes a large value if and only if the sites $j$ and $j+r$ were decimated together on both Majorana chains, in which case both expectation values $\ev{i \eta_j \zeta_{j+r}}$ and $\ev{i \zeta_j \eta_{j+r}}$ have approximately unit magnitude and opposite sign, so the sign of $C^z$ is negative.
Otherwise if this decimation did not occur in one or both Majorana chains the contribution is suppressed, arising only from higher-order terms in the perturbation theory.
Consider the correlations averaged over sites $j$ as well as over disorder realizations, which average we denote $C^z(r)$.
Nearly all terms will be vanishingly small, with rare terms of roughly unit magnitude occurring with some density; these dominate the average. 
It is a result of Ref.~\cite{fisher1994random} for the RS phase that at sufficiently large separation the likelihood of such a decimation scales as $r^{-2}$; thus for two independent Majorana chains $\eta_z = 4$.

The transverse correlations Eqs.~\eqref{eq:Cx} and \eqref{eq:Cy}, summarized as $C^\perp(j,j+r)$, are the expectation values of strings of $2r$ Majoranas.
Such operators are evaluated as the sum of $r$-fold products of expectation values of symmetry-allowed bilinear contractions, with signs arising from the signature of each permutation.
A term in the sum has a large value if and only if it contracts all Majoranas with their decimation partners in the SDRG.
This will be the case for exactly one term if all decimations of the Majoranas appearing in the string expectation value are ``internal;'' that is, if all decimation partners are also included.
If any Majoranas were decimated with orbitals which do not appear in the string, the expectation value will be small.
We again define $C^\perp(r)$ as the average over sites and disorder realizations.

If on both chains \X and \Y the sites $j$ and $j+r$ are decimation partners, then as described above, this pair contributes a large value to $C^z(r)$.
The pair also necessarily contributes a large value to $C^\perp(r)$, as pairing the extremal Majorana orbitals in a string implies that all decimations are internal to the string.
Thus, the critical exponent $\eta_\perp$ lower-bounds $\eta_z$.
As reviewed earlier, for the random XY model $\eta_\perp = 3 - \sqrt 5 \approx 0.764$; the bound is saturated in the XX model where $\eta_\perp = \eta_z = 2$ \cite{fisher1994random}.

Finally, the SDRG picture also tells us about the end-to-end spin correlations in the XX and XY models.
The expectation value $C^z(1,N) \equiv C^z(N)$ obtains large contributions if on both Majorana chains the end sites $1$ and $N$ are paired in the SDRG.
While such occurrences in the two chains are perfectly matched in the XX model and have probability $1/N$ or $\eta^\e_z = 1$, in the XY model the occurrences are independent, giving $\eta^\e_z = 2$.
On the other hand, the expectation value $C^\perp(1,N) \equiv C^\perp_\e(N)$ includes all Majorana orbitals on one chain, and all but those at sites $1$ and $N$ on the other.
This string has a large expectation value if all of these Majoranas are paired internally, which is to say that the two excluded Majoranas are decimated together.
As this is occurs on a single chain only, it has the same probability in both the random XX and XY models.
Indeed, $\eta^\e_\perp = 1$ in both cases \cite{igloi2000random}.

\section{Unbiased tensor network study}
\label{sec:unbiased_study}

\subsection{``Rigorous RG'' numerical method}
\label{sec:rrg}

The standard numerical technique for equilibrium states of many-body quantum systems in 1d is the density matrix renormalization group (DMRG) \cite{white1992density,*white1993density,schollwock2011density}, which has been remarkably effective in conjunction with matrix product state (MPS) representations of low-energy wavefunctions \cite{klumper1991equivalence,*klumper1992groundstate,*klumper1993matrix,fannes1992finitely}.
Over nearly 30 years, DMRG has seen enormous practical success in a wide range of models of physical interest.
However, for some time its effectiveness was not well explained: even as MPS attained a rigorous footing with the proof of the area law of entanglement in 1d \cite{hastings2007area,wolf2008area,arad2012improved}, the existence of an efficient algorithm for eigenstates given an area-law Hamiltonian remained unclear.
It was not until the work of \citet{landau2015polynomial} in 2015 that a polynomial-time algorithm was developed for ground states of gapped models, proving that an efficient method is possible in principle.

However, the algorithm exhibited in Ref.~\cite{landau2015polynomial} bears little resemblance in its particulars to DMRG, and a similar proof for the DMRG algorithm appears to be challenging; in fact, it is known that popular multi-site variants can be NP-hard in the worst case \cite{eisert2006computational}.
As a practical matter, in systems with strong disorder DMRG is susceptible to spurious convergence to excited states, an outcome which cannot be readily diagnosed \cite{schmitteckert1999disordered}.
This is fundamentally a consequence of performing an iterated local optimization over MPS parameters.
The rigorous algorithm is distinguished by a reliance on an \emph{approximate ground state projector} (AGSP), an operator derived from the Hamiltonian, which was introduced by \citet{arad2013area}.
The role of the AGSP is to provide global information, ensuring that intermediate states can be efficiently represented and directing the algorithm along a computationally tractable route to the ground state.

AGSP-based methods were later generalized to low-energy excited states in models with slightly relaxed conditions on the density of states \cite{arad2017rigorous}.
Based on this work, in collaboration with Vidick we introduced the \emph{rigorous renormalization group} (RRG), a numerical implementation for low-energy states of local Hamiltonians in one dimension \cite{roberts2017implementation}.
While the implemented method differs slightly from the proof construction and does not strictly satisfy the conditions of the guarantee---whose parameters are not known {\itshape a priori} regardless---it inherits the intuitive benefits of the AGSP and has been seen to be effective in practice for nontrivial low-energy spectra like those of strongly disordered systems, or in the presence of nearly degenerate manifolds \cite{roberts2017implementation,block2020performance}, where DMRG may be unreliable.

In the following sections, we perform a numerical study of the line $\tilde J^z \in [0,1]$, $\tilde J^x = \tilde J^y = 1$, in the phase diagram of Eq.~\eqref{eq:Hxyz}, using RRG.
Our objective is primarily to verify by unbiased numerics the observation of continuously varying critical exponents in the SBRG study of \citet{slagle2016disordered}, and then to shed additional light on the nature of the low-energy theory.
(Here we focus solely on the ground state properties and low-energy physics, rather than the question of MBL.)
For concreteness, we use the disorder distribution described in Eqs.~(3) and (4) of Ref.~\cite{slagle2016disordered}, namely, 
\begin{equation}
p(J_i^\alpha) = \frac{1}{\Gamma \tilde{J}^\alpha} (J_i^\alpha)^{1/\Gamma-1}~,~~J_i^\alpha \in [0,(\tilde J^\alpha)^\Gamma]~.
\label{eq:disorder_distribution}
\end{equation}
We use a milder disorder strength $\Gamma = 2$, as compared to $\Gamma=4$ for the previous work \cite{slagle2016disordered}.
Both choices lead to strong disorder physics and the specific value should have little effect on the universal low-energy physics for large enough systems.
However, we find that the logarithm of the distribution of the energy gaps depends significantly on $\Gamma$, with smaller values tending to lead to larger gaps; this eases the challenge to the numerics which in any case are limited by double-precision floating-point errors on the order of $10^{-16}$.
In RRG we are capable of accurately resolving energy scales down to $\log_{10}(\Omega/\epsilon) \sim -12$, and validate our results against the free-fermion solution at the soluble point $\tilde J^z = 0$.

To construct the AGSP for RRG we use a Trotter approximation to a thermal operator $e^{-\beta H}$.
The output of the RRG algorithm is a subspace of constant dimension approximating the low-energy states of the model.
We use an implementation based on ITensor \cite{itensor}, in which we explicitly realize the $\Z_2 \times\Z_2$ symmetry and solve for the lowest two eigenstates in each of the four symmetry sectors 
\footnote{The RRG code used in this work is available online at \url{https://www.github.com/brendenroberts/RigorousRG}.}.
In each case the MPSs generated by RRG are then further optimized using DMRG in order to minimize the overlap with high-energy states.
The RRG ``hyperparameters'' $s$ and $D$ (see Ref.~\cite{roberts2017implementation} for details) are chosen so that for the majority of disorder realizations DMRG can optimize the RRG output in a small number of sweeps.
For approximately the most challenging 1\% of realizations, DMRG requires many sweeps to converge.
In these instances we repeat the calculation, increasing the RRG hyperparameters, and find that the improved RRG states are easily converged by DMRG.
From comparison with exact free-fermion results for $\tilde J^z=0$ obtained by numerical matrix diagonalization, we find that if RRG produces states which are successfully converged by DMRG and the excitation gap is larger than the target threshold $10^{-12}$, the ground state energy and gap are numerically exact in $\gtrsim 99.5\%$ of realizations.
As we will show in the following section, at $\tilde J^z > 0$ the finite-size gaps tend to be larger than those at $\tilde J^z = 0$ and should be easier for RRG; thus we believe our results are even more reliable for these points.

\begin{table}
\begin{tabular}{c||c|c|c|c|c|c}
$\tilde J^z$ & 0.0 & 0.2 & 0.4 & 0.6 & 0.8 & 1.0 \\
\hline
$(s,D)$ & (8,14) & (8,14) & (6,10) & (6,10) & (5,8) & (5,8)
\end{tabular}
\caption{\label{tab:rrg_params} RRG hyperparameters are shown for values of $\tilde J^z$ studied numerically.
As described in the text, we optimize the output of RRG using DMRG, and for finite $\tilde J^z$ take as a measure of accuracy the number of sweeps required for convergence.
These values of $s$ and $D$ were chosen in order to accurately converge approximately 99\% of disorder realizations on $N=80$ spins.
For the small fraction of more difficult realizations which are not solved by the hyperparameters above we repeat the algorithm with increased values, finding that convergence is achieved this way.
}
\end{table}

\subsection{Results from RRG}
\label{sec:rrg_results}

\subsubsection{Critical spin correlations}

\begin{figure}[ht]
\includegraphics[width=\columnwidth]{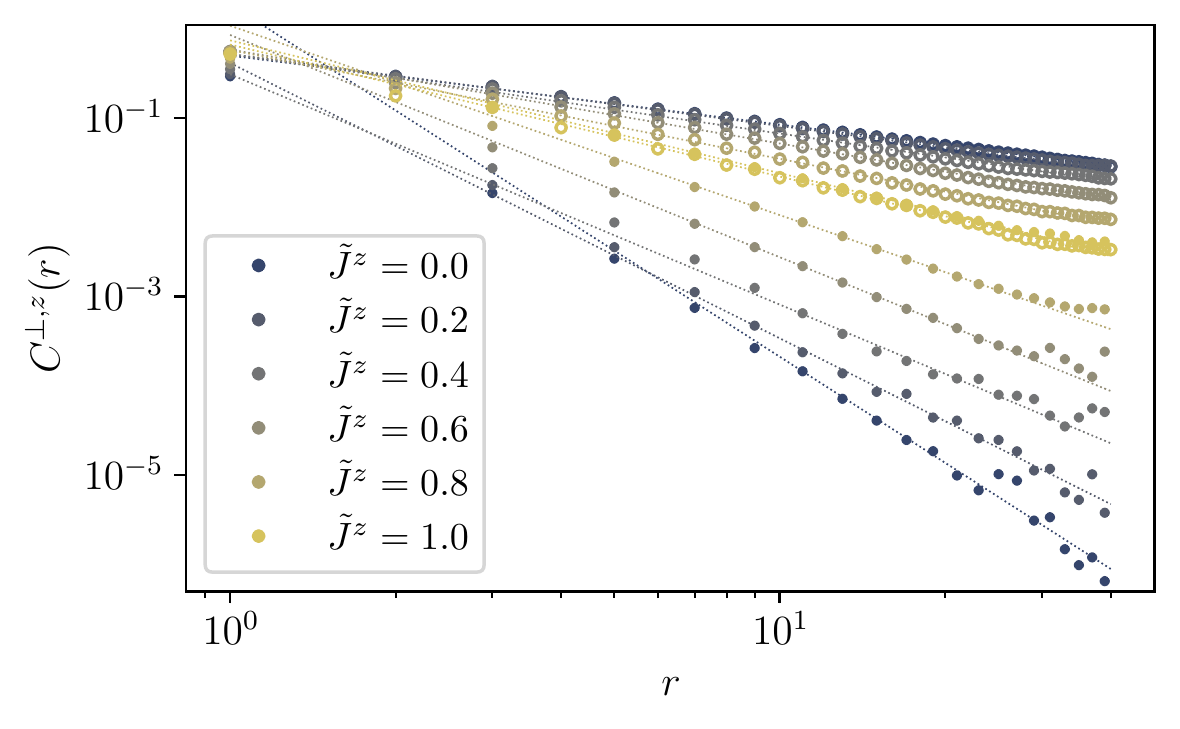}
\caption{\label{fig:rrg_bulk_correlations}
Bulk spin correlations data from RRG are shown for the random XYZ model with varying bandwidth $\tilde J^z$, up to separation $r=40$ lattice spacings, from systems of length $N=80$.
Open circles indicate $C^\perp(r)$ data, while filled circles mark $C^z(r)$.
The disorder averages for each value of $\tilde J^z$ include 1500 realizations.
In the spatial average we include only the middle half of the spin chain---that is, only sites in $\{\frac{N}{4}, \dots, \frac{3N}{4}\}$---in order to separate the bulk correlations from the ends, which exhibit different scaling laws.
See Fig.~\ref{fig:rrg_exponents} for the critical power law decay exponents extracted from this data.
In order to measure the power laws we show the absolute value of the correlations, which originally have a staggered sign pattern $(-1)^r$.
In addition, only odd $r$ are shown for $C^z$ data because the values for even $r$, though demonstrating a similar power law, are much smaller
(at $\tilde J^z = 0$ they are identically 0, see Sec.~\ref{sec:sdrg_correlations}).
}
\end{figure}

We measure spin correlations in the RRG ground state of $H[\tilde J^x = 1, \tilde J^y = 1, \tilde J^z]$ with $\tilde J^z$ ranging from 0 to 1 and microscopic disorder strength $\Gamma = 2$ throughout.
Bulk correlations in an open chain of length $N$ are measured for $r \leq \frac N2$ including only sites $j, j+r \in \{ \frac N4,\ldots,\frac{3N}{4}\}$, in order to distinguish the power law from the end-to-end correlations closer to the boundaries.
We show disorder-averaged correlations data measured in chains of length $N=80$ sites in Fig.~\ref{fig:rrg_bulk_correlations}, which includes slices at values of $\tilde J^z$ moving along the phase boundary from the free-fermion model to the tricritical point.
Already the raw data clearly shows power laws with varying exponents for both $C^\perp$ and $C^z$ in the bulk.

\begin{figure}[ht]
\includegraphics[width=\columnwidth]{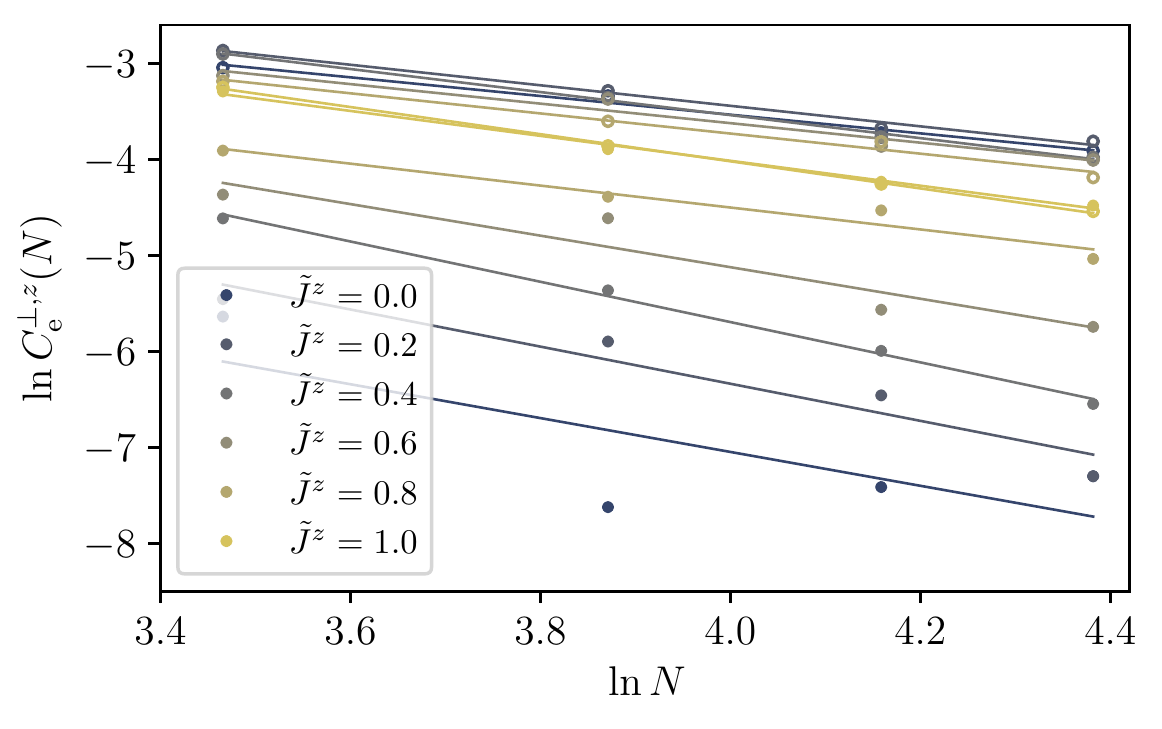}
\caption{\label{fig:rrg_end_correlations} RRG end-to-end correlations data are shown for the random XYZ model with varying bandwidth $\tilde J^z$.
System sizes $N=32,48,64,80$ are included for $C^\perp_\e(N)$ (open circles) and $C^z_\e(N)$ (filled circles).
These data are noisier than the bulk data shown in Fig.~\ref{fig:rrg_bulk_correlations} due both to reduced statistics (same number of disorder realizations but no averaging over bulk pairs) as well as the special difficulty of measuring $C^z_\e(N)$ in RRG, as described in the text.
See Fig.~\ref{fig:rrg_exponents} for the critical power law decay exponents extracted from this data.
We use the absolute value of the correlations data here; the true values all have negative sign because all $N$ are even.
}
\end{figure}

End-to-end spin correlations are measured only between the single pair of sites 1 and $N$ for each disorder realization, and exhibit correspondingly larger statistical fluctuations.
In addition, reproducing $C^z_\e(N)$ correlations presents a singular challenge for the RRG algorithm.
As discussed in Sec.~\ref{sec:sdrg_correlations}, in the SDRG the likelihood of a nonzero value of $\ev{\sigma^z_1 \sigma^z_N}$ at the XY free-fermion point is the square of the probability of an end-to-end singlet in a spin chain of length $N$ in the RS phase.
That is, the distribution is broad on a logarithmic scale, with the average being dominated by a very small tail.
More importantly, the disorder realizations located in the tail---of outsize importance in the average---are those on which sites $1$ and $N$ were decimated together on both Majorana chains, which correlate with the smallest excitation gaps in the low-energy spectrum and are the most difficult realizations for the method to solve accurately.
We show disorder-averaged end-to-end correlations as a function of $N$ in chains up to $N=80$ in Fig.~\ref{fig:rrg_end_correlations}.
One sees that the $C^\perp_\e$ correlations depend weakly on $\tilde J^z$ and have close slopes on the log-log plot, suggesting similar power law exponents.
On the other hand, the $C^z_\e$ correlations depend strongly on $\tilde J^z$ and despite evident statistical scatter appear to have varying slopes.

\begin{figure}[ht]
\includegraphics[width=\columnwidth]{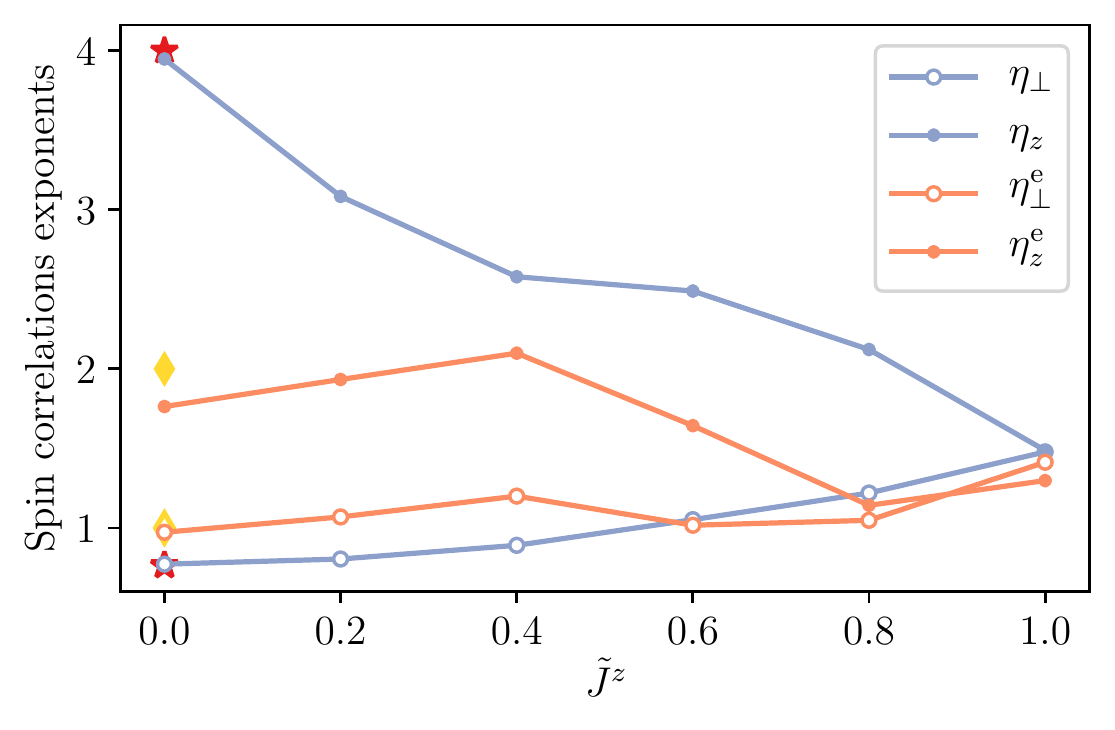}
\caption{\label{fig:rrg_exponents} Critical exponents governing spin correlations in the RRG ground states are shown, extracted from the data in Figs.~\ref{fig:rrg_bulk_correlations} and \ref{fig:rrg_end_correlations}.
Both bulk and end-to-end exponents are included, with known results for the bulk correlations in the free-fermion model at $\tilde J^z=0$ indicated by red stars, and results for end-to-end correlations by yellow diamonds.
An increase in statistical noise is evident in the end-to-end correlations as compared to the bulk.
The reason that these computations, particularly $C^z_\e(N)$, are more difficult, is discussed in the text.
}
\end{figure}

Our unbiased numerical results for the bulk correlations are in broad agreement with the finding of \citet{slagle2016disordered} of critical exponents governing the decay of spin correlations that vary continuously with $\tilde J^z$.
In contrast to the previous approach, we perform direct measurements in optimized MPS for the ground state.
We show the extracted power law exponents for the bulk and end-to-end correlations in Fig.~\ref{fig:rrg_exponents} as a function of $\tilde J^z$.
As expected, the $C^\perp$ and $C^z$ exponents approach each other at the tricritical (permutation-symmetric) point $\tilde J^z = 1$, where we estimate the bulk critical index to be $\eta_\perp = \eta_z \approx 1.48$.

\subsubsection{Entanglement structure}

We also study measures of entanglement in the RRG ground states for varying $\tilde J^z$.
The average bipartite entanglement entropy of a connected subsystem of length $\ell$ adjacent to the system boundary is known to scale according to the conformal field theory result $S_b(\ell) = \frac {\tilde c}{6} \ln \ell$, with a universal constant $\tilde c$.
In some cases the ``effective central charge'' $\tilde c$ is apparently related to the central charge of the clean model \cite{refael2004entanglement}; for example, in the critical phase of a single Majorana chain $\tilde c = \frac{\ln 2}{2} = c \ln 2$, where $c=\frac 12$ is the central charge of a clean Majorana fermion chain.
Accordingly, the XY fixed point has $\tilde c = \ln 2$, being equivalent to two decoupled critical random Majorana chains.
From finite-size scaling of the disorder-averaged half-system bipartite entanglement entropy $S_b(N/2)$ we find with fair precision that $\tilde c$ is stable at this value for any interaction strength $\tilde J^z$ along the critical line, in agreement with Ref.~\cite{slagle2016disordered}.

\begin{figure}[ht]
\includegraphics[width=\columnwidth]{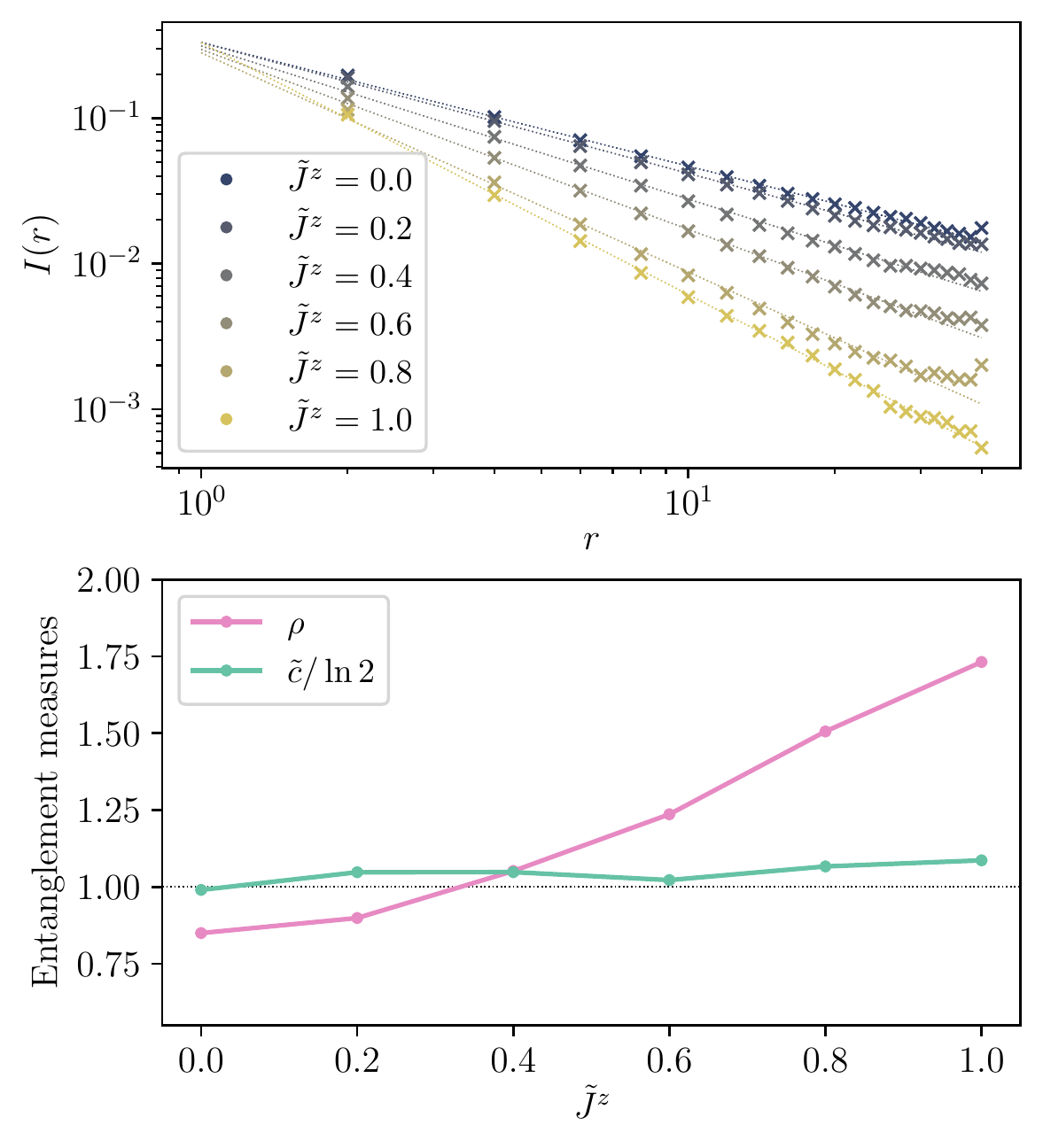}
\caption{\label{fig:rrg_entanglement} Characterizations of the entanglement structure of the ground state are shown.
We include the power-law exponent $\rho$ for decay of average long-range mutual information $I(r)$, based on the raw data shown in the upper panel.
The subsystems $A$ and $B$ considered in this case are single spins separated by a distance $r$, and the average is taken over sites in the middle half of the chain.
Also shown is the effective central charge $\tilde c$, found from finite-size scaling of the half-chain entanglement entropy.
While $\tilde c$ appears to be insensitive to the coupling between the two Majorana chains, the LRMI exponent varies continuously.
}
\end{figure}

We also measure long-range mutual information (LRMI) between disconnected regions; the formula for this entropic quantity in terms of the entanglement entropy of a subsystem is $I(A:B) = S(A) + S(B) - S(A\cup B)$.
We will take $A$ and $B$ to be single spins separated by a distance $r$; Ref.~\cite{slagle2016disordered} found that up to appropriate rescaling, the lengths of the subsystems do not affect the asymptotic behavior.
The disorder-averaged LRMI we denote $I(r)$, and this quantity will decay no faster than the slowest observable.
That is, in the symmetric ground state of an ordered phase $I(r)$ will be long-ranged; in a phase without order one expects exponential decay; and at a critical point the exponent $\rho$, $I(r) \sim r^{-\rho}$, lower-bounds the power-law decay exponent of any local observable.
We show disorder-averaged LRMI data in the upper panel of  Fig.~\ref{fig:rrg_entanglement}.
The critical exponent $\rho$ varies continuously with $\tilde J^z$, as is the case with the other critical indices measured, and is very close to the exponent $\eta_\perp$, suggesting that the correlations of the order parameters for the adjacent phases saturate the lower bound everywhere along the boundary.
Our RRG results for $\rho$ as well as the effective central charge $\tilde c$ are shown in the lower panel of Fig.~\ref{fig:rrg_entanglement}.
At $\tilde J^z = 1$ we estimate $\rho \approx 1.73$, which is somewhat larger than the estimates of $\eta^{\perp,z}$ but is in general agreement and is also similar to the SBRG estimates in Ref.~\cite{slagle2016disordered}.

\subsubsection{Scaling of excitation gap}
\label{subsubsec:gap}

Because RRG produces not only the ground state but a constant number of low-energy states, it is possible in principle to study spectral properties as well.
We focus first on the simplest of these, the energy gap to the lowest excitation in a finite system.
From the SDRG for the free-fermion point one observes that this excitation consists of flipping the parity of the complex fermion associated with the lowest-energy (i.e., the last decimated) pairing on either Majorana chain.
As we consider chains with lengths that are multiples of 4, the ground state is found in the $(g_x,g_y)=(+1,+1)$ sector of the global $(\Z_2)^2$ symmetry and the first excited state will be found in either the $(+1,-1)$ or $(-1,+1)$ sector.

\begin{figure}[ht]
\includegraphics[width=\columnwidth]{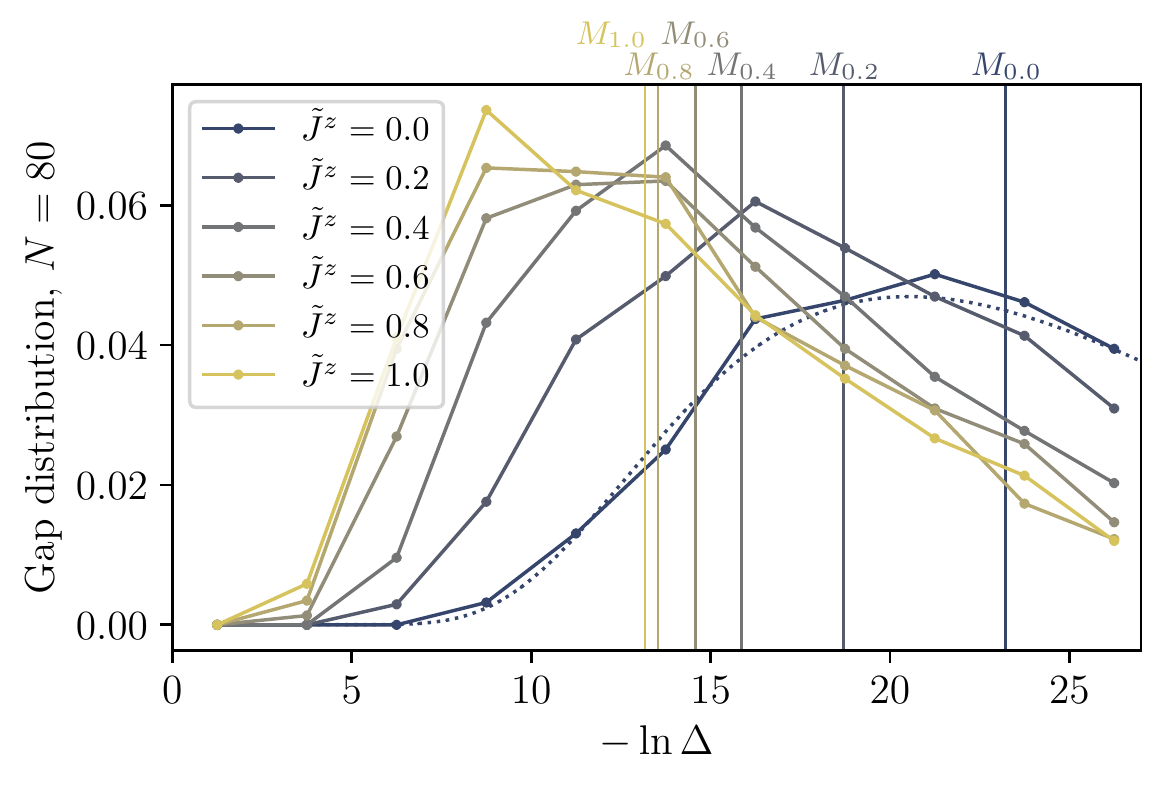}
\caption{\label{fig:rrg_gaps_distribution} Histograms of the first excitation gap are shown for the random XYZ model at system size $N=80$ sites.
Vertical lines indicate the median $M_{\tilde J^z}$ of each gap distribution.
The medians include long tails that are not shown, as they contain energy gaps too small to be accurately measured by the RRG algorithm; however the estimate of the median is not sensitive to these uncertainties.
The trace for each value of $\tilde J^z$ includes 1500 disorder realizations.
}
\end{figure}

The distribution of excitation gaps is known exactly via the mapping to two decoupled copies of the RTFIM, where the universal form of the gap distribution is known from the work of \citet{fisher1998distributions}.
The gap in the random XY model is the minimum of two independent random variables sampled from the distribution of Ref.~\cite{fisher1998distributions}.
In Fig.~\ref{fig:rrg_gaps_distribution} we show histograms of the (logarithmic) excitation gaps for the random XYZ model with varying $\tilde J^z$ for chains of length $N=80$.
The exact distribution for the $\tilde J^z = 0$ point is indicated with a dotted line.

\begin{figure}[ht]
\includegraphics[width=\columnwidth]{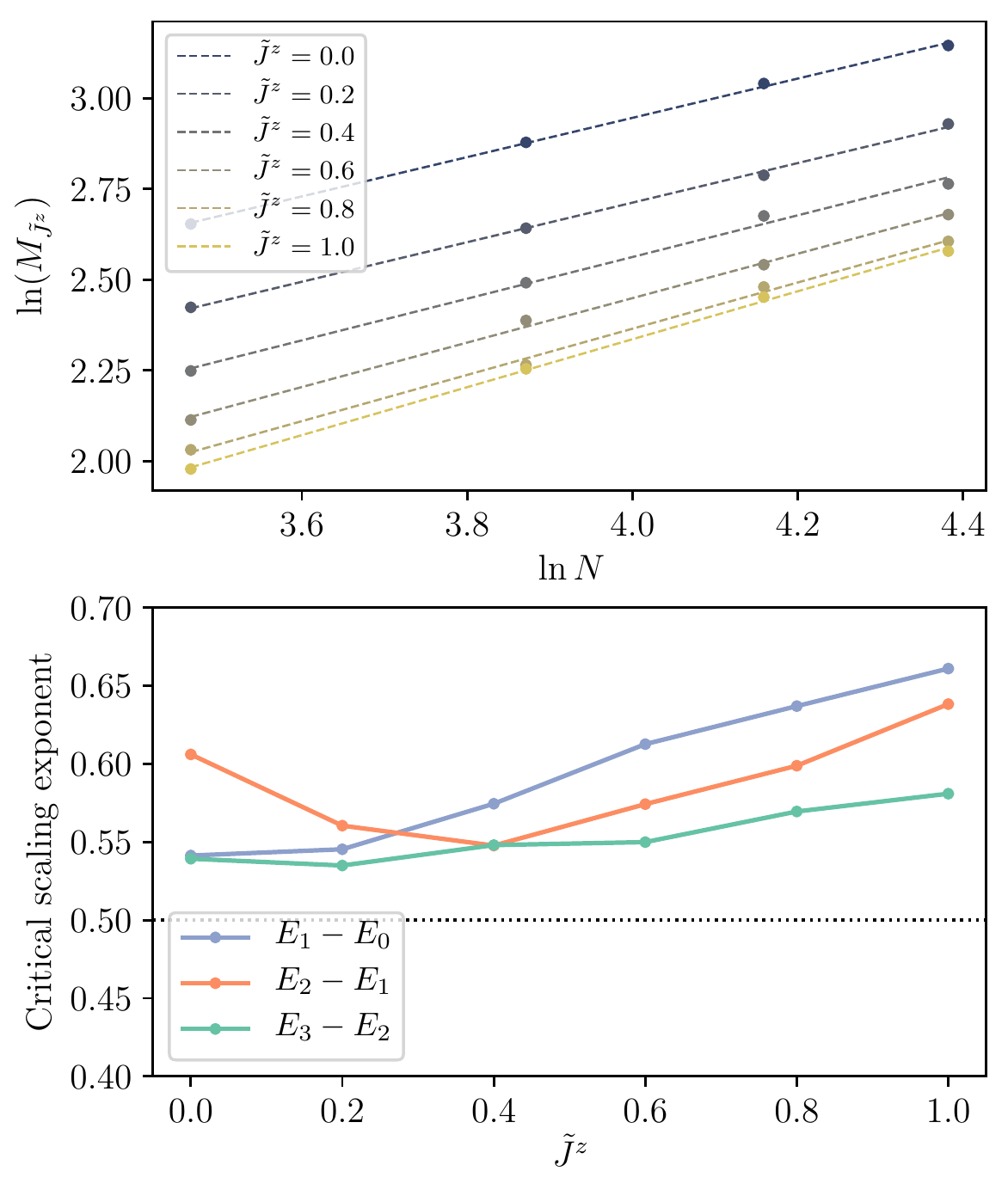}
\caption{\label{fig:rrg_psi} The value of the critical exponent $\psi$ extracted from finite-size scaling of excitation gaps in RRG is shown.
The upper panel shows the finite-size scaling of the medians $M_{\tilde J^z}$ (shown in Fig.~\ref{fig:rrg_gaps_distribution} for $N=80$), with each data point including 1500 disorder realizations.
The lower panel shows the extracted power law exponents for both the first gap, denoted $E_1-E_0$ (found from the data shown in the upper panel) as well as the second and third energy gaps.
At the free-fermion point $\tilde J^z = 0$, $\psi = \frac 12$, and the systematic deviation from the exact value is likely due to finite-size corrections.
At this point the first and third energy gaps are very often identical, both being associated with the lowest-energy decimation on one chain.
Away from this point, this is no longer necessarily the case and a drift in $\psi$ with $\tilde J^z$ is visible in the $E_1-E_0$ curve.
}
\end{figure}

Indicated on Fig.~\ref{fig:rrg_gaps_distribution} by vertical lines and the labels $M_{\tilde J^z}$ are the medians of the histograms; these are provided as a characterization of the distributions that is not overly sensitive to the tails, where the energy gaps can be close to the numerical threshold.
While the precise tails are not accessible, it is rare for RRG to make an error which would move a disorder realization out of the tail into the bulk of the distribution.
Thus, the median provides an accurate summary of the gap distribution although the mean cannot be reliably estimated.
In Fig.~\ref{fig:rrg_psi} the scaling with chain length of the median of the gap distribution is shown with varying $\tilde J^z$.
This allows an estimate of the exponent $\psi$ controlling the length-energy relationship Eq.~\eqref{eq:psi_def}, which takes the value $\psi = \frac 12$ at the free-fermion point.
The RRG scaling data suggest that there may be a systematic drift in $\psi$ as $\tilde J^z$ is varied toward the permutation-symmetric point $\tilde J^z = 1$, however it is difficult to exclude the possibility of a stable $\psi$ with a long crossover around $\tilde J^z = 1$.
In either case, this result does not support the $n=3$ Damle--Huse universality for this tricritical point.

\subsubsection{Symmetry properties of low-energy states}
\label{sec:rrg_symmetry}

As described in Sec.~\ref{subsubsec:gap}, in the non-interacting model $H_\xy$, the symmetry properties of the ground and low-lying states can be deduced from the single-particle excitations used to build the many-body states.
For convenience we relabel the $\Z_2 \times\Z_2$ symmetry sectors (always working on systems with $N \in 4\Z$): denote the free-fermion ground state sector $(g_x,g_y) = (+1,+1)$ as 0; the sector $(-1,-1)$ as 1; $(+1,-1)$ as 2; and $(-1,+1)$ as 3.
Along the critical line, $H$ has a statistical $\Z_2^\text{stat}$ symmetry exchanging sectors 2 and 3, and at the tricritical point a statistical $S_3$ relates sectors $1$, $2$, and $3$.

Beginning from a vacuum state in sector 0, the first many-body excited state---found by flipping the occupancy of the lowest-energy fermionic mode---comes from either sector 2 or 3, depending on which Majorana chain is involved.
The next excited state must also be associated with a low-energy single particle mode on one of the Majorana chains, thus will again come from sector 2 or 3.
The third many-body excited state can be of the same type, or can be associated with the simultaneous excitation of the two lowest energy single-particle states.
With a logarithmically broad disorder distribution, as at an IRFP, the third excited state is very likely to be of the latter type; thus we expect that for sufficiently long $N$, the four lowest-energy states of $H_\xy$ will most often come from the sectors $\{0,2,3,1\}$ or $\{0,2,2,0\}$, or their $\Z_2^\text{stat}$ counterparts $\{0,3,2,1\}$ and $\{0,3,3,0\}$.
The other free-fermion-allowed configurations are $\{0,2,3,2\}$, $\{0,2,3,3\}$, $\{0,2,2,2\}$, $\{0,2,2,3\}$, and $\Z_2^\text{stat}$ counterparts.

\begin{figure}[ht]
\includegraphics[width=\columnwidth]{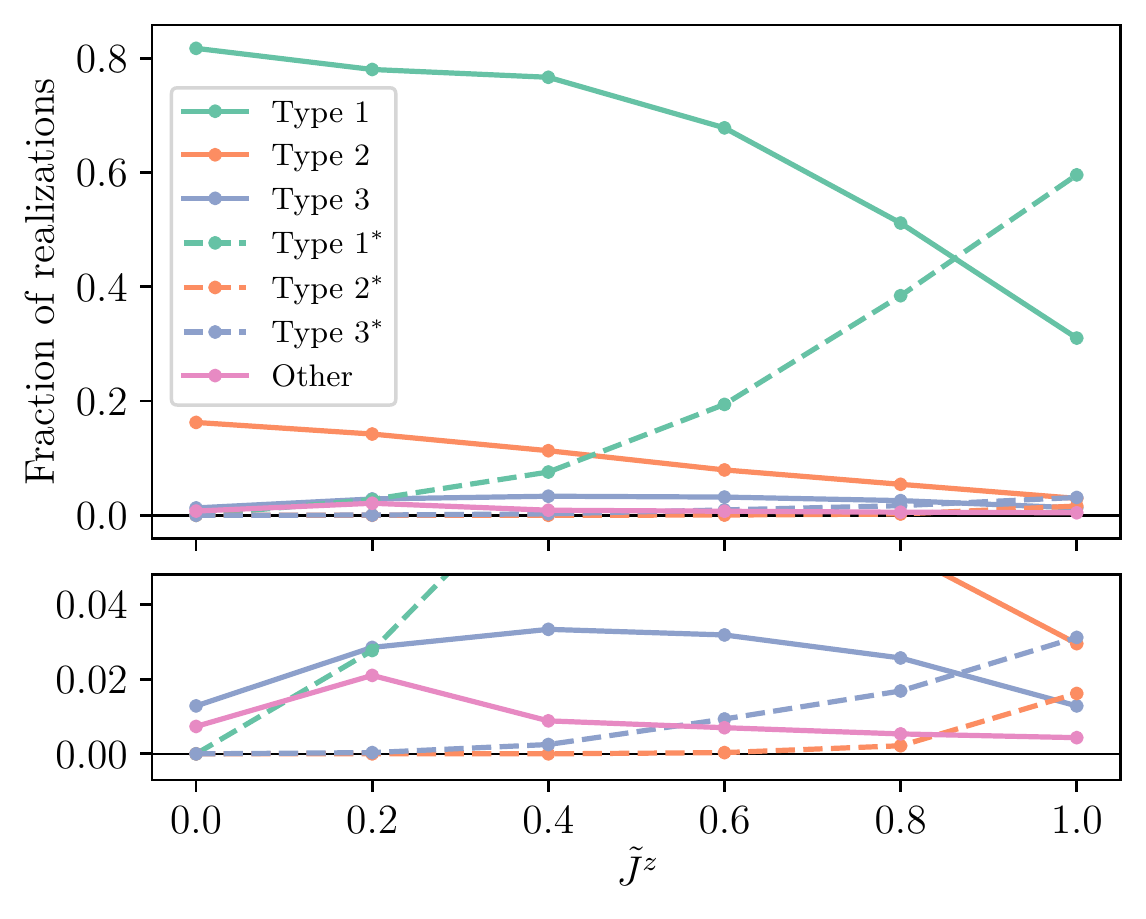}
\begin{tabular}{c|c|c}
Type 1 & Type 2 & Type 3 \\
\hline
\begin{tabular}{@{}c@{}} $\{0,2,3,1\}$, \\ $(2 \leftrightarrow 3)$
\end{tabular} &
\begin{tabular}{@{}c@{}} $\{0,2,2,0\}$, \\ $(2 \leftrightarrow 3)$
\end{tabular} &
\begin{tabular}{@{}c@{}} $\{0,2,3,2\}$, $\{0,2,3,3\}$, \\ $\{0,2,2,2\}$, $\{0,2,2,3\}$, $(2 \leftrightarrow 3)$ 
\end{tabular}
\end{tabular}
\caption{\label{fig:rrg_symmetry} 
Sampled estimates of the likelihood of the various symmetry patterns of low-energy states are shown as a function of $\tilde J^z$.
The lower panel shows the same data as the upper, zoomed in on the bottom of the $y$-axis.
The free-fermion-allowed Types 1, 2, and 3 are defined above and drawn with solid lines, and the free-fermion-disallowed Types $1^\ast$, $2^\ast$, and $3^\ast$ consist of all other partners under the action of the $S_3$ statistical symmetry, and are drawn with dashed lines.
Here we provide summary data which is averaged over system sizes $N=32,48,64,80$, with 6000 total disorder realizations for each value of $\tilde J^z$.
(In Fig.~\ref{fig:rrg_symmetry_N} we study the dependence on $N$.)
At $\tilde J^z=0$ we assume that only Types 1, 2, and 3 are present and include eigenstate permutations of the exact symmetry pattern for very small splittings $< 10^{-12}$; nevertheless there is still a low rate of ``Other'' instances.
}
\end{figure}

At the tricritical point this picture cannot apply, as the $S_3$ counterparts of the free-fermion-allowed configurations (these include, e.g., $\{0,1,2,3\}$ and $\{0,1,1,0\}$) must also occur and with equal likelihood; thus we study the critical line by tabulating occurrences of free-fermion-disallowed low-energy configurations in disorder realizations with finite $\tilde J^z$.
We classify the various configurations as described in the table in Fig.~\ref{fig:rrg_symmetry}, and their likelihood in our sample of disorder realizations is plotted.
Note that in this plot we have averaged over all system sizes, in order to provide an initial summary of the typical behavior (we will study the scaling behavior with $N$ later).

For $H_\xy$ the dominant pattern is Type 1, with a substantial minority of Type 2 and very few of Type 3.
The $S_3$ counterparts, which are forbidden in the picture of decoupled Majorana chains, are labeled Types $1^\ast$, $2^\ast$, and $3^\ast$.
The category ``Other'' includes all low-energy configurations not matching any of the types already described.
There is a very small, though finite, fraction of such instances; however these are nearly entirely associated with very small excitation gaps.
As already described, in such situations with very small splitting RRG cannot systematically identify the lowest-energy state or the exact sequence of excitations, so the precise order of symmetry sectors is not reproduced.
At $\tilde J^z=0$ we are able to ``interpret'' many such cases by assuming that the energy-permuted free-fermion-allowed symmetry pattern is the correct one, though away from this point a corrected type cannot be uniquely determined.
(At $\tilde J^z = 0$ some low-energy patterns found by RRG cannot be interpreted as one of the free-fermion-allowed configurations, and these are the realizations classified as ``Other'' at this point.) 

Moving away from $\tilde J^z = 0$, the Types $1^\ast$, $2^\ast$, and $3^\ast$ occur with increasing probability.
We find that Type 2 decreases more quickly for small $\tilde J^z$ than Type 1, which is in line with our understanding, developed in Sec.~\ref{sec:mf}, of the interaction as introducing correlations between the Majorana chains (such correlations make it less likely that the two lowest-energy single-particle states occur in the same Majorana chain).
The rate of ``Other'' instances is very low and decreasing with increasing $\tilde J^z$, suggesting that these remain attributable to errors due to small energy gaps, and the only new types of symmetry pattern appearing at low energy are those related to the free-fermion-allowed types by $S_3$.
As one expects from the definitions of each type, the frequency of Types $1^\ast$, 2, and $3^\ast$, are roughly twice those of Types 1, $2^\ast$, and 3, respectively, at $\tilde J^z = 1$.
Here the $S_3$ partners Types $1+1^\ast$ describe roughly 91\% of disorder realizations, with Types $2+2^\ast$ and $3+3^\ast$ describing roughly 4.5\% each.

\begin{figure}[ht]
\includegraphics[width=\columnwidth]{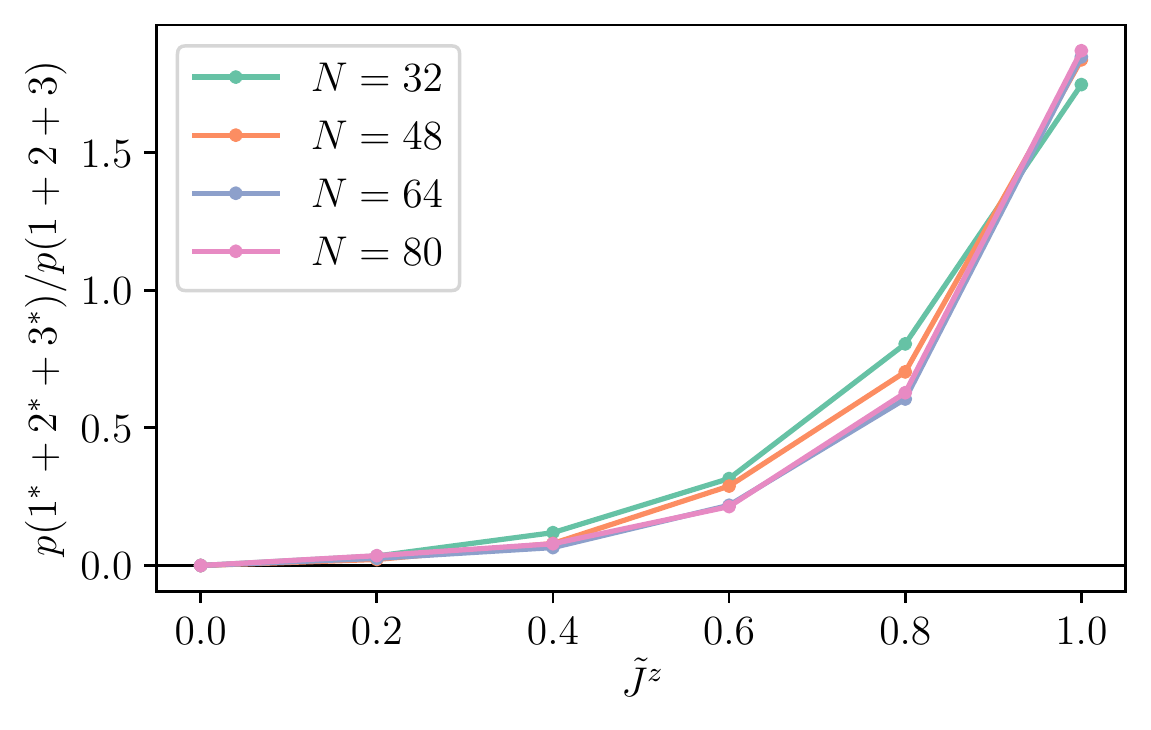}
\caption{\label{fig:rrg_symmetry_N} The ratio of the combined likelihood of the free-fermion-disallowed Types $1^\ast + 2^\ast + 3^\ast$  to the combined likelihood of Types $1+2+3$ is shown as a function of $\tilde J^z$, separately for system sizes $N=32,48,64,80$.
Each data point includes 1500 disorder realizations.
For intermediate $\tilde J^z \in (0,1)$, there is a consistent trend toward lower probabilities as the system size increases from $N=32$ to 64, meaning that the low-energy symmetry patterns of longer systems are more likely to be free-fermion-like.
The quantity $\frac{p(1^\ast+2^\ast+3^\ast)}{p(1+2+3)}$ is very similar for system sizes $N=64$ and 80 at all values of $\tilde J^z$, with the difference being within the apparent statistical scatter.
At the tricritical point $\tilde J^z = 1$ the predominant scaling behavior is reversed, and the quantity appears to be converging toward its long-distance fixed value from below with increasing system size $N$.
}
\end{figure}

From the above general picture of the low-energy states we learn that the critical line is characterized by the increasing probability of the free-fermion-disallowed symmetry partners Types $1^\ast$, $2^\ast$, and $3^\ast$ with increasing interaction strength $\tilde J^z$.
The dependence of these probabilities on system size provides a hint about the RG relevance or irrelevance of the interaction.
In Fig.~\ref{fig:rrg_symmetry_N}, we show the ratio of the combined likelihood of Types $1^\ast + 2^\ast + 3^\ast$ to that of Types $1+2+3$ as a function of $\tilde J^z$ for each system size separately \footnote{Normalizing by $p(1+2+3)$ is intended to eliminate the effect of the system size dependence of unclassifiable ``Other'' realizations, which should be associated with RRG errors.}.
While these data suffer from poorer statistics than those of Fig.~\ref{fig:rrg_symmetry}, there is a trend for all $\tilde J^z \in (0,1)$ toward lower probabilities with increasing $N$, meaning that at longer scales the disorder realizations appear more free-fermion-like.
The system sizes $N=64$ and 80 are quite similar by this measure, and the differences between these values are smaller than the apparent statistical noise.
In contrast, the dependence on system size is opposite at the tricritical point $\tilde J^z = 1$, as the likelihoods converge to their asymptotic value from below with increasing length scale.
In Secs.~\ref{sec:fp_interacting} and \ref{sec:discussion} we make a conjecture consistent with this observation, that the interactions may in fact be irrelevant but the SDRG generates a marginal perturbation (corresponding to the local correlation of renormalized terms, see Sec.~\ref{sec:mf}) which ultimately takes the system to a line of free-fermion fixed points with variable exponents.

\section{Mean field theory of interaction}
\label{sec:mf} 

Turning on $\tilde J^z >0$ introduces four-fermion interaction terms to the quadratic Hamiltonian $H_\xy$.
These terms couple the Majorana chains \X and \Y in such a way that the ground state is no longer analytically tractable under SDRG, which generates multi-fermion terms in the effective Hamiltonian that proliferate with increasing RG scale.
However, as mentioned in Sec.~\ref{sec:sdrg_xyz}, if at some point in the RG the interaction terms are typically weaker than the hopping terms then the effective higher-order descendants will be even weaker.
One might hope, then, that by beginning with a bandwidth $\tilde J^z \ll \tilde J^x,\tilde J^y$ the strength of these terms may be suppressed at all scales, leading to only a minimal effect on the criticality.

Based on this understanding, we consider the mean field theory by ``expanding'' the interaction into fermion bilinear terms.
In the Majorana language, the mean-field structure is particularly transparent; here the only symmetry-allowed bilinear terms act internally on the chains.
For $J^z_j \ll 1$,
\begin{align}
J_j^z (i\eta_j &\zeta_j)(i\eta_{j+1} \zeta_{j+1}) \approx \nonumber\\
& J_j^z \left( i\eta_j \zeta_{j+1} \ev{i\zeta_j \eta_{j+1}} + i\zeta_j \eta_{j+1} \ev{i \eta_j \zeta_{j+1}} \right).
\end{align}
This can also be seen in terms of the original spins, where the mean field theory takes the form
\begin{align}
J_j^z \sigma_j^z \sigma_{j+1}^z &= -J_j^z \sigma_j^x \sigma_{j+1}^x \sigma_j^y \sigma_{j+1}^y \nonumber\\
&\approx -J_z \left( \sigma_j^x \sigma_{j+1}^x \langle \sigma_j^y \sigma_{j+1}^y \rangle + \langle \sigma_j^x \sigma_{j+1}^x \rangle \sigma_j^y \sigma_{j+1}^y \right).
\end{align}
The effect of the allowed terms is to renormalize the existing couplings in the following way:
\begin{align}
(J^x_j)^\mathrm{mf} &= J^x_j + J^z_j \ev{i\eta_j \zeta_{j+1}} = J^x_j - J^z_j \ev{\sigma_j^y \sigma_{j+1}^y}~, \label{eq:meanfield_Jx} \\
(J^y_j)^\mathrm{mf} &= J^y_j  - J_j^z \ev{i \zeta_j \eta_{j+1}} = J^y_j  - J_j^z \ev{\sigma_j^x \sigma_{j+1}^x}~. \label{eq:meanfield_Jy}
\end{align}
With expectation values $\langle \cdot \rangle$ understood to be evaluated in the ground state of the mean field Hamiltonian with parameters $(J_j^x)^\mathrm{mf}, (J_j^y)^\mathrm{mf}$, the above represent self-consistency equations (i.e., minimization equations in the variational perspective of the mean field theory).
Because the Majorana chains remain decoupled, the mean-field theory can be solved in the analytic SDRG, at least in principle, by accounting for the distributions of effective $J_j^x$ and $J_j^y$ couplings no longer being independent.
In the following subsections we numerically investigate the universal behavior of this mean-field theory, and provide exact results from the analytic SDRG in Sec.~\ref{sec:sdrg_correlated}.

\subsection{Self-consistent Hartree--Fock treatment of interaction terms}

\begin{figure}[ht]
\includegraphics[width=\columnwidth]{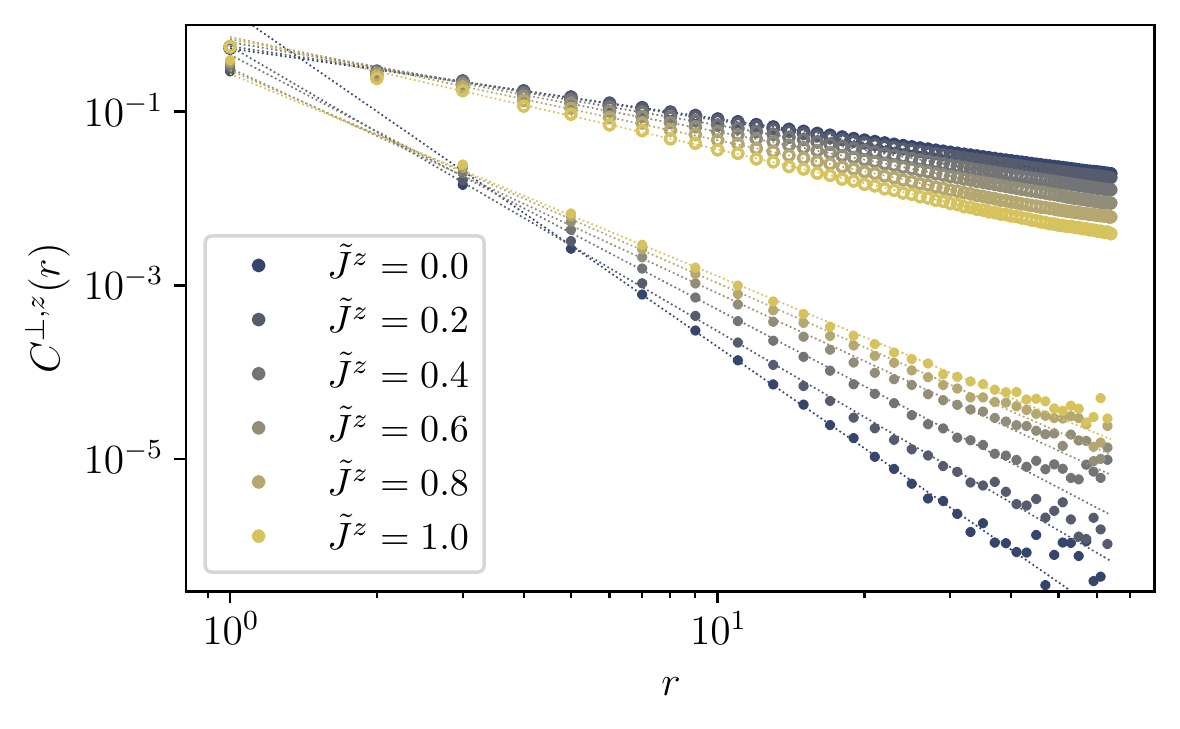}
\caption{\label{fig:mf1_bulk_correlations} Bulk correlations data from the self-consistent Hartree--Fock mean-field theory are shown with varying bandwidth $\tilde J^z$, up to separation $r=64$ in chains of length $N=128$.
Filled markers indicate $C^z(r)$ data, and open $C^\perp(r)$.
The disorder averages for each value of $\tilde J^z$ are taken over 25000 realizations and include only the middle half of the spin chain, as described in the caption to Fig.~\ref{fig:rrg_bulk_correlations}.
These simpler free-fermion calculations are cheaper to perform, and accordingly exhibit better statistics than those of Figs.~\ref{fig:rrg_bulk_correlations}--\ref{fig:rrg_symmetry_N}.
}
\end{figure}

\begin{figure}[ht]
\includegraphics[width=\columnwidth]{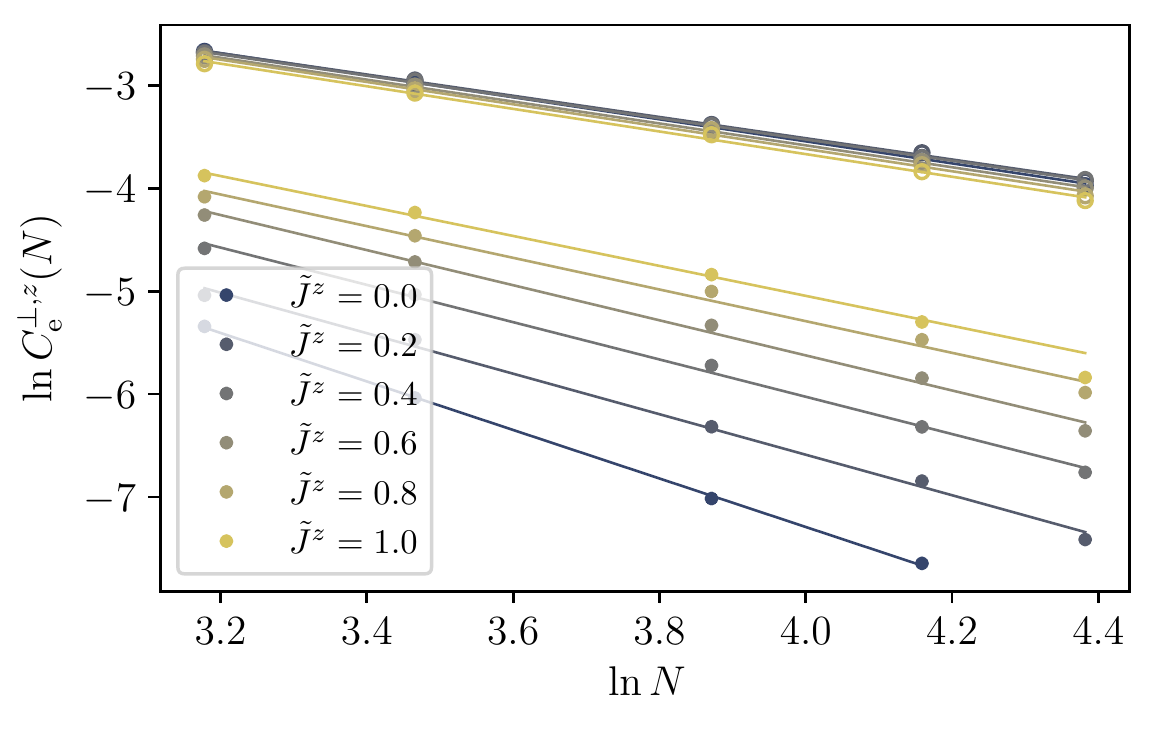}
\caption{\label{fig:mf1_end_correlations} End-to-end correlations data from the self-consistent Hartree--Fock mean-field theory are shown with varying bandwidth $\tilde J^z$.
Filled markers indicate $C^z_\e(r)$ data, and open $C^\perp_\e(r)$.
Each data point is the average end-to-end correlations from 25000 disorder realizations.
Because for small $\tilde J^z$ the likelihood of simultaneous end-to-end decimations is very low, in computing $C^z_\e(L)$ we are restricted to shorter systems in order to have reasonable statistics.
For example, in the SDRG picture, $C^z_\e(N) = e^{-7}$ corresponds to only $25000\times e^{-7} \approx 23$ important ``events.''
}
\end{figure}

\begin{figure}[ht]
\includegraphics[width=\columnwidth]{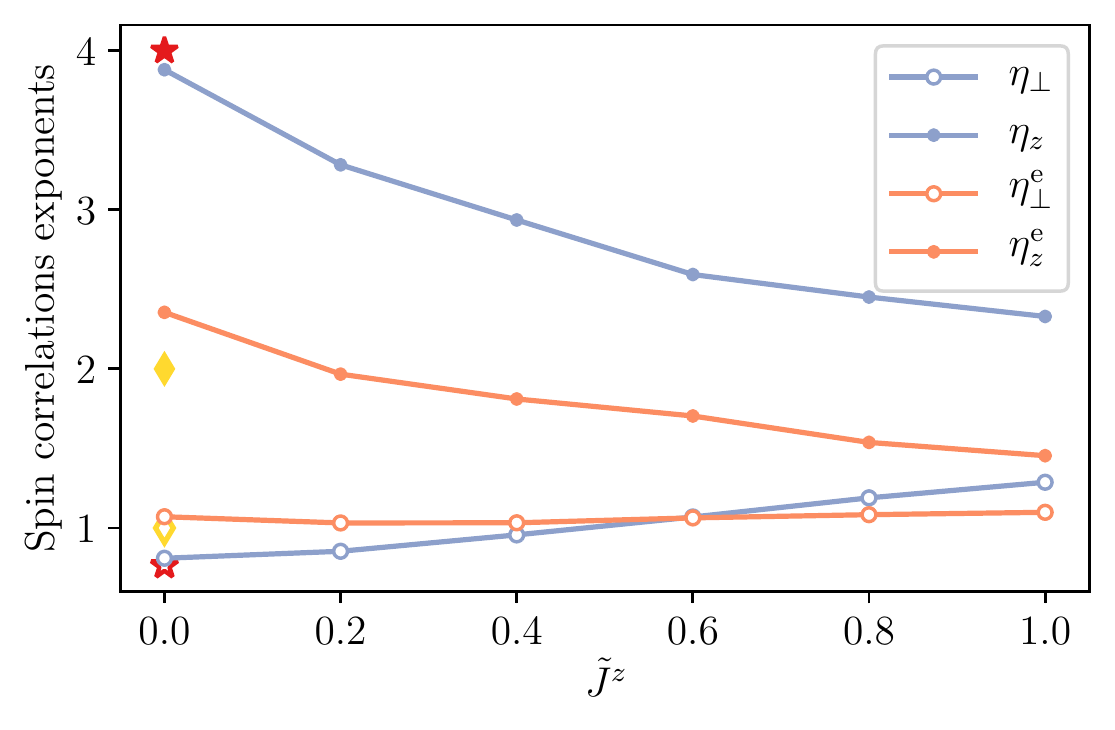}
\caption{\label{fig:mf1_exponents} Critical exponents are shown for the self-consistent Hartree--Fock mean-field theory with varying interaction strength $\tilde J^z \in [0,1]$, extracted from the correlations data in Figs.~\ref{fig:mf1_bulk_correlations} and \ref{fig:mf1_end_correlations}.
Both bulk and end-to-end exponents are included, with known results for the bulk correlations in the free-fermion model at $\tilde J^z=0$ indicated by red stars, and results for the end-to-end correlations by yellow diamonds.
The point $\tilde J^z = 1$ in this model does not feature any special symmetry.
}
\end{figure}

We first perform a self-consistent numerical study of the interaction term in the quadratic mean-field theory by directly implementing Eqs.~\eqref{eq:meanfield_Jx} and \eqref{eq:meanfield_Jy} in the BdG Hamiltonian, iteratively solving the ground state of the Hamiltonian and updating the mean-field couplings until reaching convergence.
The bulk correlations data in the thus determined mean field ground state are shown in Fig.~\ref{fig:mf1_bulk_correlations}, end-to-end correlations in Fig.~\ref{fig:mf1_end_correlations}, and a summary of the critical exponents in Fig.~\ref{fig:mf1_exponents}.

The key finding of the mean field treatment is that the power law exponents in all correlation functions do evolve with $\tilde J^z$ in a similar way to those of the interacting model.
This not necessarily expected since, e.g., in a clean XXZ model the mean field, while capturing some short-range energetics, cannot capture varying power laws in the fully interacting theory.
By understanding the features in the mean field responsible for capturing the varying power laws in the random XYZ chain, in the following sections we will be led to a plausible scenario for the physics of this system.

While the mean field theory is reasonably accurate for $\tilde J^z \leq 0.4$, it is evident from Fig.~\ref{fig:mf1_bulk_correlations} that the magnitudes of the mean-field correlation functions around $\tilde J^z = 1$ do not approach their actual values.
At the tricritical point of the interacting model the statistical $S_3$ symmetry of the Hamiltonian leads to the equivalence of the averages $C^\perp$ and $C^z$; as the mean field lacks this symmetry, it is not surprising that the distinction persists.
Moreover, there is nothing special about $\tilde J^z=1$ in the mean-field model; note also that this specific mean field does not allow any symmetry breaking, and we see that the best it can do upon increasing $\tilde J^z$ is to approach the XX chain, which is a poor approximation for $\tilde J^z \simeq 1$.

Nevertheless, buoyed by the success of the mean field at small $\tilde{J}^z$, we now examine more closely the effective parameters $(J_j^{x,y})^\mathrm{mf}$.
As the interaction strength is increased, the $J^x_j$ and $J^y_j$ terms tend to become more similar.
We can clearly see how this happens in the spin formulation of the self-consistent mean field of Eqs.~\eqref{eq:meanfield_Jx} and \eqref{eq:meanfield_Jy}: a large bare AFM $J_j^x > 0$ will tend to correlate $\sigma_j^x$ and $\sigma_{j+1}^x$ strongly antiferromagnetically (achieving $\langle \sigma_j^x \sigma_{j+1}^x \rangle \approx -1$ if this is the dominant coupling), and in the presence of AFM $J_j^z > 0$ this will lead to an increase of the effective AFM $J_j^y$ coupling, and vice versa.
However it is not clear what sort of model the full self-consistent mean field treatment actually constitutes, as the iterated nature of the solution could lead to long-range correlations effects among the couplings.
In the following section we propose a more straightforward model intended to broadly capture the features of this self-consistent Hartree--Fock mean field.
We will see that the ultra-short-range correlations among $J_j^x$ and $J_j^y$ identified above can already explain continuously varying power laws.

\subsection{Numerical study of random XY chain with locally correlated couplings}
\label{sec:mf2}

\subsubsection{Definition of locally-correlated XY model}

The rules Eqs.~\eqref{eq:meanfield_Jx} and \eqref{eq:meanfield_Jy} for the mean-field couplings modify bonds on one Majorana chain based on expectation values across the same bond on the other chain.
As a result, recalling that $J^z_j > 0$ for all $j$, the terms on a given bond---which at the mean-field level are strengthened by the interactions---develop correlations among themselves.
Terms on separate bonds also get correlated in less obvious ways, since the mean field ground state is influenced by all bonds, but we will proceed by ignoring such longer-range correlations among the couplings.
We refer to such an effective model as having ``local correlations,'' in order to distinguish from spatial correlations between terms on separated bonds.
One can mimic the behavior of the mean field theory and explore the effects of such correlations using the following parameterization of the couplings: for $A_j$, $B_j$ independent random variables and $\delta \in [0,1]$, let
\begin{align}
J^x_j &= \left(1-\frac\delta2\right) A_j + \frac\delta2 B_j~, \label{eq:mf2_jx}\\
J^y_j &= \frac\delta2 A_j + \left(1-\frac\delta2\right) B_j~. \label{eq:mf2_jy}
\end{align}
Tuning $\delta$ from 0 to 1 interpolates between fully independent couplings and the perfectly correlated case with \U1 symmetry.
That is, the parameterization runs along the line between the random XY and XX spin chains.
As mentioned in Sec.~\ref{sec:spin_chain}, \citet{fisher1994random} found that weak random anisotropy is marginal around the XX point, which is in the RS phase.
However it was not resolved whether this perturbation is truly marginal, or perhaps instead marginally relevant or irrelevant.
The mean-field numerical results in this section provide an investigation into this question, a topic which will be discussed in more detail within the analytic SDRG in Sec.~\ref{sec:sdrg_correlated}.

\subsubsection{Exact diagonalization study of locally correlated Majorana chains}

\begin{figure}[ht]
\includegraphics[width=\columnwidth]{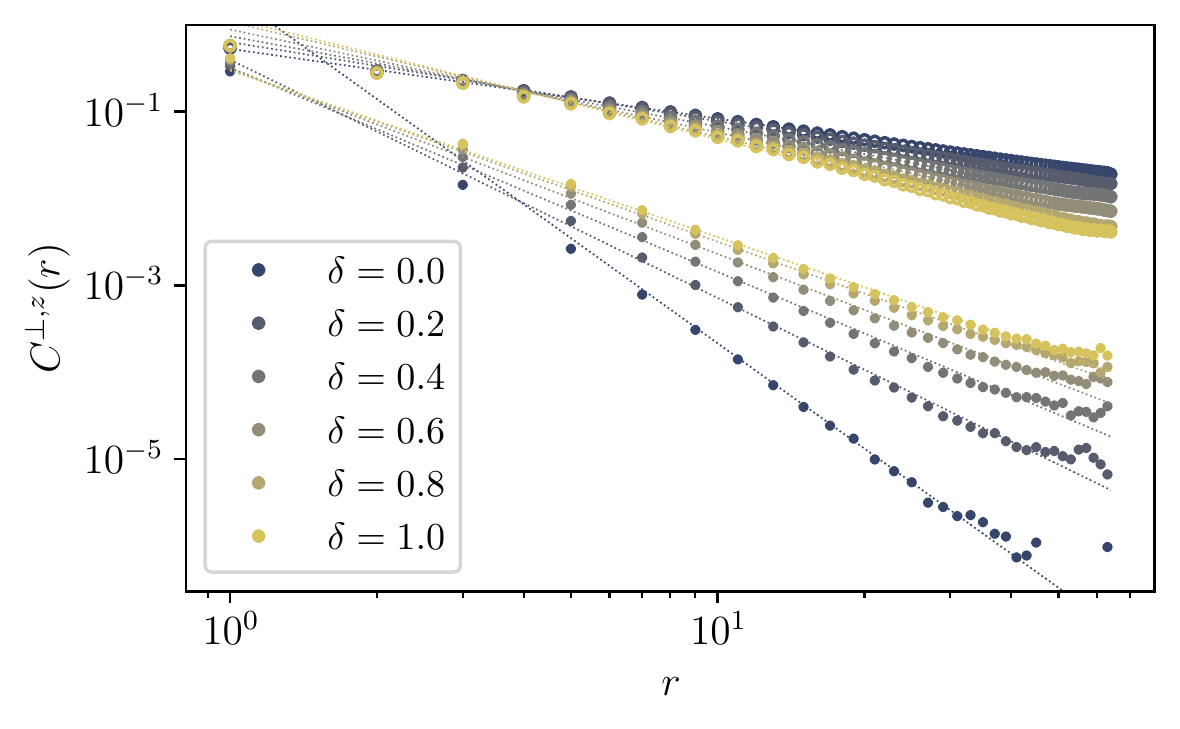}
\caption{\label{fig:mf2_bulk_correlations} Bulk correlations data from the locally-correlated effective XY model are shown with varying correlation $\delta$, up to separation $r=64$ in spin chains of length $N=128$.
Filled markers indicate $C^z(r)$ data, and open $C^\perp(r)$.
The disorder averages for each value of $\delta$ are taken over 25000 realizations.
In the average we include only the middle half of the spin chain, as described in the caption to Fig.~\ref{fig:rrg_bulk_correlations}.
}
\end{figure}

\begin{figure}[ht]
\includegraphics[width=\columnwidth]{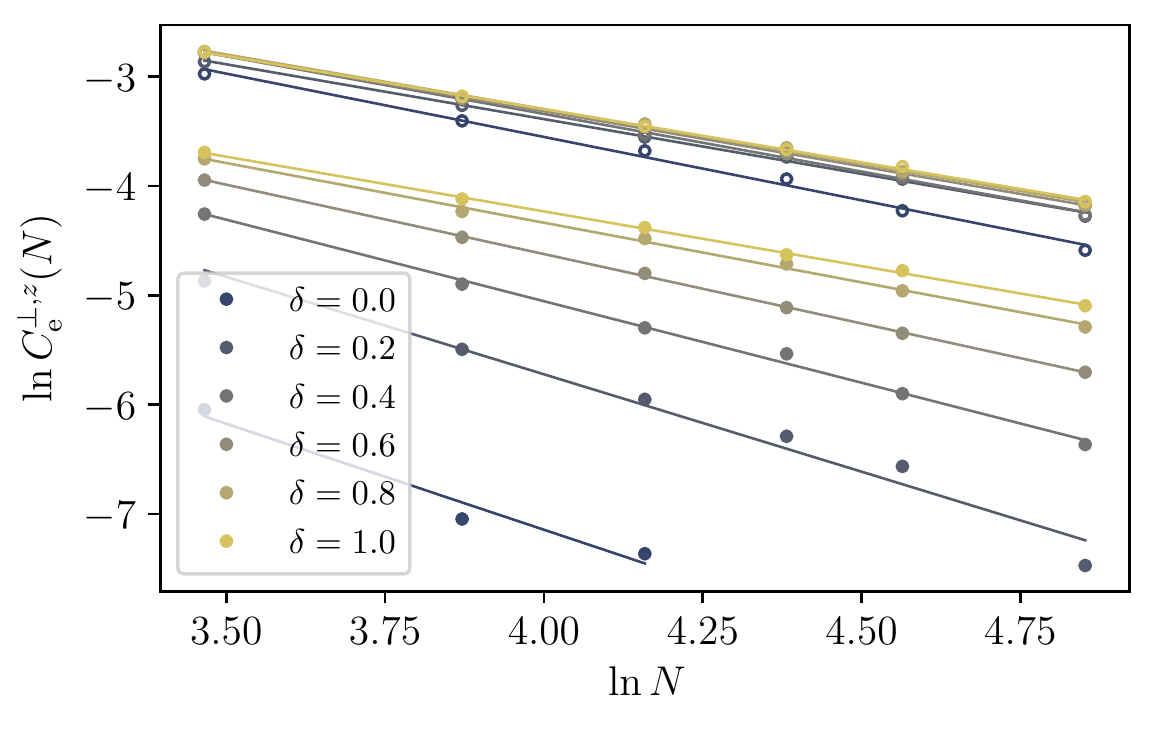}
\caption{\label{fig:mf2_end_correlations} End-to-end spin correlations data are shown in the locally-correlated effective XY model with varying coupling correlation $\delta$.
Filled markers indicate $C^z_\e(r)$ data, and open $C^\perp_\e(r)$.
System sizes $N=32,48,64,80,96,128$ are included and each data point averages over 25000 disorder realizations.
See Fig.~\ref{fig:mf2_exponents} for the critical power law decay exponents extracted from this data.
}
\end{figure}

\begin{figure}[ht]
\includegraphics[width=\columnwidth]{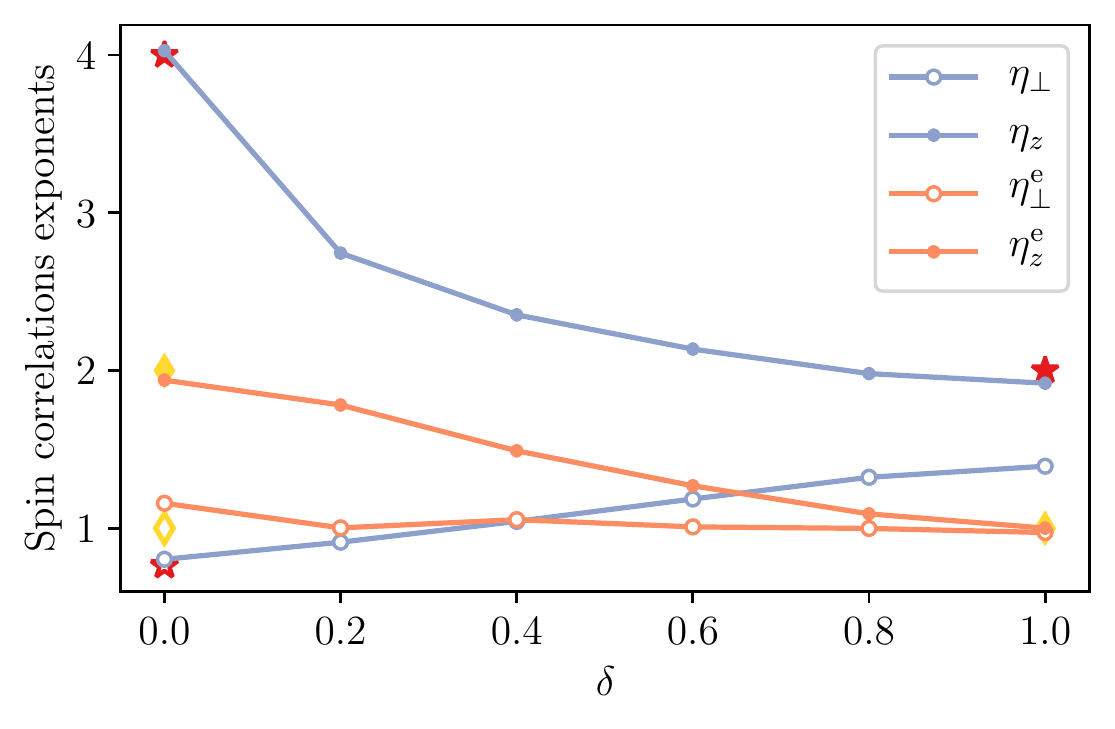}
\caption{\label{fig:mf2_exponents} Critical exponents governing spin correlations in the locally-correlated XY model with varying correlation parameter $\delta$ are shown, extracted from data shown in Figs.~\ref{fig:mf2_bulk_correlations} and \ref{fig:mf2_end_correlations}.
Both bulk and end-to-end exponents are shown, with known results for the bulk correlations in the uncorrelated XY model at $\delta=0$ indicated by red stars and known end-to-end critical spin exponents by yellow diamonds.
Known critical exponents for the \U1-symmetric XX model at $\delta=1$ are similarly indicated; in this case $\eta^\e_\perp = \eta^\e_z = 1$ and $\eta_\perp = \eta_z = 2$.
The discrepancy in $\eta_\perp$ is likely a result of a long crossover, as the disorder distribution of Eqs.~(\ref{eq:mf2_jx}) and (\ref{eq:mf2_jy}) is somewhat weaker than Eq.~\eqref{eq:disorder_distribution} for the same value $\Gamma = 2$.
}
\end{figure}

It is not immediately clear to what extent the locally-correlated free fermion effective model defined in Eqs.~\eqref{eq:mf2_jx} and \eqref{eq:mf2_jy} shares the qualitative features of the XYZ model, or indeed the self-consistent mean field theory.
We investigate this by repeating the measurements of bulk and end-to-end spin correlations in chains of similar length to the previous studies, now varying the coupling correlation parameter $\delta$.
Figures~\ref{fig:mf2_bulk_correlations}, \ref{fig:mf2_end_correlations}, and \ref{fig:mf2_exponents} demonstrate that these critical indices do vary continuously in a similar way to the interacting case.
Our observation that this mean-field approach indeed exhibits many of the qualitative features of the original case suggests that at least for small $\tilde J^z$, the primary effect of the interactions is to correlate the coefficients of the hopping terms on the two Majorana chains.
However, we emphasize that although the $\eta_z$ and $\eta_\perp$ converge to similar values at the XX point $\delta=1$ and the tricritical XYZ point $\tilde J^z = 1$, the reasons for this are not necessarily the same.
The mean field should not be taken too seriously as a picture of the interacting phase away from the perturbative regime.

\section{Locally correlated XY model in the random walk formalism}
\label{sec:sdrg_correlated}

Some types of disordered quantum Hamiltonian can be uniquely associated with a classical random walk (RW).
An alternative picture of the SDRG viewed through this connection is useful for understanding the properties of IRFP phases.
The RW formulation has previously been applied to both the RTFIM \cite{igloi1998random,igloi1998anomalous} and AFM quantum spin chains \cite{igloi2000random,motrunich2001dynamics}.
In this section we first review the RW for a single Majorana chain based on the SDRG procedure of Sec.~\ref{sec:sdrg_majorana}.
While all results for correlation functions in this case are known from Fisher's analytic solutions for flows approaching the RS fixed point, we demonstrate how to obtain some power law exponents from different arguments, which will generalize to the locally correlated XY chain where we do not have analytic flows.
We first obtain rigorous bounds in the continuum limit on the asymptotic scaling of the Majorana pairing probability (which determines the correlations of the $z$ component of spin in the random XX and XY chains) based on RW survival probability, a connection which had previously been noted in Ref.~\cite{igloi2000random}.
We then consider the problem of two locally correlated RWs, one for each Majorana chain, following the effective model developed in Sec.~\ref{sec:mf2}.
This system turns out to correspond to an anisotropic two-dimensional RW.
We again rigorously bound the likelihood of decimation using the RW survival probability, where we find that the power law exponent varies continuously with the local correlation parameter.
As a result, we are able to prove a specific form for continuously varying critical exponents of spin correlations in the locally-correlated effective model.

\subsection{RW formulation of SDRG for the Majorana chain}

Returning to the notation of Sec.~\ref{sec:sdrg_majorana}, define the logarithm of the energy associated with each bond in the Majorana chain Hamiltonian $\H_\M$ as $u_n = \ln (\tilde J/|h_n|)$, $n = 1,\ldots,N-1$.
Here $\tilde J$ is a bare bandwidth for the coupling terms, meant to evoke the parameters of the Hamiltonian Eq.~\eqref{eq:Hxyz}.
From Eq.~\eqref{eq:Hmaj} one sees that if $\tilde J^x = \tilde J^y$, in each Majorana chain of the random XY model the hopping terms are identically distributed.
Note that the signs of $h_n$ are not important for the discussion of probabilities of site pairings below, and are only needed to fix sign factors for the spin correlation functions, as discussed at the end of Sec.~\ref{sec:sdrg_majorana}.
We consider the specific disorder distribution Eq.~\eqref{eq:disorder_distribution} with $\tilde J^x = \tilde J^y = \tilde J = 1$.
Then the distribution of log-energies is exponential, with distribution parameter $\Gamma$:
\begin{equation}
\tau(u) = \frac1\Gamma e^{-u/\Gamma}~,~~u \in (0,\infty)~,
\end{equation}
which has mean $\ev u = \Gamma$ and variance $\Var(u) = \Gamma^2$.
The Majorana model $H_\M$ on $N$ sites is associated with a 1d RW $\m$, a Markov chain with state variables $(x_n,\sigma_n)$, $n=1,\ldots,N$, where $x_n \in \R$ is a cumulative log-energy defined below and $\sigma_n = (-1)^{n-1}$ is an internal $\Z_2$ variable determining the sign of the next step to be taken \footnote{That is, the RW takes alternating positive and negative steps depending on the sublattice of site $n$, and we choose step $n=1$ to be positive.
This is distinct from the alternating signs of the couplings in Eq.~\eqref{eq:Hmaj}, which are not invariant under a unitary rotation on the spins.}.
The discrete RW time $n$ matches the spatial index of the quantum chain.
A given disorder realization $\{h_j\}_{1 \leq j < N}$ corresponds to a RW step sequence $\{\sigma_j u_j\}_{1 \leq j < N}$: that is, the state of $\m$ at time $n = 1,\ldots,N$ is
\begin{equation}
\m[n] = \left(\sum_{j=1}^{n-1} \sigma_j u_j~,~\sigma_n \right).
\end{equation}
In the following we will sometimes leave the $\sigma_n$ state variable implicit, and refer to $x_n$ as $\m[n]$.
Let $\p(x,\sigma,n)$ be the distribution of $\m[n]$, which is governed by the master equation
\begin{equation}
\p(x,\sigma,n+1) = \int_0^\infty du\,\tau(u)\,\p(x - \sigma u,-\sigma ,n)~.
\label{eq:rho_master}
\end{equation}

We now consider the behavior under the SDRG of a RW \m associated with a Majorana chain $\H_\M$.
The largest local energy scale $|h_k|$, for some $k$, corresponds to the smallest log-energy $u_k$.
The effect of the Shreiffer--Wolff transformation up to second order is to eliminate the following hopping terms:
\begin{equation}
i h_{k-1} \gamma_{k-1} \gamma_k + i h_k \gamma_k \gamma_{k+1} + i h_{k+1} \gamma_{k+1} \gamma_{k+2}~,
\end{equation}
and to introduce the renormalized bond term
\begin{equation}
i h'_{k-1} \gamma_{k-1} \gamma_{k+2}~,~~h'_{k-1} = \frac{h_{k-1} h_{k+1}}{h_k}~.
\end{equation}
(There is also a shift of the leading energy scale, but this will not be important here.)
For the RW the new step is
\begin{equation}
\sigma_{k-1} u_{k-1}' = \sigma_{k-1} u_{k-1} + \sigma_k u_k + \sigma_{k+1} u_{k+1}~.
\end{equation}
In this way the SDRG transformation corresponds to a sequential ``smoothing'' of the RW, in which the global step of smallest magnitude and its neighbors are removed, and replaced by a treble step directly connecting $x_{k-1}$ and $x_{k+2}$.
For an illustration, the reader is referred to Fig.~8 in App.~B of the arXiv version of Ref.~\cite{motrunich2001dynamics}, or Fig.~1 of Ref.~\cite{igloi2005strong}.

We define an inversion operation $\mathfrak I$ acting on a RW \m of length $N$ as
\begin{equation}
\mathfrak I : \m[n] \mapsto \mathfrak I \m[n] = \m[N]-\m[N-n+1]~.
\end{equation}
That is, $\mathfrak I$ flips the spatial and time coordinates of \m.
(The constant shifts the starting point of $\mathfrak I \m$ to 0.)
We also define reflection $\mathfrak R_a$ of the spatial coordinate about the line $x=a$:
\begin{equation}
\mathfrak R_a : \m[n] \mapsto \mathfrak R_a \m[n] = 2a - \m[n]~.
\end{equation}
We will make extensive use of a ``gluing'' operation $\oplus$ which joins two RWs at their endpoints.
For RWs $\m_{1,2}$ with lengths $N_{1,2}$, then, $n = 1,\ldots,N_1 + N_2$,
\begin{equation}
(\m_1 \oplus \m_2)[n] = \begin{cases} \m_1 [n]~, & n \leq N_1 \\ \m_1[N_1]+\m_2[n-N_1]~, & n > N_1~. \end{cases}
\end{equation}
That is, the combined RW $\m_1 \oplus \m_2$ first performs the $N_1-1$ steps of $\m_1$, followed by the $N_2-1$ steps of $\m_2$.
It is assumed that the first step of $\m_2$ has opposite $\sigma$ state variable as compared to the last step of $\m_1$; this is required on the spin chain, where $\m_2$ begins and $\m_1$ ends on the same sublattice.

Using the above definitions a precise statement can be made about the decimation of a site $n=k$, which we suppose without loss of generality to be a local minimum.
For $k$ to have decimation partner $k' > k$ in the SDRG, with $k' - k = r$, a RW \m must admit a decomposition
\begin{equation}
\m = \mathfrak I \m_\mathrm{ext,L} \oplus \m_\mathrm{int} \oplus \mathfrak R_0 \m_\mathrm{ext,R}~,
\label{eq:rw_decomposition}
\end{equation}
where $\m_\mathrm{ext,L}$ has length $k$, $\m_\mathrm{ext,R}$ has length $N-k'+1$, $\m_\mathrm{int}[r] \equiv \Delta > 0$, and the following conditions hold:
\begin{enumerate}[{Condition} 1.]
\item$\m_\mathrm{int}[l]$ satisfies $x_l > 0$ for $l = 2,\ldots,r$, and attains the unique maximum $x_r = \Delta$; \label{rw:cond_int}
\item$\m_\mathrm{ext,L}$ and $\m_\mathrm{ext,R}$ reach height $x \geq \Delta$ before crossing 0. \label{rw:cond_ext}
\end{enumerate}
(For a pictorial description, see also
App.~B of the arXiv version of Ref.~\cite{motrunich2001dynamics}.)
These conditions relate the likelihood of a decimation pairing sites $k$ and $k'$ to the survival probability of the ``interior'' and ``exterior'' partial RWs on the fully bounded interval $(0,\Delta)$.
The physical interest of this quantity follows from the strong correlations shared by sites paired in the SDRG; in particular, the scaling of the decimation probability determines average spin correlations, as described in Sec.~\ref{sec:sdrg_correlations}.

Note that the writing of Eq.~\eqref{eq:rw_decomposition} is chosen so that the exterior RWs $\m_\mathrm{ext,L}$ and $\m_\mathrm{ext,R}$ have identical structure to the interior RW $\m_\mathrm{int}$.
That is, all walks evolve forward in time starting at step 1 with the first step being positive.
Implicit in this is the assumption that the inversion and reflection operations used result in identical probabilities for the RWs because the microscopic distributions for $u_n$ are identical for $n$ even and odd.

Focusing on asymptotic scaling (i.e., $n,r \gg 1$), we describe the RW in continuous time, passing from $n \to t$.
The central limit theorem specifies that a sum of random variables approaches a Gaussian distribution for sufficiently large $n$, provided only that the moments of the constituent distributions are bounded.
The variance of the continuum distribution is $\Var(x) = \Var(u) t$.
The effect of the internal state variable $\sigma$ can be accounted for by noting that sites which decimate together necessarily inhabit distinct sublattices.
This means that one additional $\sigma = +1$ step is always taken.
The mean of the probability distribution, then, is the expectation value of this step: $\ev x \equiv x_0 = \ev u$ \footnote{This can also be derived from the continuum expression of the master equation Eq.~\eqref{eq:rho_master}.}.
The asymptotic density in free space we denote by
\begin{equation}
G_\text{free}(x,t) = \frac{1}{\sqrt{2 \pi \Var(u) t}} \exp \left[ -\frac{(x-x_0)^2}{2 \Var(u) t}\right]~.
\label{eq:rho_free_space}
\end{equation}

Now the continuum limit of Eq.~\eqref{eq:rho_master} is the diffusion equation \cite{hughes1995random}
\begin{equation}
\frac{\partial}{\partial t} G(x,t) = D \frac{\partial^2}{\partial x^2} G(x,t)~,
\label{eq:diffusion}
\end{equation}
with diffusion constant $D = \Var(u)/2$.
Eq.~\eqref{eq:rho_free_space} is the Green's function of Eq.~\eqref{eq:diffusion} on $x \in \R$ with initial condition $G(x,t=0) = \delta(x-x_0)$.
This illustrates that the continuum limit of the RW can be treated as a diffusing particle initially localized at $x = x_0$.
Accordingly, in the following sections we use the language of the diffusion problem, referring to the counterparts of discrete RWs associated with particular Majorana Hamiltonians as ``paths,'' ``histories,'' or ``trajectories.''
We also sometimes write the initial condition explicitly, as $G(x,t;x_0)$.
Finally, we will use the notation defined in this section for the discrete case, e.g., $\mathfrak I$, $\mathfrak R_a$, and $\oplus$, to also refer to the counterparts of these operations in the continuum.

\subsection{Rigorous bounds on critical exponents in the Majorana chain from RW survival}
\label{sec:rw_bounds_1d}

The diffusion equation on the fully bounded interval $(0,\Delta)$, i.e., with absorbing boundary conditions at $x=0$ and $x=\Delta$, can be solved straightforwardly by harmonic expansion.
From the time-dependent solution one can directly calculate the scaling of the asymptotic decimation probability and reproduce Fisher's detailed results in Refs.~\cite{fisher1994random,fisher1995critical}.
However, in Sec.~\ref{sec:rw_bounds_2d} the fully bounded geometry for two locally correlated Majorana chains becomes too complicated to solve this way.
Instead we employ a different approach by proving upper and lower bounds with the same power-law scaling, based on the survival probability in a semi-infinite domain.
A similar method will work also for the locally correlated effective model with an arbitrary degree of correlation.

First consider the survival probability of a RW in the semi-infinite interval at time $t>0$.
As in the free case Eq.~\eqref{eq:rho_free_space}, the initial condition on the constrained density $G(x,t)$ is $G(x,t=0) = \delta(x-x_0)$, but an absorbing boundary is present at $x=0$, restricting the solution domain to $x \in (0,\infty)$ and terminating trajectories that reach $x=0$.
The boundary condition $G(x=0,t) = 0$ is accounted for by placing an ``image charge'' at $x = -x_0$ and superposing the distributions: $G(x,t) = G_\text{free}(x,t;x_0)-G_\text{free}(x,t;-x_0)$.
We generally work in a ``scaling limit,'' where
\begin{align}
G(x,t;x_0) &= \frac{1}{\sqrt{\pi D t}} e^{-(x^2 + x_0^2)/4 D t} \sinh\left(\frac{x x_0}{2 D t}\right) \\
&\approx \frac{x x_0}{\sqrt{4 \pi (D t)^3}} e^{-x^2/4 D t}~,
\end{align}
assuming in the last line $x_0 \ll \sqrt{Dt}$.
This approximation is valid at late times in integrals over the spatial coordinate, as the exponential factor strongly mitigates the error introduced, and allows us to extract leading power-law behaviors.
The survival probability in the semi-infinite geometry in the scaling limit is
\begin{equation}
S(t) = \int_0^\infty dx\,G(x,t;x_0) = \frac{x_0}{\sqrt{\pi D t}}~.
\end{equation}

\subsubsection{End-to-end decimation probability for a single finite Majorana chain}
\label{sec:rw_1d_end}

In order to support end-to-end decimation between sites $1$ and $N$, the RW $\m[n=N]$ associated with a finite Majorana chain of length $N$ need only satisfy Condition \ref{rw:cond_int} of the previous section, with $r=N$.
In the continuum limit for the RW ($N \to L$), the likelihood that the left end $t=0$ is involved in the final decimation is given by the survival probability $S(t=L) \sim 1/\sqrt L$; however Condition \ref{rw:cond_int} additionally requires that its decimation partner be the right end $t=L$.
Applying $\mathfrak I$ to \m, one sees that the requirement to reach a maximum at $t=L$ takes the same form as the absorbing boundary condition $x=0$ near $t=0$.
Thus a naive estimate of the end-to-end decimation probability $p_\e(L)$ is the independent survival of the two ends, or $S(L)^2 \sim 1/L$.
Although these events are not actually independent, we will show that the naive estimate turns out to give the correct scaling.
Some intuition for this is that surviving histories tend to be located increasingly far away from the absorbing boundary \cite{redner2001guide}: consequently, the ``special'' low-probability behavior is confined to the neighborhood of the ends, while the middle of the RW can be allowed to be nearly typical.
A precise statement of these schematic remarks is that we are able to determine the scaling of $p_\e(L)$ by considering two independent ``half-RWs'' $\m_{1,2}$ of length $t=L/2$, constructing RWs of length $L$ which satisfy Condition \ref{rw:cond_int} as $\m = \m_1 \oplus \mathfrak I \m_2$.

To be more concrete, we first give a rigorous upper bound on the end-to-end decimation probability $p_\e(L)$.
Any RW \m can be decomposed as $\m = \m_1 \oplus \mathfrak I \m_2$, that is, into two independent ``half-RWs'' running up to time $t=L/2$, one running over times $t' \in [0,L/2]$, and the other over $t' \in [L/2,L]$, with the two RWs properly glued at their respective time $t' = L/2$.
It may be the case that $\m_1$ and $\m_2$ never reach the absorbing boundary, and thus each is considered a surviving RW in the semi-infinite geometry.
Any RW instance of length $L$ producing an end-to-end pairing in the SDRG, i.e., satisfying Condition \ref{rw:cond_int}, indeed decomposes in this way, with only one absorbing boundary in each case.
The converse statement is not true, because when such two surviving trajectories are joined, we cannot guarantee that the full RW satisfies Condition \ref{rw:cond_int}.
Thus, the desired probability $p_\e(L) \leq S(L/2)^2 \sim 1/L$.

To prove a lower bound on $p_\e$ we construct a subset of all paths satisfying Condition \ref{rw:cond_int} by considering certain $\m_1$ and $\m_2$, each of length $t=L/2$, which when glued together as $\m_1 \oplus \mathfrak I \m_2$ satisfy the criterion.
Again, in the present case we can solve the problem with two absorbing boundaries, but we want to demonstrate how to extract the behavior using the semi-infinite solution, where the geometry is simpler, as this will be the only option for the locally correlated model.
Specify constants $\alpha$ and $\beta$, $0 < \alpha < \beta \leq 2 \alpha$, and define a \emph{target window} $x \in [\alpha \sqrt{Dt},\beta\sqrt{Dt}]$ for a time $t>0$.
In the problem with one absorbing boundary at $x=0$, the fraction of surviving trajectories contained in the target window at $t$ is
\begin{align}
p_\text{w}(\alpha,\beta) &= \frac{1}{S(t)}\int_{\alpha\sqrt{Dt}}^{\beta\sqrt{Dt}} dx\,G(x,t) = e^{-\alpha^2/4}-e^{-\beta^2/4}~.
\label{eq:p_window_uncorrected}
\end{align}
That is, a constant fraction $p_\w(\alpha,\beta)$ of the surviving density of RWs at time $t$ is located within the target window.

The above calculation Eq.~\eqref{eq:p_window_uncorrected} leads to an overcounting of valid paths which can be glued to satisfy Condition \ref{rw:cond_int}, because it includes ``dangerous'' histories which take an excursion to large $x$ values before returning to the target window at time $t$.
Half-RWs $\m_1$ and $\m_2$ constrained in this way and glued as $\m_1 \oplus \mathfrak I \m_2$ may cross the eventual decimation log-energy scale $\Delta$ prematurely, which would spoil the lower bound.
To account for the dangerous cases, we exclude those histories which ever cross $x = \beta \sqrt{Dt}$ and then return to the target window.

The way we achieve the exclusion is the following.
Suppose that a history $\m[t']$, $t' \in [0,t]$, performs $q$ crossings of the line $x = \beta \sqrt{Dt}$ at times $\{t_1,t_2,\ldots,t_q\}$ before returning to the target window at $t' = t$.
Immediately after $t_q$, the history must travel downwards and remain below $x = \beta \sqrt{Dt}$ until $t' = t$.
We apply the following transformation:
\begin{equation}
\mathfrak T : \m = \m_{t'\leq t_q} \oplus \m_{t' > t_q}~\mapsto~\m_{t'\leq t_q} \oplus \mathfrak R_{\beta\sqrt{Dt}}\,\m_{t' > t_q}~,
\end{equation}
where as indicated by the subscripts $\m_{t'\leq t_q}$ describes the RW up to time $t' = t_q$ and $\m_{t' > t_q}$ the section $t' \in (t_q,t]$.
$\mathfrak T$ does not change the earlier partial RW but reflects the later about the line $x = \beta \sqrt{Dt}$.
Because $\m[t] \in [\alpha\sqrt{Dt},\beta\sqrt{Dt}]$, the transformed endpoint $\mathfrak T \m[t]$ necessarily lies in a ``shadow window'' $x \in [\beta \sqrt{Dt},(2\beta - \alpha)\sqrt{Dt}]$.
Moreover, the likelihood of the trajectory is unaffected by $\mathfrak T$.
Now every dangerous path with $q \geq 1$ crossings can be identified with a transformed partner terminating in the shadow window and having the same probability.
Thus the density in the shadow window at time $t$ upper bounds the contribution to the density in the target window arising from dangerous histories.
(The upper bound is not saturated, because a trajectory included in the shadow window could deviate above $x = 2\beta\sqrt{Dt}$ for some $t' \in (t_q,t]$, and this RW would have no $\mathfrak T^{-1}$ counterpart due to the absorbing boundary at $x=0$.)

From the previous calculation, the fraction of the surviving density contained in the shadow window is $p_\sw(\alpha,\beta) = e^{-\beta^2/4} - e^{-(2\beta-\alpha)^2/4}$.
Consequently a lower bound on the density of \emph{valid} surviving histories in the target window at time $t$ is given by
\begin{align}
p^\mathrm{corr}_\w(\alpha,\beta) &= p_\w(\alpha,\beta) - p_\sw(\alpha,\beta) \\
&= e^{-\alpha^2/4}-2e^{-\beta^2/4} + e^{-(2\beta-\alpha)^2/4}~.
\end{align}
There is an extended region of $(\alpha,\beta)$ for which the coefficient is positive; for example, $p^\mathrm{corr}_\w(\alpha=2,\beta=4) \approx 0.33$.

\begin{figure}[ht]
\includegraphics[width=\columnwidth]{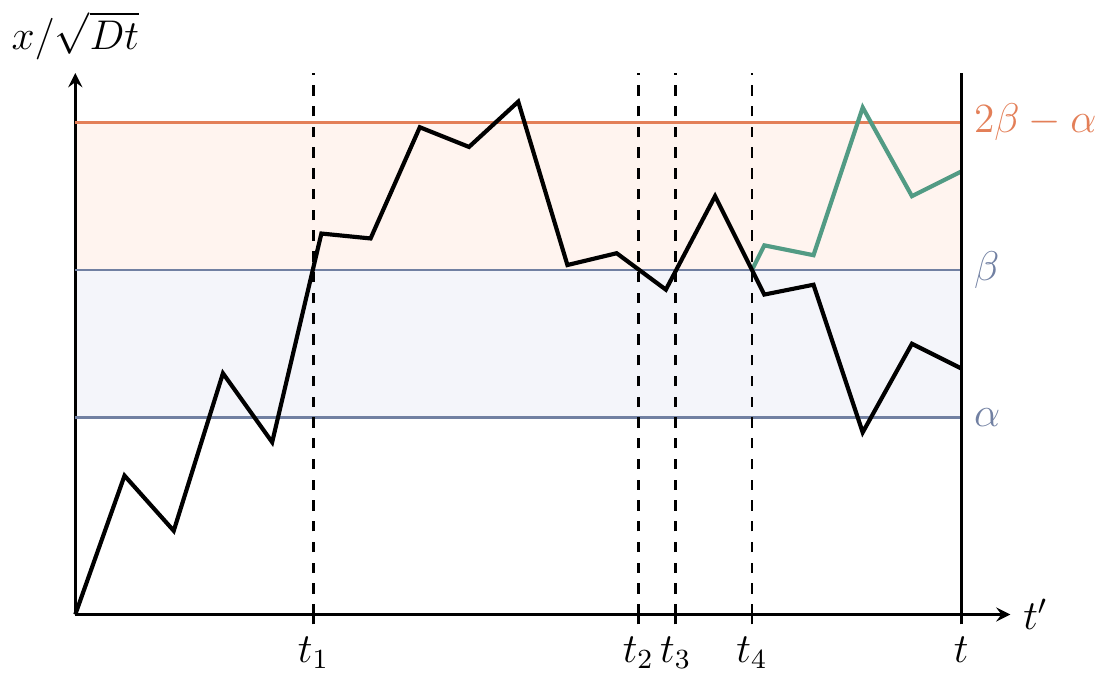}
\caption{\label{fig:shadow_window} A dangerous trajectory contributing to the counting $p_\w$ of the density in the target window, colored in blue, is illustrated.
The shadow window used to eliminate these trajectories is also shown, colored in orange.
The particular history \m shown has $q=4$ crossings of the upper limit of the target window and the reflected partial path $\mathfrak R_{\beta\sqrt{Dt}}\,\m_{t' > t_q}$, terminating in the shadow window, is shown in green.
Because the diffusion is unbiased, both \m and the transformed $\mathfrak T \m$ path have the same probability, and as any such dangerous trajectory has a counterpart under the transformation, the density in the shadow window upper-bounds the associated contribution to the density in the target window.
}
\end{figure}

Now take $t = L/2$.
Two RWs $\m_1$ and $\m_2$ fulfilling the criteria above are suitable for constructing a RW of length $L$ which satisfies Condition \ref{rw:cond_int} as $\m = \m_1 \oplus \mathfrak I \m_2$.
The result is a trajectory of length $L$ reaching a maximum at $t=L$ (assured by taking $\beta \leq 2\alpha$) without crossing $x=0$.
Not all RWs of length $L$ which support end-to-end decimation in the SDRG can be constructed this way, only those with $\m[L/2]$ lying in the target window and $\m[t' \leq L/2]$ below the upper limit of the target window, but every RW coming from this construction evidently satisfies Condition \ref{rw:cond_int}.
Thus this probability is a lower bound on $p_\e(L) \geq [p^\mathrm{corr}_\w(\alpha,\beta)S(L/2)]^2 \sim 1/L$.

Together with the upper bound, this establishes the scaling of end-to-end decimation probability $p_\e(L)$---and thus the power law for end-to-end correlations in a single random Majorana chain---as $1/L$.

\subsubsection{Bulk decimation probability in a single Majorana chain}
\label{sec:rw_1d_bulk}

Guaranteeing decimation away from the edges of a Majorana chain requires satisfying both Conditions \ref{rw:cond_int} and \ref{rw:cond_ext}.
To find the probability $p_\b(r)$ of decimation at scale $r$ in the bulk---i.e., that two fixed sites separated by $r$ are decimated as a pair---we decorate interior RWs $\m_\mathrm{int}$ by gluing exterior RWs to the left and right.
We showed that the probability of such an $\m_\mathrm{int}$ is $p_\e(L=r) \sim 1/r$, so we need only find suitable exterior RWs satisfying Condition \ref{rw:cond_ext} (while bearing in mind conditions involving both interior and exterior RWs).

For the probabilities associated with the exterior walks, we are interested in the likelihood $\omega(x;A)$ that a RW with spatial coordinate $x'$ starting from $x' = x \geq 0$ eventually reaches a value $x' = A$ before being absorbed at the domain boundary $x' = 0$.
We require the consistency condition $\omega(x;A) = \ev{\omega(x-dx;A)}$, where the average is taken over sufficiently small displacements $dx$, and $\ev{dx} = 0$, $\ev{(dx)^2} \neq 0$ (reflective of the microscopic step distribution) \cite{redner2001guide,bray2013persistence}.
Taylor expanding leads to Laplace's equation $\nabla^2 \omega = 0$ which, together with the boundary conditions $\omega(0)=0$ and $\omega(A)=1$, has solution $\omega(x;A) = x/A$.

A lower bound on $p_\b(r)$ is now straightforward based on $\m_\mathrm{int}$ as defined in Sec.~\ref{sec:rw_1d_end}, coming from a subset of all RWs of length $L=r$ supporting end-to-end decimation.
Any such $\m_\mathrm{int}$ is constructed from two glued half-RWs, each terminating at $t = r/2$ inside of a target window $x \in [\alpha\sqrt{D r/2},\beta\sqrt{D r/2}]$; thus the total and maximum deviation at $t=r$ is bounded above by $\Delta(r) = \beta\sqrt{2Dr}$.
Given $\m_\mathrm{int}$, the probability of a suitable exterior RW $\m_\mathrm{ext,L}$ or $\m_\mathrm{ext,R}$ is greater than or equal to $\omega(x_0;\Delta(r))$; writing a full RW satisfying all conditions, we find that $p_\b(r) \geq [p^\mathrm{corr}_\w(\alpha,\beta)S(r/2)]^2 \omega(x_0;\Delta(r))^2 \sim r^{-2}$.

In the same spirit as the upper bound on end-to-end decimation probability, consider $\m_\mathrm{int} = \m_1 \oplus \mathfrak I \m_2$; that is, decomposed as two half-RWs surviving until $t=r/2$, with final spatial deviations $\Delta_1$ and $\Delta_2$ and likelihoods $G(\Delta_1,r/2;x_0)$ and $G(\Delta_2,r/2;x_0)$, respectively.
All RWs with end-to-end decimation are of this form.
Now incorporating the probability of exterior RWs which must reach a height $\Delta_1 + \Delta_2$, the likelihood of the full RW provides an upper bound on the probability of bulk decimation:
\begin{align}
p_\b(r) &\leq \int_0^\infty \int_0^\infty d\Delta_1 d\Delta_2\, G(\Delta_1,r/2;x_0)\,G(\Delta_2,r/2;x_0) \nonumber \\
&\qquad\qquad\qquad\qquad\qquad\times\omega(x_0; \Delta_1 + \Delta_2)^2~.
\end{align}
Making use of $\omega(x_0; \Delta_1 + \Delta_2)^2 \leq \frac12  \omega(x_0; \Delta_1)\omega(x_0; \Delta_2)$ the integrals factorize, and we find
\begin{align}
p_\b(r) &\leq \frac 12\left[\int_0^\infty d\Delta_1\,G(\Delta_1,r/2;x_0)\,\omega(x_0;\Delta_1)\right]^2 \\
&= \frac{x_0^4}{2(D r)^2}.
\end{align}
Again these upper and lower bounds exhibit the same scaling, proving that $p_\b(r) \sim r^{-2}$ for a single Majorana chain, in agreement with known results (see the XX case in Sec.~\ref{sec:sdrg_correlations}).

\subsection{Locally-correlated Majorana chains as a two-dimensional RW}

To make statements about locally correlated Majorana chains requires dealing simultaneously with two RWs (returning for the moment to the discrete formulation) $\m_x[n]$ and $\m_y[n]$, associated respectively with the $\mathcal{X}$ and $\mathcal{Y}$ Majorana hopping chains.
In the general case, the steps taken by each at time $n$ are not independent, being instead drawn from a joint distribution $\mu(u,v)$.
If the full state of the system is specified by variables $(x_n,y_n,n)$, the master equation for the probability distribution $\p(x,y,n)$ is
\begin{equation}
\p(x,y,n+1) = \int du \int dv\,\mu(u,v)\,\p(x-u,y-v,n)~.
\end{equation}
This is however just the master equation for a RW in two dimensions (2d).
In the natural 2d vector notation with $\bm x = (x,y)^\top$ and $\bm u = (u,v)^\top$,
\begin{equation}
\p(\bm x,n+1) = \int d^2\bm u\,\mu(\bm u)\,\p(\bm x-\bm u,n)~.
\label{eq:rho_master_corr}
\end{equation}
The continuum limit of the master equation Eq.~\eqref{eq:rho_master_corr} is determined by the details of the microscopic distribution $\mu$, and does not in general reduce to the simple Laplacian.
As a remedy we begin by transforming the problem into isotropic diffusion.

Let $\mu$ be centered, with covariance matrix \footnote{The central limit theorem allows us to ignore higher-order moments, provided only that they are finite, so for our purposes all acceptable microscopic distributions are fully characterized by this one-parameter family of covariance matrices.}
\begin{equation}
\Sigma = \sigma^2 \begin{bmatrix} 1 & \delta \\ \delta & 1 \end{bmatrix}~,
\end{equation}
where $\mathrm{corr}(u, v) = \mathrm{cov}(u,v)/\sigma^2 \equiv \delta \in [0,1]$, with fixed $\sigma^2 = \mathrm{Var}(u) = \mathrm{Var}(v)$.
(The value of $\delta$ here is related to, but not necessarily the same as, the bare $\delta$ defined in Sec.~\ref{sec:mf2}.
$\delta > 0$ implies positive correlation between $u$ and $v$, as observed in the mean field for the AFM spin chain.)
The continuum limit of evolution driven by $\mu$ is anisotropic diffusion along the eigenvectors of $\Sigma$, $\hat e_\pm = \frac{1}{\sqrt 2}(1,\pm 1)^\top$, with diffusion coefficients $D_\pm = \frac{\sigma^2}{2} (1\pm\delta)$.

The 2d RW evolves by isotropic diffusion under a linear transformation of the plane $\mathcal W : \bm x \mapsto \tilde{\bm x} \equiv W \bm x$, with 
\begin{equation}
W = \frac{1}{\sqrt 2}\begin{bmatrix} \frac{1}{\lambda} & -\frac{1}{\lambda} \\ \lambda & \lambda \end{bmatrix},~~\lambda \equiv \left(\frac{1-\delta}{1+\delta}\right)^{1/4}~.
\end{equation}
$\mathcal W$ performs a rotation about the origin by $\pi/4$, followed by a $\delta$-dependent anisotropic rescaling.
There is a divergence at $\delta = 1$, where $\Sigma$ is rank-deficient; this reflects the fundamentally one-dimensional nature of the perfectly correlated case.
We will refer to the $(x,y)$ coordinates of the original problem as the ``physical geometry,'' and the image $(\tilde x,\tilde y)$ of $\mathcal W$ as the ``solution geometry,'' where the governing equation is isotropic diffusion, now with coefficient $D \equiv \sqrt{D_+ D_-} = \frac{\sigma^2}{2}\sqrt{1-\delta^2}$:
\begin{equation}
\frac{\partial}{\partial t} G = D \left(\frac{\partial^2}{\partial \tilde x^2} + \frac{\partial^2}{\partial \tilde y^2} \right) G~.
\label{eq:2d_solution_diffusion}
\end{equation}

\subsection{Rigorous bounds on critical exponents in the locally correlated model}
\label{sec:rw_bounds_2d}

\subsubsection{End-to-end decimation probability for two locally correlated finite Majorana chains}

Investigating end-to-end decimation directly in the exact solution for the fully bounded geometry would necessitate solving Eq.~\eqref{eq:2d_solution_diffusion} in a parallelogram.
A harmonic decomposition is not possible here, and as far as we are aware the solution requires a prohibitively complicated Schwarz--Christoffel conformal transformation usually performed numerically \cite{driscoll2002schwarz}.
Nevertheless, analytic results for two Majorana chains with arbitrary local correlations are possible by utilizing the connection to the survival probability in the simpler semi-infinite geometry.

As was the case for the single Majorana chain, we employ a semi-infinite domain, now bounded by the lines $x=0$ and $y=0$.
The origin is evidently fixed by $\mathcal W$, and the boundaries map to the lines $\tilde y = \pm\lambda^2\tilde x$, where $\tilde x$ lies in the \eM direction and $\tilde y$ in \eP.
These boundaries delimit an absorbing wedge geometry with opening angle $\Theta$ given by $\cos\Theta = -\delta$, which runs from $\Theta = \pi/2$ at $\delta=0$ to $\Theta = \pi$ at $\delta=1$.
In terms of the wedge half-angle $\theta \equiv \Theta/2$, the domain boundaries are $\tilde y = \pm (\cot\theta)\tilde x$.
For easy reference, we collect some relationships between these geometric parameters:
\begin{gather}
\cos\Theta = -\delta~,~\sin\Theta = \sqrt{1-\delta^2}~, \\
\cos\theta = \sqrt{\frac{1-\delta}{2}}~,~\sin\theta = \sqrt{\frac{1+\delta}{2}}~,~\lambda = \sqrt{\cot\theta}~.
\end{gather}

The Green's function in the infinite wedge can be found from the free-space distribution by the method of images for opening angles $\Theta = \pi/m$, with $m$ a positive integer.
This entails $2m-1$ image charges with alternating sign, arranged symmetrically around the wedge apex.
However this approach is of limited use, as we need $\Theta \in [\frac\pi2,\pi)$, and instead we will use the Green's function known for arbitrary opening angle from an alternative solution.
In polar coordinates, with the wedge apex at radius $\rho=0$ and solution domain bounded by absorbing walls $G(\rho,\phi = 0,t) = G(\rho,\phi = \Theta,t) = 0$ (i.e., the angle $\phi$ is defined relative to one of the absorbing boundaries), we have \cite{carslaw1986conduction}
\begin{align}
&G(\rho,\phi,t;\rho_0,\phi_0) = \nonumber \\
&\quad\frac{e^{-(\rho^2+\rho_0^2)/4Dt}}{\Theta D t} \sum_{l=1}^\infty I_{l\nu}\!\left(\frac{\rho \rho_0}{2 D t}\right) \sin(l\nu\phi)\sin(l\nu\phi_0)~,
\label{eq:2d_greens}
\end{align}
where $\nu = \pi/\Theta$ and $I_{l\nu}$ is a modified Bessel function of the first kind:
\begin{equation}
I_s(x) = \sum_{m=0}^\infty \frac{(x/2)^{s+2m}}{m!\,\Gamma(s+m+1)}~.
\end{equation}
In the physical geometry the initial condition is $(x_0,y_0) = (\ev u,\ev v)$, where $\ev u = \ev v$ is again the result of each 1d RW taking one additional positive step according to the discrete microscopic distribution.
In the solution geometry this point maps to $\rho_0 \hat{e}_+$, where $\rho_0 = \sqrt2 \lambda \ev u$.
In polar coordinates the source point is $(\rho_0,\phi_0 = \theta)$.
Consequently, in Eq.~\eqref{eq:2d_greens} the factor $\sin(l\nu\phi_0)$ vanishes for even $l$ and for odd $l$ is equal to a sign $(-1)^{(l-1)/2}$.
As in the 1d case, we work in the scaling regime at late times $t$, where we are able to extract the leading power-law behavior.
Again, spatial integrals are regulated by the exponential factor, which decays fast enough to suppress errors arising at large $\rho$.
Because $\nu \in (1,2]$ the leading behavior requires only the $l=1$, $m=0$ term in the double sum, and sets $e^{-\rho_0^2/4Dt} \to 1$.

The survival probability is determined from the Green's function by integration over the wedge.
Explicitly, in the scaling limit
\begin{align}
S(t) &= \int \rho\,d\rho\,d\phi \,G(\rho,\phi,t;\rho_0,\phi_0=\theta) \\
&= \frac{\int_0^\Theta d\phi \sin(\nu\phi)}{\Theta \Gamma(\nu+1) Dt}  \int_0^\infty \rho\,d\rho\,e^{-\rho^2/4Dt}\left(\frac{\rho \rho_0}{4 D t}\right)^\nu \\
&= \frac{2\,\Gamma(\frac\nu2)}{\pi \Gamma(\nu)} \left(\frac{\rho_0}{\sqrt{4 D t}}\right)^\nu~.
\end{align}
The survival exponent depends on the opening angle as 
\begin{equation}
S(t) \sim t^{-\pi/2\Theta}~.
\end{equation}
This result for a RW in a 2d wedge is in fact well known \cite{fisher1988reunions,redner2001guide,bray2013persistence}.
As $\Theta$ is a function of the correlation coefficient $\delta$, continuously varying behavior of this type is in agreement with the numerical observations in Sec.~\ref{sec:mf2}.
Specifically, again relying on the naive assumption that the two ends of the chain decimate independently, the likelihood of this pairing scales as $[S(L)]^2 \sim L^{-\pi/\Theta}$, which matches the known end-to-end scaling exponents $\eta^\e_z = 2$ for the uncorrelated model at $\delta=0$ and $\eta^\e_z = 1$ for $\delta=1$.

Our strategy for rigorously bounding the probability of end-to-end decimation occurring on both chains using the infinite wedge results is analogous to that of Sec.~\ref{sec:rw_bounds_1d}.
From the Green's function we establish that at late times a constant fraction of surviving RWs are suitable for subsequent gluing to contribute to this probability, being found in a specified target window, using a shadow window to exclude dangerous trajectories.
By gluing the ends of two RWs at time $t=L/2$ we establish bounds on the power law.
We will use the notation of the previous section, namely $\mathfrak I$ and $\oplus$, to refer to the generalizations of the relevant transformations to 2d.

In particular, we can write an upper bound immediately.
Any 2d RW of length (duration) $L$ corresponding to two locally correlated Majorana chains can be decomposed into half-chains of length $L/2$ as $\m = \m_1 \oplus \mathfrak I \m_2$, as in the 1d case.
$\m_1$ and $\m_2$ may be valid surviving trajectories in their semi-infinite wedge, and some will produce end-to-end decimations on both physical Majorana chains described by the 2d RW \m.
Trajectories that do not decompose in this way into surviving half-chains will not satisfy Condition \ref{rw:cond_int}.
Because not every pair of surviving $\m_1$ and $\m_2$ will do so either, the probability is upper-bounded as $p_\e(L) \leq S(L/2)^2 \sim L^{-\pi/\Theta}$.

\begin{figure}[ht]
\includegraphics[width=\columnwidth]{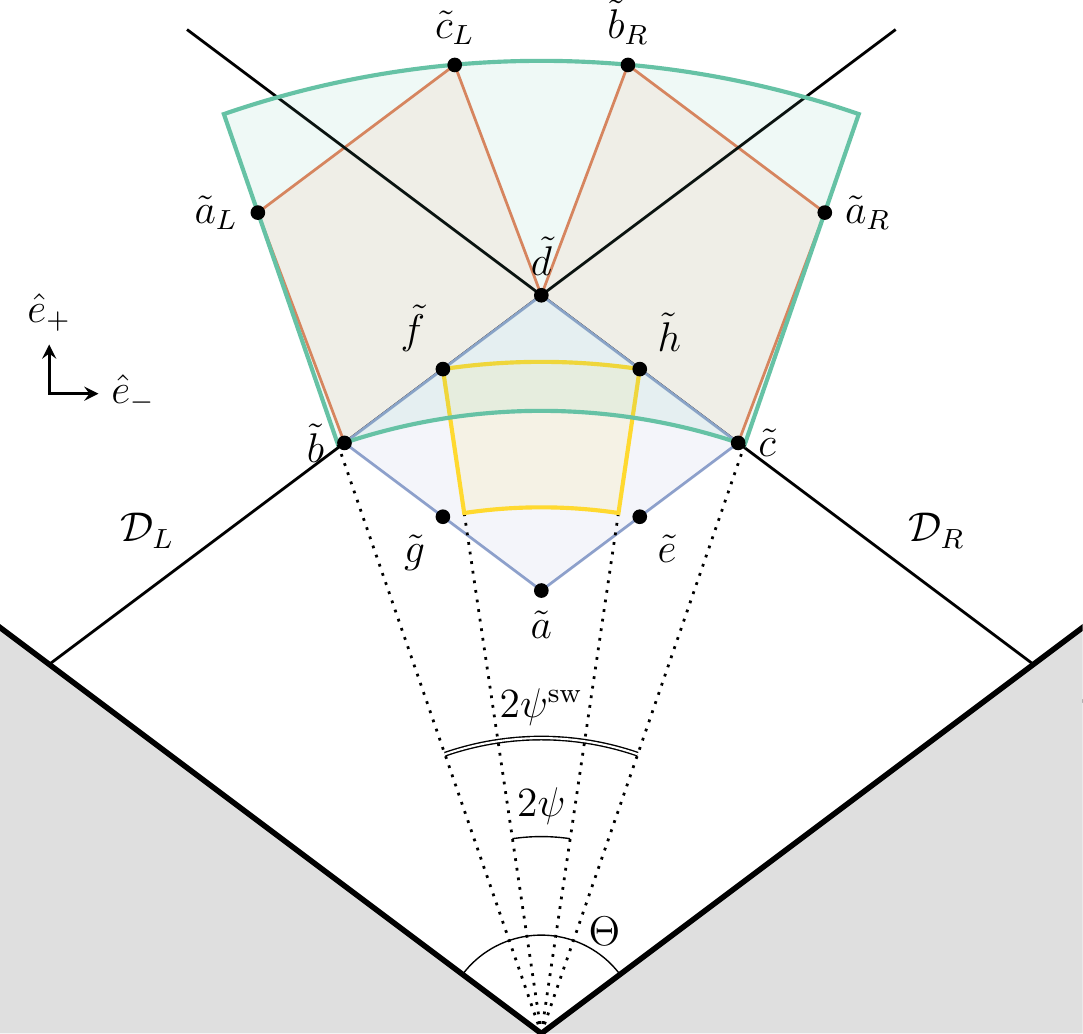}
\caption{\label{fig:shadow_window_2d} The solution geometry is illustrated for the 2d RW problem in the wedge with opening angle $\Theta$, found from the correlation coefficient by $\cos \Theta = -\delta$.
The exact target window is colored in blue, and the sector defining the easier integration subregion for the target in yellow.
The two components of the shadow window are found by reflecting the exact target window across the lines $\mathcal D_L$ and $\mathcal D_R$ and are colored in orange, with the easier bounding shadow integration region, which necessarily covers these areas, in green.
}
\end{figure}

Now in order to prove a lower bound on $p_e(L)$, let $\alpha$ and $\beta$ be positive constants, $\alpha < \beta \leq 2\alpha$, and define the target window for a 2d RW at time $t$ to be the square $(x,y) \in [\alpha\sqrt{Dt},\beta\sqrt{Dt}]\times [\alpha\sqrt{Dt},\beta\sqrt{Dt}]$.
In the physical geometry the window is a square; however, when mapped to the solution geometry the window becomes a parallelogram.
The corners $\{a,b,c,d\}$ map to
\begin{align}
\{\tilde a,\tilde b,\tilde c,\tilde d\} = \sqrt{\frac{Dt}{2}}\, \Big\{&2\alpha\lambda\eP~,~\frac{\alpha-\beta}{\lambda}\eM+(\alpha+\beta)\lambda\eP~, \nonumber \\
&\frac{\beta-\alpha}{\lambda}\eM + (\alpha+\beta)\lambda\eP~,~2\beta\lambda\eP\Big\}~.
\label{eq:solution_window}
\end{align}
as illustrated in Fig.~\ref{fig:shadow_window_2d}.
Treating this exact shape in the polar coordinates of Eq.~\eqref{eq:2d_greens} is complicated; instead we define an integration volume that is a subset of the target window, with the same $t$ scaling, but which leads to a simpler bound.
Consider the midpoints of the edges of the target window in the solution geometry, which we denote $\{\tilde e,\tilde f,\tilde g,\tilde h\}$.
They describe the four corners of a rectangle, symmetric about the line $\phi = \theta$, with edges in the directions $\eM$ and $\eP$ (see Fig.~\ref{fig:shadow_window_2d}).
We define an integration domain bounded by radial values $\rho_+$ (of points $\tilde f$ and $\tilde h$) and $\rho_-$ (of $\tilde e$ and $\tilde g$), and the angular deviation $\psi$ of points $\tilde f$ and $\tilde h$ from the midline $\phi = \theta$.

The proof that this ``sector'' geometry is indeed a subvolume of the target domain for any opening angle $\Theta < \pi$ can be seen by drawing a picture.
The specific integration bounds can be found straightforwardly from Eq.~\eqref{eq:solution_window}, but the crucial property is their scaling with $t$.
Define the radial limits as $\rho_\pm = C_\pm(\alpha,\beta,\delta) \sqrt{Dt}$; the angular integration half-width $\psi = \psi(\alpha,\beta,\delta)$ turns out to be purely geometric, with no $t$ dependence.
Again extracting the leading behavior for late times $t$, the fraction of surviving paths whose position at time $t$ is in the integration window is
\begin{align}
p^\td_\w(\alpha,\beta,\delta) &= \frac{1}{S(t)} \int_{\rho_-}^{\rho_+} \rho\,d\rho\int_{\theta-\psi}^{\theta+\psi} d\phi\, G(\rho,\phi,t;\rho_0,\theta) \\
&= \frac{4 \sin(\nu\psi)}{\nu \Gamma(\frac\nu2)} \mathcal I(\alpha,\beta,\delta)~,
\label{eq:p_window_uncorrected_2d}
\end{align}
where 
\begin{equation}
\mathcal I(\alpha,\beta,\delta) = \int_{C_-/2}^{C_+/2}du\,e^{-u^2} u^{\nu+1}~.
\end{equation}
So $p^\td_\w$ is indeed a constant, determined only by the correlation coefficient $\delta$ and the constants $\alpha$ and $\beta$.

As was the case for the 1d RW, the calculation above includes a ``dangerous'' contribution which should be subtracted in order to lower-bound the decimation probability by subsequent gluing of half-chains $\m_1$ and $\m_2$.
Again we upper-bound this contribution by calculating the fraction in a shadow window.
We consider those paths to be dangerous which ever cross the lines $x = \beta\sqrt{Dt}$ or $y = \beta\sqrt{Dt}$ in the physical space before returning to the target window at time $t$.
In the solution geometry these lines map to
\begin{align}
\mathcal D_R &:~\lambda\,\tilde x + \frac{1}{\lambda}\tilde y - \beta\sqrt{2 Dt} = 0~, \\
\mathcal D_L &:~-\lambda\,\tilde x + \frac{1}{\lambda}\tilde y - \beta\sqrt{2 Dt} = 0~.
\end{align}
We define the boundary for dangerous trajectories piecewise as (see Fig.~\ref{fig:shadow_window_2d})
\begin{equation}
\mathcal D(\phi) = \begin{cases} \mathcal D_R~,& 0 < \phi \leq \theta \\ \mathcal D_L~,& \theta < \phi < \Theta~. \end{cases}
\end{equation}

Suppose a trajectory with time parameter $t'$ makes $q$ crossings of $\mathcal D$ at times $\{t_1,\ldots,t_q\}$ at various points $\{(\rho_1,\phi_1),\ldots,(\rho_q,\phi_q)\}$ before returning to the target window at time $t' = t$.
After its last crossing at $(\rho_q,\phi_q)$, it must stay within the allowed region for times $(t_q,t]$.
We transform the trajectory by reflecting the partial RW for times $t' \in (t_q,t]$ about the component of $\mathcal D$ that was crossed at $t'=t_q$, either $\mathcal D_R$ if $\phi_q \in (0,\theta]$ or $\mathcal D_L$ if $\phi_q \in (\theta,\Theta)$.
This is the counterpart in 2d to the 1d transformation $\mathfrak T$.
Because the step distribution in the solution geometry is isotropic, the transformed path has the same probability as the dangerous original.
(The reflection must be performed in the solution geometry, and does not commute with $\mathcal W$.)
The shadow window in this case has two components, which are disconnected for $\Theta < \frac{2\pi}{3}$ but overlap for $\Theta > \frac{2\pi}{3}$.
Note that overlap of the mapped regions does not introduce the possibility of double-counting, as the full dangerous and transformed trajectories are uniquely related.

The corners $\tilde c$ and $\tilde d$ of the target window lie on line $\mathcal D_R$, and $\tilde b$ and $\tilde d$ on line $\mathcal D_L$.
Thus we need only reflect $\tilde a$ and $\tilde b$ about $\mathcal D_R$, and $\tilde a$ and $\tilde c$ about $\mathcal D_L$.
The coordinates of the points reflected about $\mathcal D_R$ are
\begin{align}
\{ \tilde a_R,&\,\tilde b_R\} = \sqrt{2 Dt}\,\times \nonumber \\
&\Bigg\{\frac{\lambda (\beta-\alpha)}{\cosh(2\ln\lambda)}\eM + \left( \frac{\beta - \alpha}{\lambda\cosh(2\ln\lambda)} + \alpha \lambda\right)\eP~, \nonumber \\
&\qquad\left( \frac{\lambda(\beta-2\alpha)}{\cosh(2\ln\lambda)} + \frac{\alpha-\beta}{\lambda} \right)\eM \nonumber \\
&\qquad\quad+ \left( \frac{\beta-2\alpha}{\lambda\cosh(2\ln\lambda)} + \lambda(\alpha+\beta) \right)\eP \Bigg\},
\end{align}
with similar forms for $\tilde a_L$ and $\tilde c_L$.
The four-sided figures described by the exact shadow window are evidently complicated.
As with the target window, we bound the area using a sector which scales in the same way, however in this case an upper bound is required.
The upper limit $\rho^\sw_+$ is the radial coordinate of points $\tilde c_L$ and $\tilde b_R$, and the lower limit $\rho^\sw_-$ is that shared by the corners $\tilde b$ and $\tilde c$.
The angular half-width is the maximum of the angular half-widths of points $\tilde c$ and $\tilde a_R$; this depends on the specific value of $\Theta$.
Again we find integration limits $\rho^\sw_\pm = C^\sw_\pm(\alpha,\beta,\delta)\sqrt{Dt}$, and $\psi^\sw = \psi^\sw(\alpha,\beta,\delta)$.

Based on the previous calculation, $p^\td_\sw(\alpha,\beta,\delta) = \frac{4 \sin(\nu\psi^\sw)}{\nu \Gamma(\frac\nu2)} \mathcal I^\sw(\alpha,\beta,\delta)$ and the corrected fraction is
\begin{align}
p^{\td,\mathrm{corr}}_\w(\alpha,\beta,\delta) &= p^\td_\w(\alpha,\beta,\delta) - p^\td_\sw(\alpha,\beta,\delta) \\
&= \frac{4}{\nu\Gamma(\frac\nu2)} \big(\sin(\nu\psi)\mathcal I-\sin(\nu\psi^\sw)\mathcal I^\sw\big)~.
\end{align}
By working explicitly through the algebra one can verify that $p^{\td,\mathrm{corr}}_\w$ is positive for all values of $\delta \in [0,1)$, e.g., for the choice $\alpha=1$, $\beta=2$.

Now, taking $t=L/2$, for any such $\m_1$ and $\m_2$ we can construct a RW which satisfies Condition \ref{rw:cond_int} for end-to-end decimation in the quantum chain as $\m = \m_1 \oplus \mathfrak I \m_2$.
Therefore a lower bound on the simultaneous end-to-end decimation probability is given by $p_\e \geq [p^{\td,\mathrm{corr}}_\w S(L/2)]^2 \sim L^{-\pi/\Theta}$.
In combination with the upper bound, this shows that the power law exponent controlling end-to-end decimation probability (and consequently $\eta^\e_z$) varies continuously with $\delta$ as
\begin{equation}
\eta^\e_z = \pi/\arccos(-\delta)~.
\label{eq:boundary_z_exp}
\end{equation}

\subsubsection{Bulk decimation probability in two locally correlated Majorana chains}

Once again we can extend the result for end-to-end decimation $p_\e$---requiring that both Majorana chains satisfy Condition \ref{rw:cond_int}---to the bulk likelihood $p_\b(r)$ (for two fixed spins separated by $r$) by considering also Condition \ref{rw:cond_ext}.
We first write a lower bound on the bulk pair decimation probability by identifying exterior RWs which are guaranteed to satisfy Condition \ref{rw:cond_ext} when properly adjoined to an interior RW of the type used for the lower bound on $p_\e$ in the previous section.
Specifically, we restrict to exterior RWs with endpoints at time $t \equiv r$ (for concreteness, but any constant multiple of $r$ would do as well)
within a particular sector (specified below) in the solution geometry.
In the physical geometry, $\Delta(r) = \beta\sqrt{2 D r}$ is an upper bound on the total deviation of each of the 1d RWs $\m_x$ and $\m_y$ described by the 2d interior RW $\m_\mathrm{int}$.

One way to guarantee the bulk decimation is to require that each of the physical 1d RWs described by each of the exterior 2d RWs $\m_\mathrm{ext,L}$ and $\m_\mathrm{ext,R}$ survive, and exceed $\Delta(r)$ at $t=r$.
A point $(\rho,\phi)$ in the solution geometry corresponds to
\begin{equation}
x = \frac{\rho \sin(\Theta-\phi)}{\sqrt{\sin\Theta}}~,~y = \frac{\rho \sin(\phi)}{\sqrt{\sin\Theta}}
\label{eq:2d_physical_coords}
\end{equation}
in the physical geometry.
Employing angular integration limits $\phi \in (\theta-\psi,\theta+\psi)$, where $\psi$ can be chosen to be the same value used for $\m_\mathrm{int}$, sufficient radial limits for our purposes are $\rho^\mathrm{ext}_- = \Delta(r)\sqrt{\sin\Theta}/\sin(\theta-\psi)$ and $\rho^\mathrm{ext}_+ \to \infty$ (noticing that $\sin(\theta-\psi) \leq \sin(\theta+\psi)$ for all $\psi \in [0,\theta]$).
From the calculation of the previous section there is a constant probability $\kappa(\alpha,\beta,\delta)$ that any surviving RW lies in a window bounded by $\rho \in [\rho^\mathrm{ext}_-,\rho^\mathrm{ext}_+]$ and $\phi \in [\theta-\psi,\theta+\psi]$ at $t=r$.
Such a RW has deviation at least $\Delta(r)$ in the physical $x$ and $y$ coordinates and thus as either $\m_\mathrm{ext,L}$ or $\m_\mathrm{ext,R}$ is suitable for satisfying Condition \ref{rw:cond_ext} for bulk decimation when properly adjoined to $\m_\mathrm{int}$ as constructed previously; thus $p_\b(r) \geq p_\e(r) [\kappa S(t=r)]^2 \sim r^{-2\pi/\Theta}$.

Similar to the case of a single Majorana chain, for an upper bound we make use of the probability $\omega(\rho,\phi;A)$ of a RW with spatial coordinates $(\rho',\phi')$ reaching radius $\rho' = A$ in the wedge given a starting point $(\rho,\phi)$.
This probability follows Laplace's equation $\nabla^2 \omega = 0$, now with boundary conditions $\omega(\rho,\phi=0) = \omega(\rho,\phi=\Theta) = 0$, $\omega(\rho = A,\phi) = 1$.
Assuming a separable solution $\omega(\rho,\phi) = R(\rho)T(\phi)$, we find that for the angular coordinate the solutions are $T_n(\phi) = \sin (n \nu \phi)$, $n = 1,2,3,\ldots{}$, where as before $\nu = \pi/\Theta$.
For the radial coordinate
\begin{equation}
\rho^2 \frac{\partial^2 R}{\partial\rho^2} + \rho \frac{\partial R}{\partial\rho} - (n \nu)^2 R = 0~,
\end{equation}
which has solutions of the form $R_n(\rho) = \rho^{\pm n \nu}$.
Determining the constants from the boundary conditions,
\begin{equation}
\omega(\rho,\phi;A) = \sum_{\substack{n=1\\n~\text{odd}}}^{\infty} \frac{4}{n \pi} \left(\frac \rho A\right)^{n \nu} \sin(n \nu \phi).
\end{equation}
Along the relevant line $\phi = \theta$, the probability simplifies to 
\begin{equation}
\omega(\rho,\phi=\theta;A) = \frac 4\pi \arctan\left[\left(\frac{\rho}{A} \right)^\nu\right] \leq \frac{4}{\pi} \left(\frac{\rho}{A} \right)^\nu ~.
\end{equation}

In order to write an upper bound on the bulk decimation probability, we consider a full RW satisfying both Conditions assembled from an $\m_\mathrm{int} = \m_1 \oplus \mathfrak I \m_2$, where each of $\m_1$ and $\m_2$ must survive until $t \equiv r/2$, along with exterior RWs $\m_\mathrm{ext,L}$ and $\m_\mathrm{ext,R}$ which must reach a particular radial coordinate (determined from $\m_\mathrm{int}$ as specified below) without being absorbed.
Suppose that $\m_1$ and $\m_2$ terminate at coordinates $(\rho_1,\phi_1)$ and $(\rho_2,\phi_2)$, which define the deviations of the physical RWs $(\Delta_{x,1}, \Delta_{y,1})$, and $(\Delta_{x,2}, \Delta_{y,2})$ according to Eq.~\eqref{eq:2d_physical_coords}.
The full deviation of the interior walk $\m_\mathrm{int}$ in the physical coordinates is $(\Delta_x, \Delta_y) = (\Delta_{x,1}+\Delta_{x,2}, \Delta_{y,1}+\Delta_{y,2})$ and the physical 1d RWs described by $\m_\mathrm{ext,L}$ and $\m_\mathrm{ext,R}$ must exceed the corresponding $\Delta_x$ or $\Delta_y$ before being absorbed.
For this to be the case it is necessary, but not sufficient, that the 2d exterior RWs each survive in the wedge until reaching radial coordinate $A \equiv \sqrt{\sin\Theta} \min(\Delta_x, \Delta_y)$ in the solution geometry.
Defining for $\m_1$ and $\m_2$ similar $A_1 \equiv \sqrt{\sin\Theta} \min(\Delta_{x,1}, \Delta_{y,1})$ and $A_2 \equiv \sqrt{\sin\Theta} \min(\Delta_{x,2}, \Delta_{y,2})$, we note that $A \geq A_1, A_2$.
The probability of finding two such $\m_\mathrm{ext,L}$ and $\m_\mathrm{ext,R}$ given the terminating locations of $\m_1$ and $\m_2$ is
\begin{align}
p(\mathrm{ext}|\rho_1,\phi_1,\rho_2,\phi_2) &\leq \omega(\rho_0,\theta;A)^2 \\
&\leq \left(\frac4\pi\right)^2 \left(\frac{\rho_0}{A}\right)^{2\nu} \\
&\leq \left(\frac4\pi\right)^2 \left(\frac{\rho_0}{A_1}\right)^\nu \left(\frac{\rho_0}{A_2}\right)^\nu.
\end{align}

Then, integrating over the distribution of the interior half-chain coordinates,
\begin{align}
p_\b(r) &= \int \rho_1\,d\rho_1\,d\phi_1\, G(\rho_1,\phi_1,r/2;\rho_0,\theta)\times \nonumber \\
&\qquad \int \rho_2\,d\rho_2\,d\phi_2\, G(\rho_2,\phi_2,r/2;\rho_0,\theta)\times \nonumber \\
&\qquad\qquad p(\mathrm{ext}|\rho_1,\phi_1,\rho_2,\phi_2) \\
&\leq \Bigg[\frac8\pi \int_0^\infty \rho_1\,d\rho_1\int_0^\theta d\phi_1\times \nonumber \\
&\qquad\qquad G(\rho_1,\phi_1,r/2;\rho_0,\theta) \left(\frac{\rho_0}{\rho_1 \sin \phi_1}\right)^\nu \Bigg]^2 \\
&= \left(\frac{16 \mathcal I_\phi}{\pi^2 \Gamma(\nu)} \right)^2 \left( \frac{\rho_0^2}{2 D r}\right)^{2\nu}.
\end{align}
We restrict to the right half-wedge, as the integrand is symmetric about $\phi=\theta$.
The angular integral is
\begin{align}
\mathcal I_\phi = \int_0^\theta d\phi_1 \frac{\sin(\nu\phi_1)}{(\sin \phi_1)^\nu}~,
\end{align}
which converges for $\Theta > \pi/2$, equivalently $\delta > 0$.
(The exponent we are bounding is known at $\delta = 0$, and follows from the result of Sec.~\ref{sec:rw_1d_bulk}.)

Combining the upper and lower bounds, we prove that $p_\b(r) \sim r^{-2\nu}$, and the bulk correlations exponent for two locally correlated Majorana chains with parameter $\delta$ is
\begin{equation}
\eta_z = 2\pi/\arccos(-\delta)~.
\label{eq:bulk_z_exp}
\end{equation}

\subsection{Numerical SDRG study}
The final results of this section, Eqs.~\eqref{eq:boundary_z_exp} and \eqref{eq:bulk_z_exp}, are in qualitative agreement with the quantum simulations of Sec.~\ref{sec:mf2} for relatively short Majorana chains, and are consistent with previously-known results at the points $\delta=0,1$, where the locally-correlated model describes the random uncorrelated XY and perfectly correlated XX IRFPs.
For further verification we implement the SDRG update Eq.~\eqref{eq:sdrg_xy} directly for two Majorana chains with locally-correlated terms, and are able to access larger system sizes.
This also allows us to study the bulk $C^\perp(r)$ power laws, which are not analytically tractable in the mapping to RWs used in the preceding subsections.

\begin{figure}[ht]
\includegraphics[width=\columnwidth]{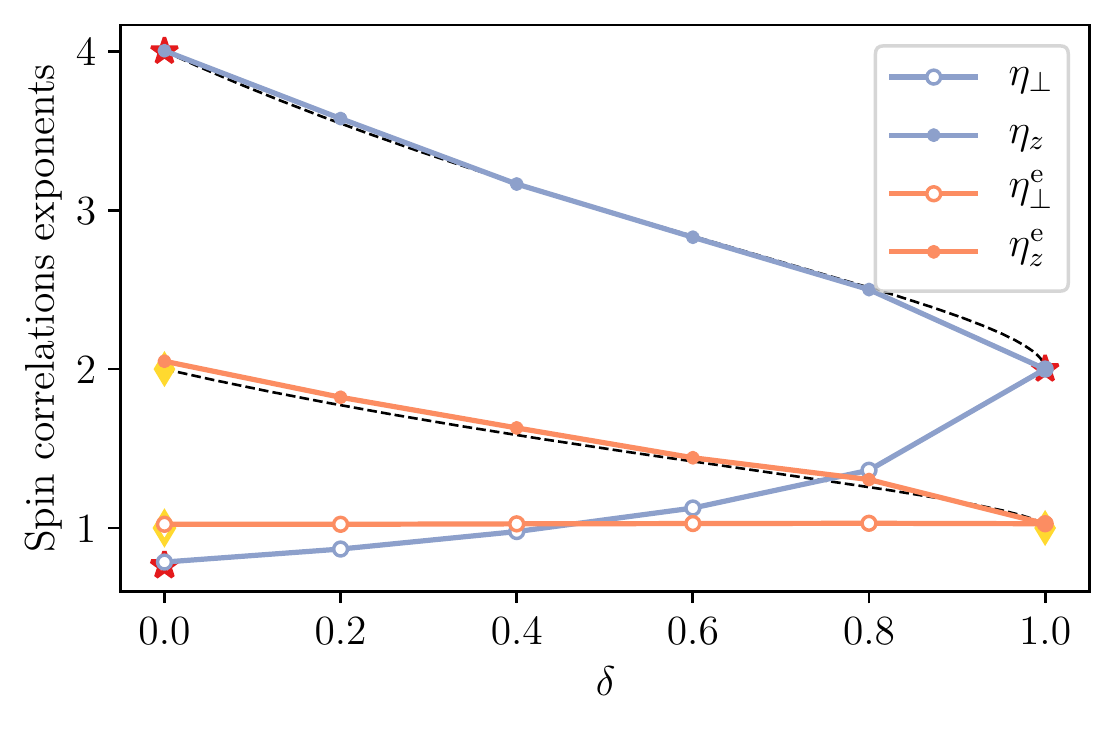}
\caption{\label{fig:sdrg_exponents} Numerical SDRG data are shown for two locally-correlated Majorana chains, with the end-to-end and bulk decimation probability exponents---equivalent to $\eta^\e_z$ and $\eta_z$, respectively, in the quantum model---compared to the analytic forms Eqs.~\eqref{eq:boundary_z_exp} and \eqref{eq:bulk_z_exp} (dashed lines).
Also shown are critical exponents $\eta_\perp^\e$ and $\eta_\perp$ measured in the numerical SDRG, as well as red stars indicating known values of bulk correlations exponents at $\delta=0$ and $1$, and yellow diamonds indicating known values of end-to-end correlations exponents.
The end-to-end correlations data were taken from 1\,000\,000 disorder realizations each for system sizes up to $N=128$, and the bulk correlations data were taken from 100\,000 disorder realizations at system size $N=256$, utilizing the middle half of each of the two Majorana chains.
}
\end{figure}

The numerically extracted exponents are shown in Fig.~\ref{fig:sdrg_exponents}.
The bare correlation coefficient $\delta$ may become slightly renormalized from the lattice scale definition in Eqs.~\eqref{eq:mf2_jx}--\eqref{eq:mf2_jy} compared to the meaning in the continuum 2d RW treatment in Sec.~\ref{sec:rw_bounds_2d}, but these simulations are in good agreement with the analytic forms for $\eta_z^\e(\delta)$ and $\eta_z(\delta)$ we obtained.
While we have precise analytical knowledge only of the critical exponents $\eta_z^\e$ and $\eta_z$, we observe that $\eta_\perp$ also varies continuously.
In contrast, $\eta_\perp^\e = 1$ for any value of $\delta$, by the argument presented in Sec.~\ref{sec:sdrg_correlations}.

\section{Fixed points for the interacting model}
\label{sec:fp_interacting}

In Sec.~\ref{sec:sdrg_correlated} we performed a study of the behavior of critical exponents under a varying degree of local correlations in a random free-fermion model.
Despite the lack of tractable SDRG flow equations, we showed that the local correlation controlled by $\delta$ is a marginal perturbation which tunes along a line of IRFPs.
In the present section we advance the perspective that this line of non-interacting fixed points in fact also controls the long-distance behavior of the interacting model for small $J^z$ strength below the transition to the $z$-AFM phase.

To do so requires a study of the SDRG at intermediate stages, taking into account more general terms produced by the interactions.
Equation \eqref{eq:sdrg_xyz_h} describes the result of an initial decimation, but eventually descendant terms will be frequent and must also be taken into account.
We change our conventions here from those of Sec.~\ref{sec:sdrg_xyz} for convenience: namely, we denote the Majorana chains by $\I,\II$ rather than $\X,\Y$; and by a gauge transformation (described at the end of Sec.~\ref{sec:sdrg_majorana}) we set the signs of $h^\I_n,h^\II_n > 0$ for all $n = 1,\ldots,N-1$, and $K_n \equiv K_{n,n} < 0$.

In order to capture the effect of iterated decimations, we observe that in Eq.~\eqref{eq:sdrg_xyz_h} descendants of the form $K_{n,m} (i \gamma^\I_n \gamma^\I_{n+1}) (i \gamma^\II_m \gamma^\II_{m+1})$ are produced, which generalize the $K_n$ of Eq.~\eqref{eq:sdrg_h_int}.
We enlarge the space of couplings to include all such terms, with initial distribution $K_{n,m} = 0$, $n \neq m$.
If the average $K \equiv |\ev{K_{n,m}}|$ can be considered to be a small parameter (for weak interactions $K < |\ev{h_n}|$), the higher-fermion term in Eq.~\eqref{eq:sdrg_xyz_h} appears at order $O(K^2)$ and can thus be neglected.
We will demonstrate that the space of couplings including all $K_{n,m}$ is closed under RG flow up to $O(K)$, and that the structure of the signs is preserved.
Furthermore, we will show that the strength of the $K$ terms decreases in some sense relative to the $h$ terms, suggesting that interactions are irrelevant, at least in the neighborhood of the free-fermion fixed point.

Following the approach of Sec.~\ref{sec:sdrg}, denote the largest term as $H_0 = ih^\I_k \gamma^\I_k \gamma^\I_{k+1}$ and associate with the eigenstates of this term a complex fermion $f^\dag_0 = \frac 12 (\gamma^\I_k + i \gamma^\I_{k+1})$ with projectors $\pi^+ = f_0 f^\dag_0$ and $\pi^- = f^\dag_0 f_0$ into the even- and odd-parity sectors, or the high- and low-energy eigenstates, respectively, of $H_0$.
The off-diagonal terms in the Schrieffer-Wolff treatment share exactly one Majorana operator with $H_0$:
\begin{align}
V_\od &= i h^\I_{k-1} \gamma^\I_{k-1} \gamma^\I_k + i h^\I_{k+1} \gamma^\I_{k+1} \gamma^\I_{k+2} \\
&\quad+\sum_{m=1}^{N-1} \big( K_{k-1,m} (i\gamma^\I_{k-1} \gamma^\I_k) (i\gamma^\II_m \gamma^\II_{m+1}) \nonumber \\
&\qquad\qquad+ K_{k+1,m} (i\gamma^\I_{k+1} \gamma^\I_{k+2})(i\gamma^\II_m \gamma^\II_{m+1}) \big)~.
\end{align}
Separating $V_\od$ into symmetry sectors, we find that
\begin{widetext}
\begin{align}
\pi^+ H \pi^- &= \left[ \left(h^\I_{k-1} + \sum_m K_{k-1,m} (i\gamma^\II_m\gamma^\II_{m+1})\right)i\gamma^\I_{k-1} + \left(h^\I_{k+1} + \sum_m K_{k+1,m}(i \gamma^\II_m \gamma^\II_{m+1})\right) \gamma^\I_{k+2} \right] f_0 \\
&\equiv ( i h^{\I,\text{int}}_{k-1} \gamma^\I_{k-1} + h^{\I,\text{int}}_{k+1} \gamma^\I_{k+2} ) f_0~, \\
\pi^- H \pi^+ &= \left[ \left(h^\I_{k-1} + \sum_m K_{k-1,m} (i\gamma^\II_m\gamma^\II_{m+1})\right) i\gamma^\I_{k-1} - \left(h^\I_{k+1} + \sum_m K_{k+1,m}(i \gamma^\II_m \gamma^\II_{m+1})\right) \gamma^\I_{k+2} \right] f^\dag_0 \\
&\equiv ( i h^{\I,\text{int}}_{k-1} \gamma^\I_{k-1} - h^{\I,\text{int}}_{k+1} \gamma^\I_{k+2} ) f^\dag_0~.
\end{align}
\end{widetext}
We make use of the ``interacting couplings'' notation used also in Sec.~\ref{sec:sdrg_xyz} to connect with the non-interacting case, but here it is not evident that these couplings---which are really operators---all commute.
Nevertheless, a suitably generalized version of Eq.~\eqref{eq:sdrg_xy} implements the Schrieffer--Wolff transformation:
\begin{widetext}
\begin{align}
H' &= H_0 + V_\d + \frac{(h^{\I,\text{int}}_{k-1})^2 + (h^{\I,\text{int}}_{k+1})^2}{2 h^\I_k} (i \gamma^\I_k \gamma^\I_{k+1}) + \frac{h^{\I,\text{int}}_{k-1} h^{\I,\text{int}}_{k+1} + h^{\I,\text{int}}_{k+1} h^{\I,\text{int}}_{k-1}}{2 h^\I_k} (i \gamma^\I_{k-1} \gamma^\I_{k+2}) \\
&= H_0 + V_\d + (i \gamma^\I_k \gamma^\I_{k+1}) \left[\frac{(h^\I_{k-1})^2 + (h^\I_{k+1})^2}{2 h^\I_k} + \frac{h^\I_{k-1}}{h^\I_k}\sum_m K_{k-1,m} (i \gamma^\II_m \gamma^\II_{m+1}) + \frac{h^\I_{k+1}}{h^\I_k}\sum_m K_{k+1,m} (i \gamma^\II_m \gamma^\II_{m+1}) \right] \nonumber \\
&\quad+ (i\gamma^\I_{k-1} \gamma^\I_{k+2}) \left[\frac{h^\I_{k-1} h^\I_{k+1}}{h^\I_k} + \frac{h^\I_{k-1}}{h^\I_k} \sum_m K_{k+1,m} (i \gamma^\II_m \gamma^\II_{m+1}) + \frac{h^\I_{k+1}}{h^\I_k} \sum_m K_{k-1,m} (i \gamma^\II_m \gamma^\II_{m+1}) \right] + O(K^2)~.
\label{eq:sdrg_h_int_full}
\end{align}
\end{widetext}
The effective terms in the first line of Eq.~\eqref{eq:sdrg_h_int_full} (and the first term of the second line) are $h$-type, with positive coefficients in the low-energy sector of $H_0$ where $\ev{i \gamma^\I_k \gamma^\I_{k+1}} = -1$.
Conversely, the remaining terms in the second line are $K$-type (recalling that $\gamma^\I_{k-1}$ and $\gamma^\I_{k+2}$ become adjacent after the decimation of $\gamma^\I_k$ and $\gamma^\I_{k+1}$), and have coefficients with negative signs.
One sees that the signs of the initial distributions, namely $h_n^{\I,\II} > 0$ and $K_{n,m} < 0$, are maintained during the RG flow, and it is evident from Eq.~\eqref{eq:sdrg_h_int_full} that these types of terms are closed under the SDRG up to $O(K)$.

As a measure of the evolution of the relative strength of $K$ terms to $h$ terms under this RG step, we compare the renormalized $K^\mathrm{eff}_{k-1,m}$ to the geometric mean of the proximate $h$ terms $h^{\I,\mathrm{eff}}_{k-1}$ and $h^\II_m$:
\begin{equation}
\frac{K^\mathrm{eff}_{k-1,m}}{\sqrt{h^{\I,\mathrm{eff}}_{k-1} h^\II_m}} = \sqrt\frac{h^\I_{k-1}}{h^\I_k} \frac{K_{k+1,m}}{\sqrt{h^\I_{k+1} h^\II_m}} + \sqrt\frac{h^\I_{k+1}}{h^\I_k} \frac{K_{k-1,m}}{\sqrt{h^\I_{k-1} h^\II_m}}
\end{equation}
We see that if such ratios are small to begin with, i.e., $K_{k+1,m}/\sqrt{h^\I_{k+1} h^\II_m}, K_{k-1,m}/\sqrt{h^\I_{k-1} h^\II_m} \ll 1$ before the decimation, they will likely become even smaller under the RG flow if the disorder in the Majorana hoppings is strong, so that $h_{k-1}^\I, h_{k+1}^\I \ll h_k^\I$.
This suggests that if the $h$ terms are dominant initially, they will be even more so during the SDRG and will asymptotically constitute the entirety of the decimations.

The diagonal terms which contain both decimated Majoranas are
\begin{equation}
\sum_m K_{k,m} (i\gamma_k^\I \gamma_{k+1}^\I) (i\gamma_m^\II \gamma_{m+1}^\II) ~.
\label{eq:K2h}
\end{equation}
Upon decimation, setting $\ev{i \gamma^\I_k \gamma^\I_{k+1}} = -1$ in the ground state gives $O(K)$ contributions to the Majorana hoppings in the other chain, $h_m^{\II,\text{eff}} = h_m^\II - K_{k,m}$. Given the opposite signs of the $h$ and $K$ couplings, this increases the overall strength of the remaining Majorana hoppings.
This is the local SDRG analog of the ``mean field'' of Eqs.~\eqref{eq:meanfield_Jx} and \eqref{eq:meanfield_Jy} where the $J^z$ interactions renormalize the $J^x$ and $J^y$ couplings by strengthening and correlating them, as was already noted in Sec.~\ref{sec:sdrg_xyz} and discussed in Sec.~\ref{sec:mf}.
Here we note that including these renormalizations of the $h$ couplings only improves our arguments for the persistence of the dominance of these couplings over the $K$ couplings.

The terms omitted from Eq.~\eqref{eq:sdrg_h_int_full} at $O(K^2)$ are the following:
\begin{widetext}
\begin{align}
\frac{1}{2 h^\I_k} &(i \gamma^\I_k \gamma^\I_{k+1}) \left[\sum_m (K_{k-1,m}^2 + K_{k+1,m}^2) + \sum_{m,l\neq m,m\pm1} (K_{k-1,m} K_{k-1,l} + K_{k+1,m} K_{k+1,l}) (i \gamma^\II_m \gamma^\II_{m+1}) (i \gamma^\II_l \gamma^\II_{l+1})\right] \nonumber \\
&+ \frac{1}{h^\I_k} (i \gamma^\I_{k-1} \gamma^\I_{k+2}) \left[ \sum_m K_{k-1,m}K_{k+1,m} + \sum_{m,l\neq m,m\pm1} K_{k-1,m}K_{k+1,l} (i \gamma^\II_m \gamma^\II_{m+1}) (i \gamma^\II_l \gamma^\II_{l+1})\right]~.
\end{align}
\end{widetext}
The first terms in each line are corrections to the ground-state energy and the strength of the renormalized bond coupling on chain \I [which again preserves the sign structure and strengthens this hopping compared to the leading contribution in Eq.~\eqref{eq:sdrg_h_int_full}].
Along with these, four-fermion terms within chain \II and six-fermion inter-chain terms appear at $O(K^2)$.
The former are expected to be ultimately irrelevant, based on previous studies of a single Majorana chain realized in the quantum Ising model \cite{fisher1994random}.
However these four-fermion terms and the six-fermion terms will produce yet more complicated descendants in subsequent RG steps, and there will also be ``degradation'' processes leading to fewer-fermion terms, including renormalization of the two-fermion terms, similar to the discussion after Eq.~\eqref{eq:K2h} \cite{monthus2018strong}.
In this case we must rely on the perturbative argument to justify dropping them, viewing them as irrelevant other than feeding into strictly marginal correlations among the effective Majorana hoppings in the two chains.

Together with the understanding of the locally correlated XY model in the previous section, this leads us to propose the following picture for the critical XYZ chain along the line separating the $x$-AFM and $y$-AFM phases.
For small $\tilde J^z$, this critical line is actually controlled by the line of free Majorana fixed points with locally correlated hoppings characterized in Sec.~\ref{sec:sdrg_correlated}.
The effect of the interactions $J^z$ in the original model with no correlations among the couplings ($\delta=0$) is to develop such correlations among the renormalized $J^x$ and $J^y$ couplings under RG while the $J^z$ couplings flow to zero.
The ultimate degree of such correlations (i.e., the fully renormalized parameter $\delta_\text{eff}$) then determines the long-distance power laws in the average spin correlation functions.
We further conjecture that this persists for all $\tilde J^z < \tilde J^z_\text{crit} = 1$ below the transition to the $z$-AFM phase.
While we do not have perturbative control close to this transition, any alternative would require yet another transition below $\tilde J^z_\text{crit}$ which we did not observe and consider to be less natural.
Note that in this scenario the transition to the $z$-AFM phase is controlled by a different \emph{non-free-fermion} fixed point, and we do not have access to this $S_3$-symmetric fixed point in the present study.
We will further discuss the above conjecture, its corollaries and possible tests, as well as open questions in the concluding section.

\section{Discussion}
\label{sec:discussion}

In this paper, motivated by the observations of \citet{slagle2016disordered}, we have performed a study of the low-energy properties of the random XYZ model using unbiased numerics.
We focus on the line separating the $x$-AFM and $y$-AFM phases, which exhibits statistical symmetry between $J^x$ and $J^y$ couplings.
At all points allowing comparison our results are in general agreement with the previous findings of Ref.~\cite{slagle2016disordered} which used SBRG and presumed critical MBL physics at arbitrary energy density.
Our results strongly suggest that---regardless of the behavior of highly excited states---there is quantum critical behavior in the ground state and the critical line is described by IRFPs with continuously varying critical exponents in the disorder-averaged correlation functions.
Perhaps surprisingly, a Hartree--Fock mean-field theory treating the $J^z$ interaction terms as perturbations around the random XY (free-fermion) fixed point yielded results that are qualitatively rather consistent with the full interacting model at small to moderate $J^z$ couplings, including continuously varying power laws.
This is in contrast to the clean case, where the mean field model is not qualitatively accurate due to divergences in the perturbation theory \cite{giamarchi2004quantum}.

The locally correlated XY effective model, introduced with the idea of distilling the essential feature of the mean field theory, again exhibited continuously varying critical exponents, which we were able to establish numerically in larger sizes than for the XYZ chain.
Because of the particular free-fermion form of this effective model, we were able to treat it in the SDRG using the random walk formulation in two dimensions.
By making use of a connection between survival probability and the structure of decimation in the RG, we showed analytically that critical exponents for end-to-end and bulk $C^z$ spin correlations vary continuously as the coupling correlation parameter $\delta$ is tuned, and we also observed varying exponents in the bulk $C^\perp$ correlations by running the SDRG numerically.
This result singles out and proves one of the scenarios of \citet{fisher1994random} that random anisotropy is strictly marginal along the critical line connecting the random XX and random XY fixed points; that is, there is a line of fixed points connecting the XX and XY IRFPs as sketched in Fig.~\ref{fig:rg_cartoon1}.

Motivated by the successful understanding of the locally correlated XY model, we revisited the SDRG for the full interacting XYZ chain in the regime of small interactions and proposed a scenario where these interactions are irrelevant, but during the initial flows they generate effective correlations between the local $J^x$ and $J^y$ couplings (i.e., Majorana hopping amplitudes on the two chains).
Such flows are sketched in Fig.~\ref{fig:rg_cartoon2}.
These local correlations in the free-fermion couplings then lead to non-universal power laws in the average spin correlations: this is our story for the continuously varying criticality in the XYZ spin chain.

We note that continuously varying critical exponents were previously observed in IRFPs associated with correlated disorder by \citet{rieger1999random}, however in a qualitatively different setting than ours.
Specifically, disordered fixed points perturbed by the introduction of long-range correlations $\sim r^{-a}$ to the disorder in the random transverse-field Ising chain exhibit critical indices varying continuously with $a$ for $a < 1$.
Their setting has only one Majorana chain and the correlated disorder is within the chain.  
Also, in their case the $\psi$ exponent varies continuously, which reflects a different character of the corresponding ``random walker'' imprinted by the long-range correlations in the disorder.

Non-universal exponents at IRFPs were also observed in cases with very broad (singular) distributions of random couplings~\cite{karevski2001random, krishna2020beyond}.
This again occurs already in a single chain and has varying exponent $\psi$, and the variation can be traced directly to the singularity in the probability distribution of the microscopic couplings, while the exponents are universal for non-singular probability distributions.

The XYZ chain studied here is different from the above examples with varying exponents in that there are no long-range correlations or singular distributions input into the microscopic disorder.
In this way the continuously varying exponents are intrinsic to this system rather than imprinted extrinsically.
What is important in the XYZ chain is that we have two simultaneously critical Majorana chains whose couplings become locally correlated.
This insight may be useful when looking for other IRFPs with intrinsic continuously varying critical indices.

We conclude by returning to the discussion of the proposed scenario for the fully interacting XYZ chain.
This scenario is based on the conjecture that the four-fermion and higher terms are irrelevant other than feeding into correlations between the Majorana hoppings.
While this is plausibly justified for small interactions in Sec.~\ref{sec:fp_interacting}, we have not fully proved it and the status for intermediate interactions is less certain.
In this respect, it would be useful to carry out a systematic numerical SDRG study of the fully interacting problem (e.g., using the scheme of \citet{monthus2018strong}) keeping track of all generated interactions as well as allowing decimations of the interaction terms when they happen to be the strongest.
If our scenario is correct, we should see the interaction terms progressively decreasing relative to the Majorana hoppings.
One should be able to perform such a study also directly in the spin variables using the SBRG approach of \citet{slagle2016disordered} projected onto the ground state branch, e.g., as used in Ref.~\cite{duque2021topological} in a different problem.
Employing the insights gained here, it should be helpful to interpret various Pauli string terms generated under the SBRG as either Majorana hoppings or specific multi-fermion interactions.
The SBRG can also be indispensable for studying the putative $S_3$-symmetric fixed point describing the transition to the $z$-AFM phase, as a possible new IRFP that is not tractable with available analytical tools.

Thinking about a broader phase diagram, our work suggests that it could be fruitful to add another parameter ``axis'' and study the XYZ chain with locally correlated $J^x$ and $J^y$ couplings in the bare model (analogous to parameter $\delta$ in the correlated XY model), in addition to the interactions $J^z$.
Figure~\ref{fig:rg_cartoon2} shows this parameter space, and constitutes a mild abuse inasmuch as it serves as both a phase diagram and a picture of RG flows, the latter of which occur in space not captured by just the two parameters.
In the space shown, the bare $\delta = 0$ corresponds to the present XYZ chain, with the transition from the critical phase to the $z$-AFM phase at the $S_3$ symmetric point, marked XYZC in Fig.~\ref{fig:rg_cartoon2}.
On the other hand, $\delta = 1$ corresponds to the XXZ chain studied in the original work by \citet{fisher1994random}.
For $\tilde J^z$ below some threshold value, the XXZ spin chain is critical and controlled by the free-fermion XX point, while for larger $\tilde J^z$ it undergoes a transition to the $z$-AFM phase.
Fisher concluded that this transition is controlled by the so-called XXZC fixed point which is essentially random singlet--like, also marked in Fig.~\ref{fig:rg_cartoon2}.
An interesting question is the nature of the transition to the $z$-AFM phase driven by the $\tilde J^z$ coupling as we vary the disorder correlation parameter from $\delta=1$ (XXZC fixed point) to the statistically isotropic XYZC fixed point.
This line is marked with a question mark in Fig.~\ref{fig:rg_cartoon2}, and one possibility is that it is also described by a line of fixed points, but we cannot at present exclude other scenarios.
We leave these questions for future investigations, noting that the possibility of novel IRFPs is quite tantalizing and worth further exploration.

\acknowledgements

We acknowledge helpful discussions with Jason Alicea, Matteo Ippoliti, Cheng-Ju Lin,  Sanjay Moudgalya, Gil Refael, Kevin Slagle, and Christopher White.
We are also grateful for earlier collaboration with Thomas Vidick on the RRG which led us to look for new applications of this method.
O.M.~is also grateful for previous collaborations with Kedar Damle, David Huse, and Daniel Fisher on the IRFPs which provided important background for this project.
This work was supported by National Science Foundation through grant DMR-2001186. 
Part of this work was performed at Aspen Center for Physics, which is supported by National Science Foundation grant PHY-1607611.

\bibliography{refs}

\end{document}